\pdfoutput=1

\newcommand{\degr}{$^\circ$}
\newcommand{\caproman}[1]{\uppercase\expandafter{\romannumeral#1}}
\newcommand{\bs}{$\backslash$}
\newcommand{\BGsite}{http://users.physics.harvard.edu/\raisebox{-4pt}{$\tilde{\;\;}$}gottschalk}
\newcommand{\LPPCsite}{http://lppc.physics.harvard.edu/gottschalk/}

\newcommand{\BGmail}{bgottsch\,@\,fas.harvard.edu}

\documentclass{article}

\title{\bf Nuclear halo of a 177\,MeV proton beam in water:\\theory, measurement and parameterization}
\author{Bernard Gottschalk\thanks{Harvard University Laboratory for Particle Physics and Cosmology, 18 Hammond St., Cambridge, MA 02138, USA (corresponding author, \BGmail)}, ~Ethan W. Cascio\thanks{Francis H. Burr Proton Therapy Center, Mass. General Hospital,
30 Fruit Street, Boston, MA, USA 02114}, ~Juliane Daartz\thanks{Francis H. Burr Proton Therapy Center} ~and Miles S. Wagner\thanks{Mevion Medical Systems Inc., 300 Foster St.
Littleton, MA 01460, USA}}

\usepackage{amsmath}
\usepackage{latexsym}
\usepackage{graphicx}

\begin{document}

\maketitle


\begin{abstract}
\noindent
The dose distribution of a monoenergetic pencil beam in a water tank may conveniently be divided into a {\em core}, a {\em halo} and an {\em aura}. The core consists of primary protons which suffer multiple Coulomb scattering (MCS) and slow down by multiple collisions with atomic electrons (Bethe-Bloch theory). Their number slowly decreases because of nuclear interactions, which feed the halo and aura. 

The halo consists of charged secondary particles, many but not all of them protons, from elastic interactions with H, elastic and inelastic interactions with O, and nonelastic interactions with O. After reviewing these reactions, we show by kinematics that the maximum radius of the halo must be roughly 1/3 the beam range. 

The aura, extending many meters, consists of neutral secondaries (neutrons and $\gamma$-rays) and the charged particles they set in motion.

We measure the core/halo using a 177\,MeV test beam offset in a water tank. The beam monitor is a fluence calibrated plane parallel ionization chamber (IC) and the field chamber, a dose calibrated Exradin T1, so the dose measurements are absolute (MeV/g/p). We perform depth-dose scans at ten displacements from the beam axis ranging from 0 to 10\,cm, adjusting the beam current and the sensitivity of the Keithley electrometer as required. The dose spans five orders of magnitude, and the transition from halo to aura is clearly visible.

We perform model-dependent (MD) and model-independent (MI) fits to the data. The MD fit separates the dose into core (electromagnetic), elastic nuclear, nonelastic nuclear, and aura components. It reveals roughly how much each process contributes. It has 25 parameters, and the goodness of fit (global rms ratio measurement/fit) is 15\%. The MI fit uses cubic splines in depth and radius. The goodness of fit is 9\%, and the fit is conceptually simpler and much more portable. Some of the residual error is probably due to small errors in positioning the T1 chamber.

We review the literature to see what other measurements and fits exist, and what Monte Carlo simulation has shown so far. All others use (in varied notations) the parametrization of Pedroni et al. \cite{pedroniPencil}. Several have improved on his Gaussian transverse distribution of the halo, but all retain his $T(w)$, the radial integral of the depth-dose, multiplying both core and halo terms and motivating measurements with large `Bragg peak chambers' (BPCs).

We argue that this use of $T(w)$, which by its definition includes energy deposited by nuclear secondaries, is incorrect. $T(w)$ should be replaced in the core term, and in at least one part of the halo, by $S_\mathrm{em}/\rho_\mathrm{water}$\,, a purely electromagnetic mass stopping power. It follows that BPC measurements are unnecessary, and irrelevant to parameterizing the pencil beam.

In clinical practice, comprehensive measurements and fits to the pencil beam are likely to be most important in the dosimetry of small fields involving relatively few pencils. For larger fields the already common `field size factor' approach should be adequate.
\end{abstract}

\clearpage
\tableofcontents

\clearpage
\section{Introduction}
The dose distribution of a proton pencil beam in solid or liquid matter has three parts which we will call the {\em core}, the {\em halo}
(after Pedroni et al. \cite{pedroniPencil}) and the {\em aura}. They overlap, but each has distinct spatial characteristics.

The {\em core} consists of primary protons which have suffered only electro\-magnetic (EM) interactions: slowing down by multiple collisions with atomic electrons, and multiple Coulomb scattering (MCS) by atomic nuclei. Its cross section is Gaussian, typically of the order of a centimeter, and increases with depth according to Fermi-Eyges theory \cite{preston,transport2012}. For an ideal pencil in homogeneous matter the transverse spreading can be written in closed form, and its maximum value in any material is proportional to range. For protons in water, $\sigma_\mathrm{max}=0.022\,\times$\;range \cite{preston}. Non-ideal pencils of finite initial size, angular divergence and emittance are easily accommodated in the Gaussian approximation \cite{transport2012}. The number of primaries in the core falls nearly linearly with depth \cite{CascioFC} due to nuclear interactions, which feed the halo and aura.

The {\em halo} consists of the charged secondaries from those interactions: elastic scattering on H, elastic and inelastic scattering on O and nonelastic scattering on O. The radius of the halo (which we will find is roughly one-third the beam range) is comparable to typical targets in radiotherapy, so it deposits dose both where it is wanted and unwanted. Being rather diffuse, it hardly affects the shape of the high-dose region, but the total energy residing in the halo must be taken into account to avoid field-size dependent errors in computing absolute dose \cite{pedroniPencil}.  

We will include multiple Coulomb scattered protons from H and O (the `single scattering tail' of the Moli\`ere distribution \cite{mcsbg}) in the halo even though, strictly speaking, these are primary protons which have only undergone EM interactions. Thus an alternative definition of the halo might be `anything arising from charged particles which falls outside the Gaussian of Fermi-Eyges theory'.

We exclude dose from beam contamination (such as degraded and scattered protons from beam profile monitor wires \cite{Sawakuchi2010}) entering from upstream. Such contamination can be avoided by appropriate beam line design and, if present, is more properly treated as a `secondary core' to be transported using Fermi-Eyges theory.

Lastly, the {\em aura} consists of neutral secondaries from inelastic and nonelastic nuclear interactions: $\gamma-$rays and neutrons. It extends many meters, pervading the patient, treatment room, shielding and facility and depositing unwanted dose in all. Its relative contribution to the high-dose region is negligible. Our measurements clearly reveal the transition between the halo and the aura.

At 177\,Mev about 12\% of the beam energy is lost to nonelastic nuclear reactions (Appendix \ref{app:energyLost}). It is thought \cite{pedroniPencil} that about one-third of that or 4\% goes into the aura.

The halo exists, of course, in both scattered and magnetically scanned beams, but by tradition it is treated differently. In scattering, it undoubtedly affects measurements and semiempirical models of the beam monitor output factor, but is usually not broken out explicitly. Only recently \cite{DaartzFieldSize2009} has its role in the field-size dependence of scattered beams been considered. By contrast, in pencil beam scanning (PBS) the desired absolute dose distribution is computed by summing over all pencils sufficiently near the point of interest. The halo must be explicitly included, to some adequate approximation, in the model of the pencil, and `sufficiently near' must be defined accordingly.

Therefore, with the advent of PBS facilities, a number of groups have published measurements and/or Monte Carlo studies of the halo. These have tended to focus narrowly on a clinical goal, correcting field size dependence of the absolute dose. Often, energy-dependent parameterizations (all complicated, and no two alike) are reported, at the expense of actual measurements. Beams and techniques designed for PBS treatment, not experiment, are used in the measurements. Though something is learned from each study, no complete overall picture of the halo has yet emerged.

Our goal is the opposite. Laying aside clinical concerns for the moment, we wish to obtain as complete as possible a picture of the halo at a single energy. As we define it, the halo in water must, after all, be universal: beyond some small radius, it is little affected by the core \cite{Peeler2012}. The measurement can be done with conventional equipment in a test beam. Eventually, of course, one will need to measure a number of energies similarly.

In addition to extensive raw data, which span five orders of magnitude, we will present a model-dependent (MD) and a model-independent (MI) fit. The MD fit is not very good; it is rather complicated and not very portable. It does, however, confirm what basic nuclear reactions are involved and how they determine the shape of the halo. It is complemented by the MI fit which is better, easy to apply to any data set, and yields the most convenient function $D(r,z)$ for exportation and further computation. By no means do we claim that {\em either} fit is the most efficient for any given task, and we report all our absolute dose measurements to invite further work. 

In the model-dependent fit and elsewhere we require the range-energy relation of protons in water. We will use the 1982 data of Janni \cite{janni82} because a growing body of experimental evidence \cite{moyers,Siiskonen2011,cascio07} suggests these are quite accurate, and better than ICRU Publication\,49 \cite{icru49} which gives ranges in water some 0.9\% smaller.

Ultimately, the best use of our measurements may be as a benchmark for Monte-Carlo (MC) electromagnetic (EM) and nuclear models. As we shall see, different reaction types are rather cleanly separated in the halo, which may make it easier to diagnose problems (if any) with the MC. The crucial advantage of the halo as an MC test lies in the fact that the {\em entire} halo (outside the core) is of nuclear origin. By contrast, in the large-field depth-dose distribution (Bragg peak) nuclear effects contribute \cite{pedroniPencil,bortfeld} but are of secondary importance.

The present work is organized as follows. Section \ref{sec:medium} is an overview of the medium energy reactions that dominate the halo. Their kinematics, also discussed, predict the halo size rather well. Section\,\ref{sec:measurement} describes our measurement. Key features are the use of a test beam whose  current can be adjusted over a wide range, an integral monitor rather than a reference chamber for normalization, and using scans in depth rather than transverse scans. Section \ref{sec:MD} describes the model-dependent fit and Section \ref{sec:MI}, the model-independent fit. 

Section \ref{sec:review} is a review of the literature. Section \ref{sec:discussion} is a general discussion. In it we take issue with the prevailing parameterization of the halo, and not just the Gaussian transverse profile, which has already been improved by others. More fundamentally, we dispute the role of integral depth dose in the parameterization and therefore, the pervasive role of the `Bragg peak chamber' (BPC). 

Throughout, we discuss each topic in rather general terms, leaving mathematical details to the Appendices.

\section{Medium-Energy Nuclear Reactions in Water}\label{sec:medium}

The halo (look ahead to Figs.\;\ref{fig:dmlg} and \ref{fig:dmlin}) will be found to have two distinctive components. The first is a Bragg-like peak due to elastic and inelastic scattering on H and O. It persists to quite large radii, as already implied by Fig.\;5 of Pedroni et al. \cite{pedroniPencil}. The second is a broad sideways dose enhancement, generally due to a mix of charged secondaries from nonelastic reactions on O. We expect sideways protons from the relatively simple reactions $^1$H(p,p)$^1$H and $^{16}$O(p,2p)$^{15}$N to form the outer envelope of this region. The final states of these two- and three-body processes are easily characterized. Doing so, to estimate the halo size, is the main object of this section. First, however, we will provide some background. The reader familiar with medium energy nuclear physics should skip to Section\,\ref{sec:haloRadius}.

Physics in the 70 to 300\,MeV regime was studied extensively in the 1950s and 60s, the era of the synchrocyclotron. Radiotherapy with protons in this energy range, already proposed in 1946 \cite{wilson}, was just beginning. At these energies the proton is an elementary particle. Its internal states cannot be excited; still less are its constituent quarks evident. Its de Broglie wavelength is smaller than the nucleus, pion production is negligible, and the nucleus is relatively transparent. It can easily knock protons, neutrons and clusters (such as $\alpha$\;particles) out of the nucleus, their binding energies being of order 10\,MeV. Relativistic kinematics are appropriate, though the nonrelativistic approximation is not terrible.

Coherent and incoherent reactions take place. In the former, elastic and inelastic scattering, the proton interacts with the nucleus as a whole and the recoiling residual nucleus is the same as the target, though it may be in an excited state. In the latter, the proton interacts with constituents of the target nucleus. The residual nucleus differs from the target nucleus but may nevertheless (through the magic of quantum mechanics) be left in its own ground state. These are nonelastic reactions, with the important subgroup of quasi-elastic (QE) (`quasi-free' or `knockout') reactions, in which just a single proton, neutron or cluster is knocked out. 

In water, elastic scattering on free hydrogen also occurs. Kinematically, it can be treated either as a special case of elastic or of QE scattering. In reactions with O, the recoil nucleus has a small fraction of the total energy and a very short range. By contrast, in scattering on H the projectile and target have the same mass. In the final state, they are indistinguishable, and separated by 90\degr.\footnote{~Relativistically, a few degrees less.}

We cite just two experimental papers from an extensive literature.

\subsection{Elastic and Inelastic Scattering on O}\label{sec:elastic}
\begin{figure}[p]
\centering\includegraphics[width=5.00in,height=3.5in]{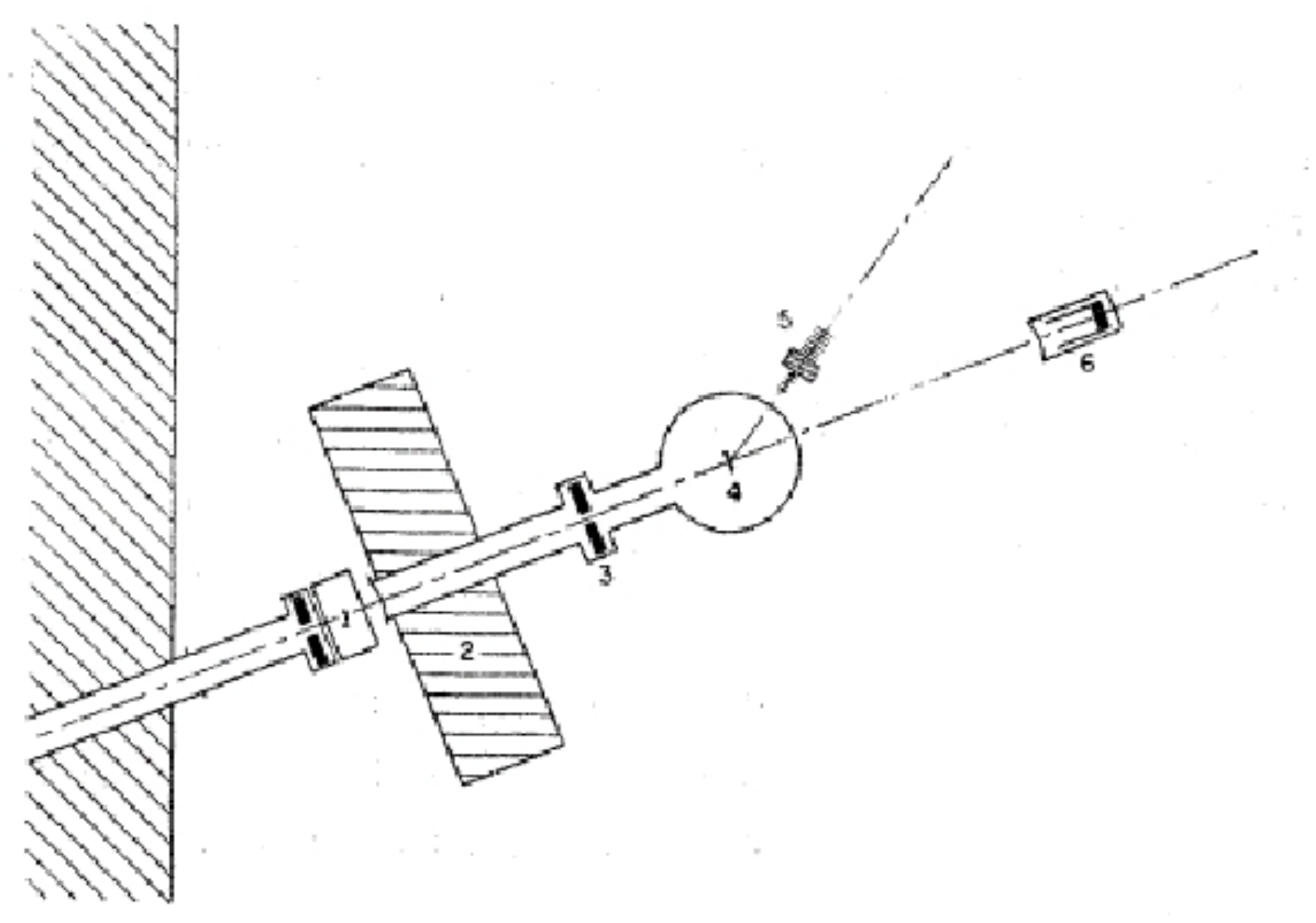}
\caption{Gerstein et al. \cite{Gerstein1957} Fig.\,1: `Experimental arrangement. (1) Ionization chamber; (2) lead shield; (3) defining slit; (4) scattering chamber and target; (5) telescope; (6) Faraday cup.'\label{fig:Gerstein1}}
\end{figure}

\begin{figure}[p]
\centering\includegraphics[width=2.91in,height=3.5in]{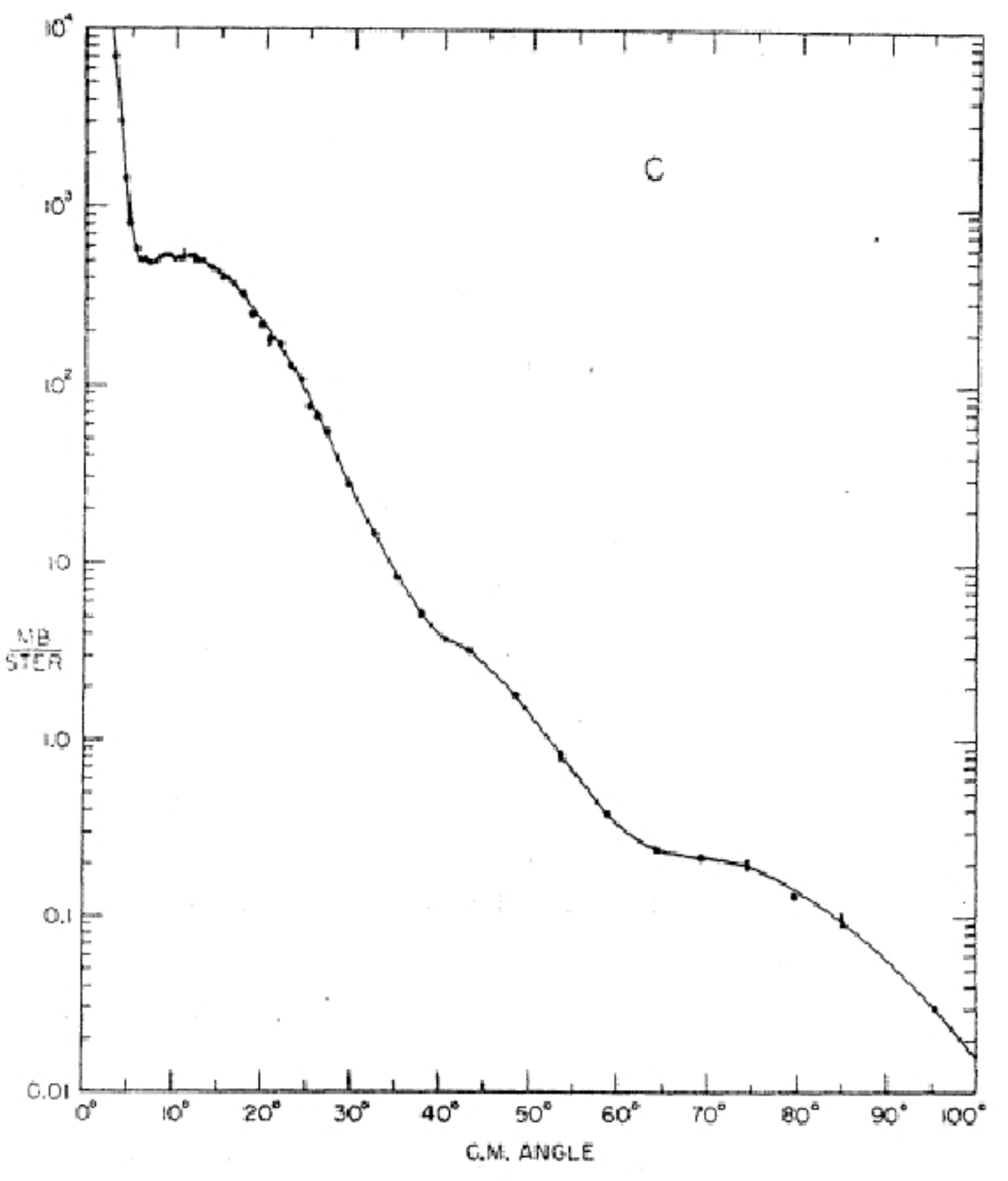} 
\caption{Gerstein et al. \cite{Gerstein1957} Fig.\,3: `Carbon elastic scattering differential cross section.'\label{fig:Gerstein3}}
\end{figure}

Gerstein et al. \cite{Gerstein1957} studied elastic scattering on various targets at the Harvard Cyclotron before it acquired a proper 160\,MeV external beam and long before it was converted to medical use. Their experiment was performed in a scattered-out 96\,MeV beam of 10$^7$ protons/sec and a collimated cross section of $0.5$\,in$\times1$\,in (Fig.\,\ref{fig:Gerstein1}). They did not study O but their C results (Fig.\,\ref{fig:Gerstein3}) will suffice since we are only interested in the general nature of the angular distribution. 

Elastic scattering obeys an optical model. The nucleus behaves like an absorbing disk with a diffuse boundary. The angular distribution has diffraction minima and maxima whose sharpness increases with atomic number. The sharp drop at small angles seen in Fig.\,\ref{fig:Gerstein3} is the single scattering tail of the Moli\`ere distribution \cite{mcsbg}. Here the proton interacts coherently with the nucleus, but via the Coulomb rather than the nuclear force. Since the final states of the two processes are indistinguishable, they interfere quantum mechanically leading to the `Coulomb interference' dip seen near 5\degr.  Thereafter the nuclear force takes over and the cross section falls at a more leisurely rate, with shallow diffraction features.

As concerns the halo, we need take away only two facts from Fig.\,\ref{fig:Gerstein3}. The angular dependence is exponential (not Gaussian), and it extends to angles far greater than multiple or single Coulomb scattering.

The systematics of inelastic scattering are similar. The residual nucleus is left in one of its many excited states, and usually decays by $\gamma$ emission. Excitation tends to happen at larger momentum transfers (larger angles) and will shift and broaden the Bragg-like peak in the halo at large radii. 

Elastic scattering results are always presented in the center-of-mass (CM) frame rather than the lab frame. For completeness Appendix\,\ref{app:CM} presents the transformation, even though the details do not matter much in the present context since we merely require the general shape of the angular distribution. For elastic scattering on O, the lab and CM angles are nearly the same.

\subsection{Quasi-elastic Scattering on O}\label{sec:QE}
In quasi-elastic (QE) scattering experiments e.g. $^{16}\mathrm{O}~(\mathrm{p},2\mathrm{p})~^{15}\mathrm{N}$ \ ($Q=-12.1$\,MeV) the final state consists of two protons and a recoiling residual nucleus. (The $Q$ of a reaction equals the rest energy of the reactants minus the rest energy of the products. If $Q$ is negative, kinetic energy is lost in the reaction.) Measuring the energies and directions of the outgoing protons allows us to compute the kinetic energy lost in the reaction (the binding energy of the target proton) as well as the momentum of the recoil nucleus. The latter is interesting for the following reason.

Nucleons confined to the nucleus are not at rest. They have finite momenta corresponding to a few MeV of kinetic energy. Their momentum distributions are given by Fermi's model or, more accurately for light nuclei, nuclear shell model waveforms. QE experiments are performed to study these  momentum distributions. In a simple picture of the reaction, the target nucleus is viewed as a two-body system, the target proton and the recoil-nucleus-to-be, with equal and opposite momenta since the whole is at rest in the lab. If we now assume that, during the reaction, neither the incident nor the ejected proton interact with the rest of the nucleus, the momentum of the residual nucleus is equal and opposite to that of the target proton before the reaction. The residual nucleus is often called a `spectator'.

Neglecting secondary interactions is an approximation, albeit a fairly good one for light nuclei and higher incident energies. Because secondary interactions do play some role, the observed distributions are called `distorted' momentum distributions.

\begin{figure}[p]
\centering\includegraphics[width=4.62in,height=3.5in]{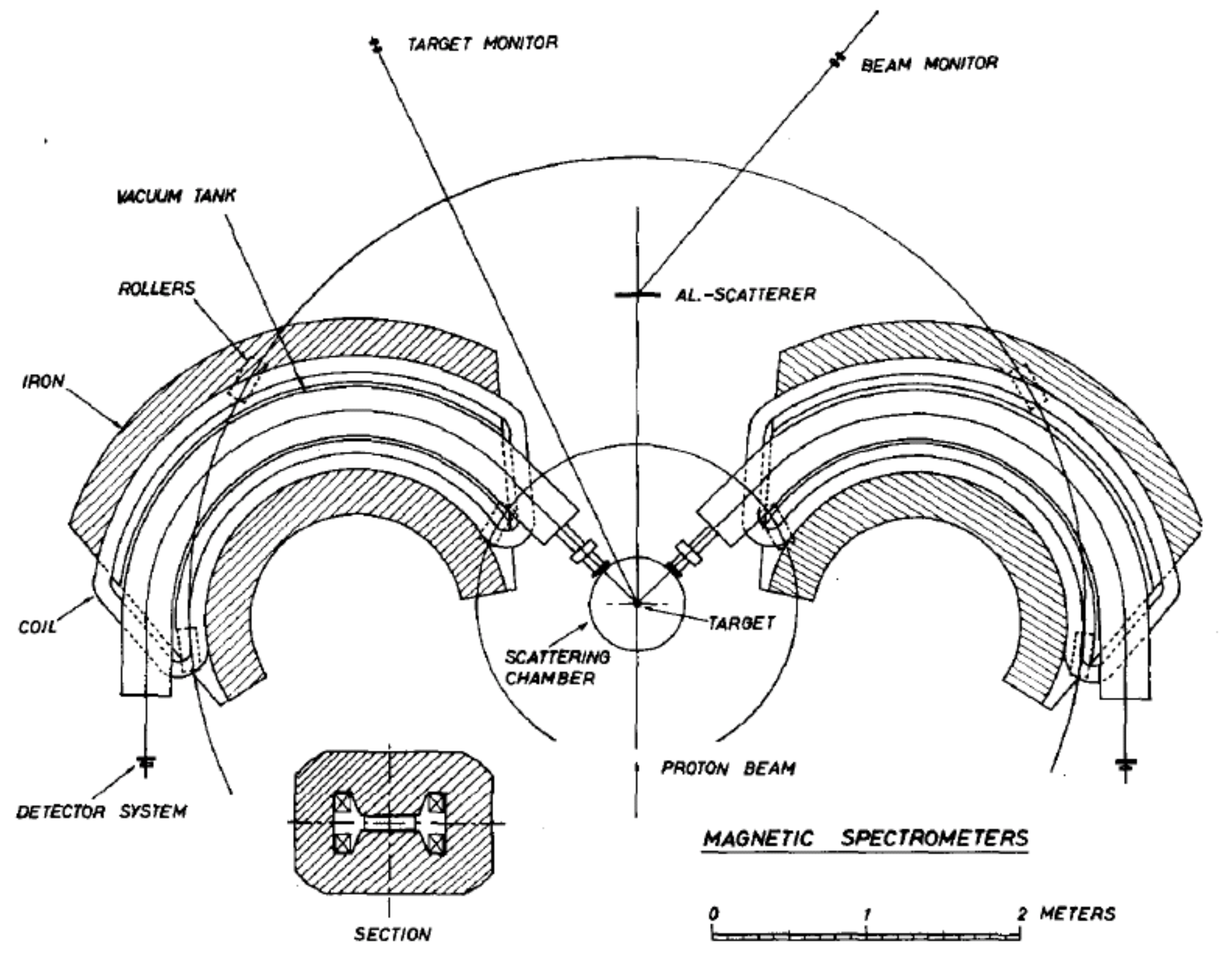}
\caption{Tyr\'en et al. \cite{Tyren1966} Fig.\,1: `Magnetic pair spectrometer for momentum analysis of protons in quasi-free proton-proton scattering.'\label{fig:Tyren1}}
\end{figure}

\begin{figure}[p]
\centering\includegraphics[width=6.06in,height=3.5in]{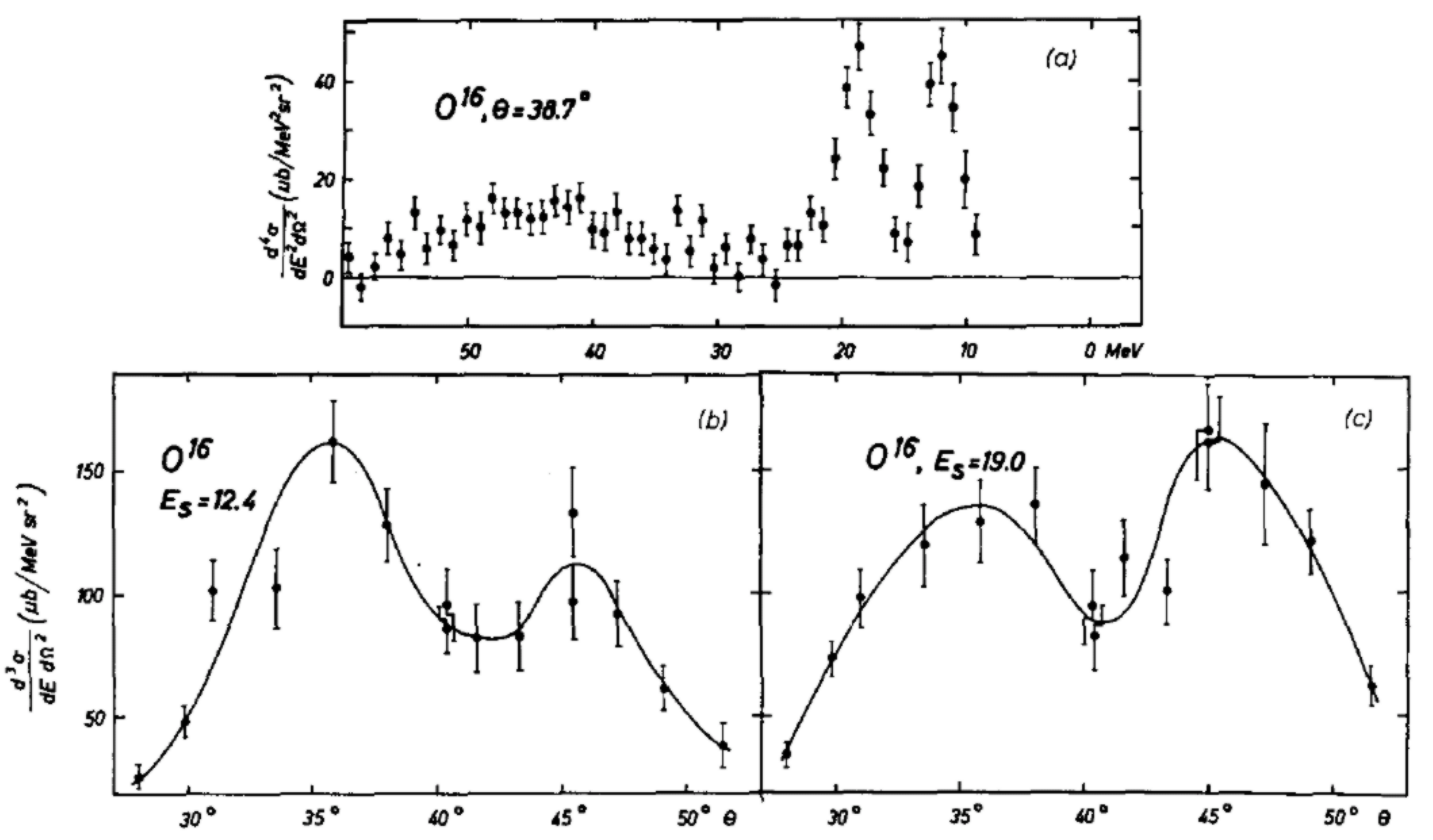} 
\caption{Tyr\'en et al. \cite{Tyren1966} Fig.\,12: `The summed-energy spectrum reveals the closed-shell structure of $^{16}$O. We see the well separated and sharp peaks corresponding to the knocking out of a p$_{1/2}$ proton (at 12.4\,MeV) and a p$_{3/2}$ proton (at 19.0\,MeV). The s-shell peak is very broad.'\label{fig:Tyren12}}
\end{figure}

The best data on $^{16}$O come from a massive experiment by Tyr\'en et al. \cite{Tyren1966} using magnetic spectrometers at the Chicago synchrocyclotron (Fig.\,\ref{fig:Tyren1}). According to the nuclear shell model $^{16}$O should have two protons in a p$_{1/2}$ angular momentum state, four in a p$_{3/2}$ state, and two in an s state. The distribution of missing kinetic energy (Fig.\,\ref{fig:Tyren12}, top) indeed shows three well separated groups. In Tyr\'en's geometry, the distribution of the separation angle of the two protons is equivalent to a distribution of the recoil nucleus momentum, with zero momentum near 40\degr. The dip there shows that p state protons, by virtue of their orbital angular momentum, have zero probability of being found with zero momentum. However, the dip is considerably filled in by secondary interactions and experimental resolution.

Later, we will need the binding energies and most probable momenta that can be derived from these results. They are $E_\mathrm{B}=12.4$\,MeV, $(pc)_5=75$\,MeV for the $p_{1/2}$ state protons, and  $E_\mathrm{B}=19.0$\,MeV, $(pc)_5=75$\,MeV for the $p_{3/2}$ protons. These numbers agree well with an earlier experiment at 160\,MeV \cite{BGthesis}.

\subsection{Elastic Scattering on H}
We will find that elastic scattering on H has the same general kinematics as (p,2p) scattering on O, and that is all we really need to know as far as the halo is concerned. Elastic scattering on H was studied extensively to determine the nucleon-nucleon interaction itself (rather than nuclear structure) and is therefore found in a separate category of papers. Wilson's monograph \cite{wilson1} includes a good experimental account. The mathematics of the center-of-mass (CM) transformation (which Wilson takes for granted) is covered in our Appendix\,\ref{app:CM}.

\subsection{Estimate of the Halo Radius}\label{sec:haloRadius}
We are now in a position to estimate the maximum radius of charged secondaries from nonelastic reactions. Though there are many such reactions, maximum range will obtain for the lightest (and incidentally most common) charged secondaries, protons, and for those reactions that distribute the available kinetic energy among the fewest secondaries, namely (p,2p) reactions. We now turn the reasoning of Section\,\ref{sec:QE} around, assuming the numbers given there as the most likely parameters of the recoil nucleus and deducing the energy and angle, hence transverse range, of one of the protons.\footnote{~The other behaves in a complementary way. Incoming and target protons are indistinguishable in the final state.} Elastic scattering on free H is a special case of the same calculation. Appendix\,\ref{app:QE} gives the mathematical details.

\begin{figure}[p]
\centering\includegraphics[width=4.57in,height=3.5in]{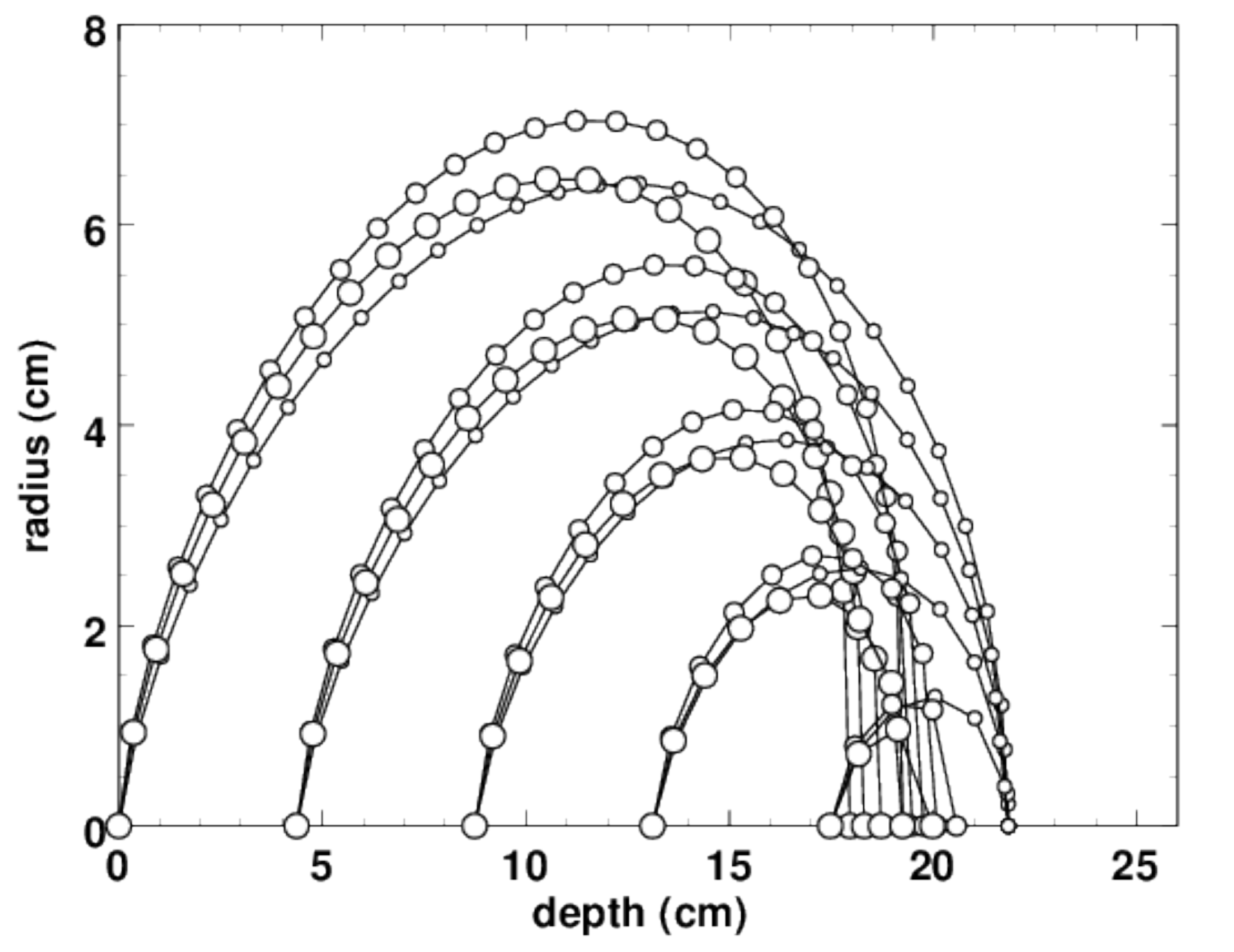}
\caption{Secondary proton stopping points in water for $T_1=180$\;MeV, reactions at five depths and $\theta_5=300$\degr.  Small, medium, large circles represent elastic, QE with $E_\mathrm{B}=12.4$\,MeV, $(pc)_5=75$\,MeV, and QE with $E_\mathrm{B}=19.0$\,MeV, $(pc)_5=75$\,MeV.\label{fig:haloBump}}
\end{figure}

Fig.\,\ref{fig:haloBump} shows the loci of secondary proton stopping points, using a 180\,MeV incident pencil beam, for elastic scattering on H and QE scattering on O using parameters given above and in the caption. We assume all reactions occur on the $z$ axis, ignoring the size of the pencil beam. In the QE case we have chosen the direction of the recoil nucleus to maximize the transverse range of the more energetic outgoing proton. In Fig.\,\ref{fig:haloBump} the reaction is assumed to take place at one of five points along the incident proton's path, corresponding to five `incident energies'. 

Since the target proton has a distribution of momenta of which we have only used the most likely, we must imagine a broad band around each of the QE cases. Still, the overall picture is clear. The most probable QE scatter adds only slightly to elastic scattering. We also observe that, because of the loss of kinetic energy to $E_\mathrm{B}$, elastic scattering on H eventually dominates at the largest radii.   

\begin{figure}[p]
\centering\includegraphics[width=4.57in,height=3.5in]{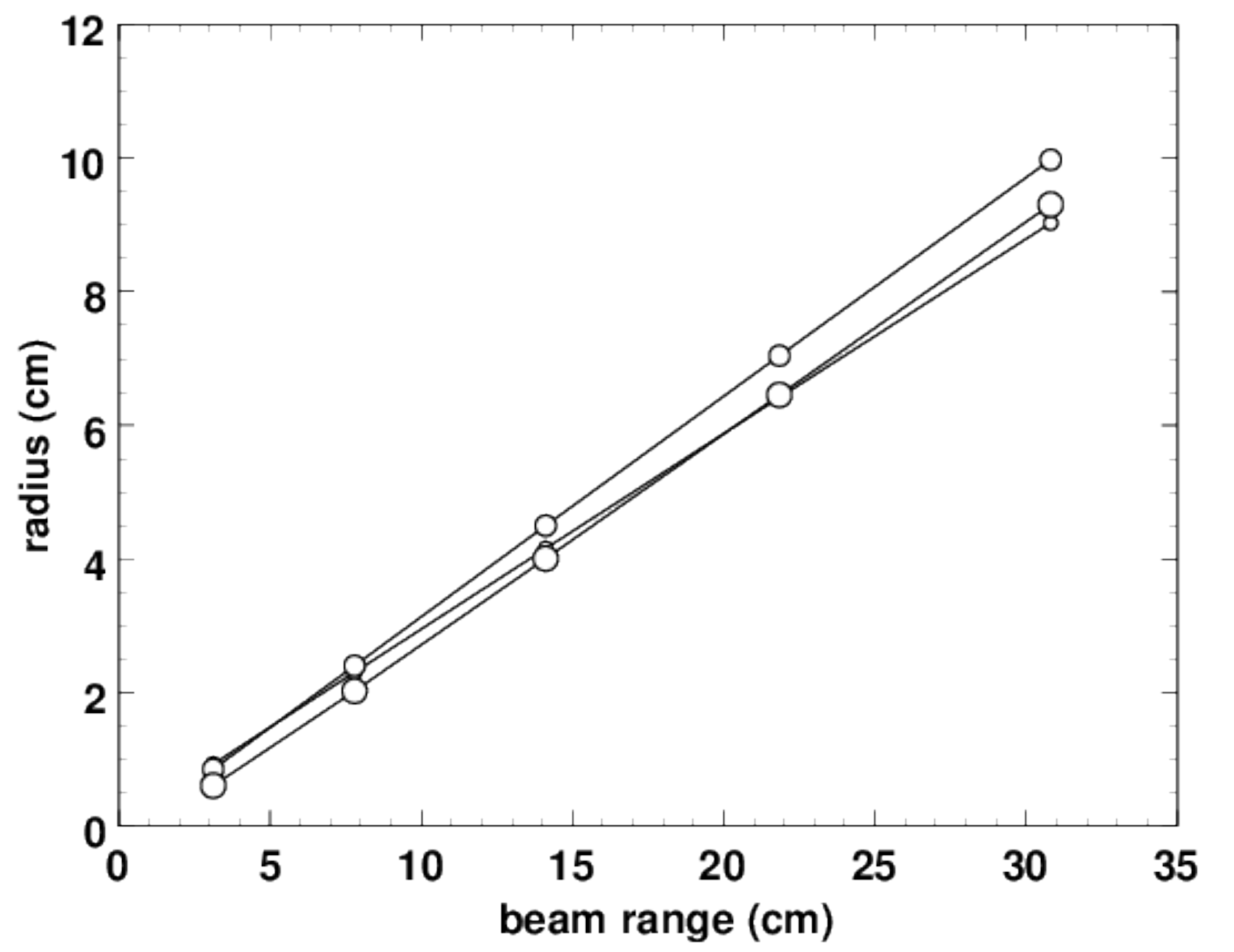} 
\caption{Maximum stopping radii of secondary protons in water v. incident beam range. Conditions and symbols same as Fig.\,\ref{fig:haloBump} except $T_1=60,\,100,\,140,\,180,\,220$\;MeV.\label{fig:haloMax}}
\end{figure}

We now perform the same calculation at several incident energies, finding the maximum radius at each.\footnote{~We have kept the same angle for the recoil nucleus throughout, not troubling to retune it each time. That has a very small effect.}  That leads to Fig.\;\ref{fig:haloMax}. The maximum radius, without allowing for beam size, is seen to be roughly one-third the range of the incident beam. That has implications, which we will discuss later, for the size of the BPC often used in characterizing the halo.

In summary, we have tried to convey the flavor of relevant nuclear physics at radiotherapy beam energies and have used kinematic arguments to estimate the radius of the halo. It was Fig.\,\ref{fig:haloBump} that prompted us to do longitudinal rather than radial scans because it suggests that, at large radii, the `nonelastic bump' should come and go with depth, as proves to be the case.

\section{Measurement}\label{sec:measurement}

\subsection{Method}
We used the experimental beam line at the Burr Center. Emerging from a 0.008\,cm Kapton window the nominal 177\,MeV beam passed through a 150\,cm helium bag, a 30\,cm air gap and a beam monitor before entering the water tank. The beam size corresponded to $\sigma_x=0.6$\,cm as determined by fits to the data discussed later. 

The beam monitor was an air filled plane parallel ionization chamber (IC) of in-house design whose gain had previously been determined as 96.8 at 177\,MeV. Though hermetically sealed, it has slack windows and therefore behaves as a vented chamber.  Recombination is negligible under the beam conditions used here. The electronics chain (also in-house) consisted of a recycling integrator (10\,pC/count), a divide-by-10$^4$ prescaler and a presettable counter that turned off the beam, for an overall calibration of $6.448\times10^9$\,p/MU (MU = monitor unit).

The CRS water tank \cite{crs} was used for positioning only. The beam was offset to one side of the tank. A lateral extension to the ionization chamber holder allowed us to position the dosimeter at distances up to 10\,cm from the nominal beam center. The dosimeter was an Exradin T1 (0.053\,cm$^3$ collection volume, internal diameter 0.4\,cm) biased to 300\,V, read out by  Keithley Model 6512 electrometer. The T1 calibration factor was 76.37\,cGy/nC $\pm1.4$\,\% at STP.

Since both the monitor and field IC were effectively vented, it would have sufficed to know both calibrations at STP (independent of the atmospheric pressure on the day of measurement) to compute the absolute dose per proton. Unfortunately the multiplication of 96.8 mentioned above was measured at an unrecorded pressure. (The experimental beam line is normally used for non-clinical measurements where a few percent absolute error is unimportant.) Unfortunately, before the monitor IC could be recalibrated at known pressure, it was physically damaged and had to be rebuilt. The resulting uncertainty, say $\pm3$\,\%, dominates the absolute dose per proton.

The main experiment was performed by positioning the T1 at radii ranging from 0 to 10\,cm and taking, at each radius, measurements at selected points between 1.53\,cm water equivalent (the lowest obtainable) to 25\,cm (well beyond the Bragg peak). At each radius the number of MU was set (between 2 and 500) to obtain a convenient Keithley reading and from time to time the beam current was adjusted (between 1 and 33\,nA) to obtain a convenient counting time ($<1$\,s to 15\,s). The entire experiment took about 6 hours.
 
Drift current was measured to be 0.5\,pC/15\,s just after the $r=10$\,cm (lowest dose rate) run. Since this was only 2\% of the lowest signal observed, we have ignored it. 

Recombination in the T1 chamber was explored by taking data at the Bragg peak over a wide range of beam currents. The signal per MU changed less than 1\%.

Two or more readings at each point were averaged and converted to absolute dose in MeV/g/p (1\,MeV/g = 0.1602\,nGy).

Subsequent analysis would have been easier had all measurements been taken over the full range in $z$. For the MI fit we had to invent points at some values of $r$ to fill out the $z$ range. (They are not shown in our graphs, and do not affect the results.) Nonuniform spacing in $z$, appropriate to the rate of variation of the dose, causes no problems.

\subsection{Results}

\begin{figure}[p]
\centering\includegraphics[width=4.78in,height=3.5in]{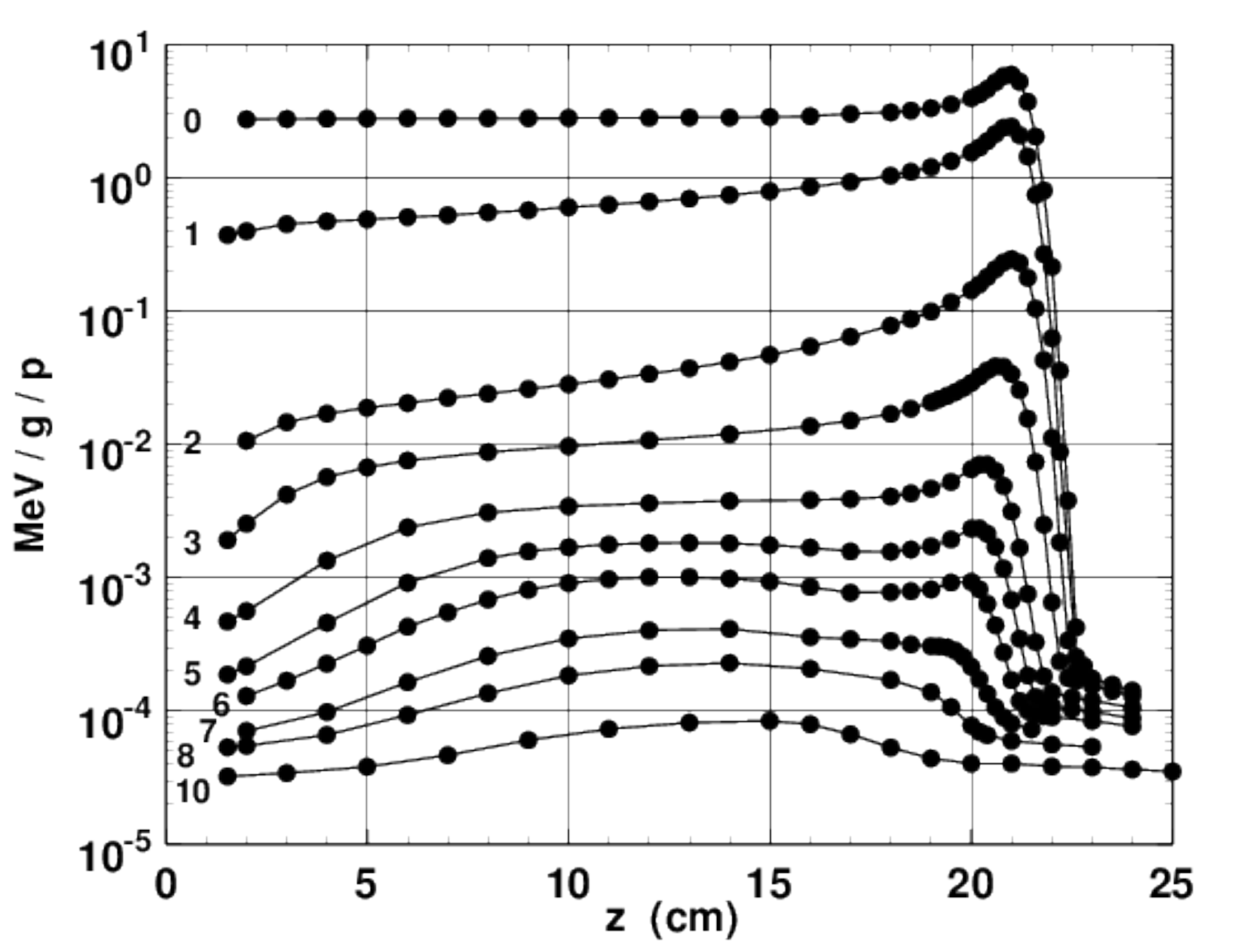}
\caption{Measured doses. Numbers indicate the distance $r$ (cm) of each longitudinal scan from the beam center line. The lines serve only to guide the eye.\label{fig:dmlg}}
\end{figure}

\begin{figure}[p]
\centering\includegraphics[width=4.66in,height=3.5in]{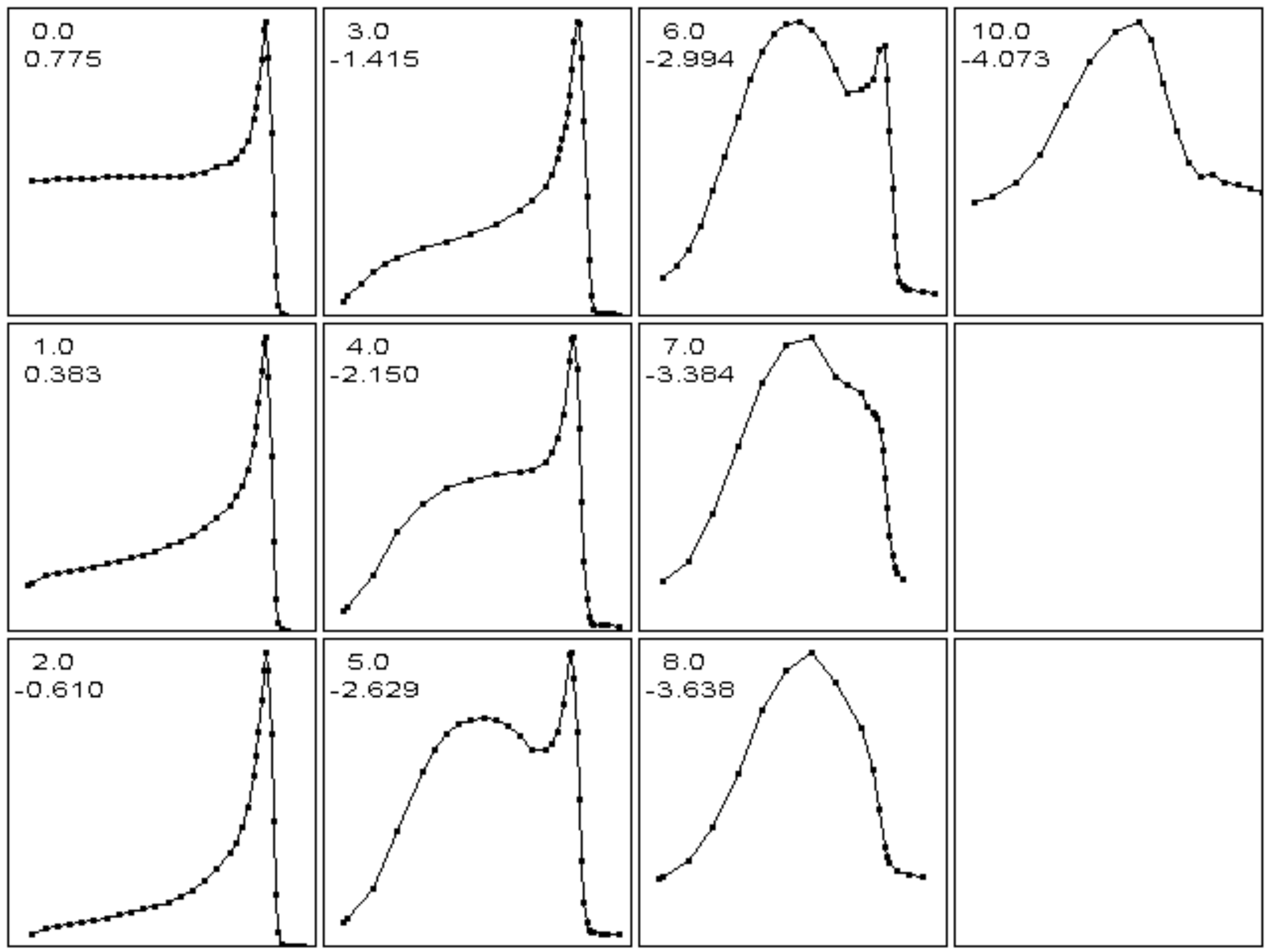}
\caption{The same data as Fig.\,\ref{fig:dmlg} in auto normalized linear presentation. The upper number in each frame is $r$ (cm), the lower is log$_{10}$(dose/(MeV/g/p)) for the highest point in the frame (see also Table\,\ref{tbl:dmlg}).\label{fig:dmlin}}
\end{figure}

Table\,\ref{tbl:dmlg} gives our data (some 300 points) in the form log$_{10}$(dose/(MeV/g/p)). Fig.\,\ref{fig:dmlg} shows them in semilog format, and Fig.\,\ref{fig:dmlin} in linear format. Fig.\,\ref{fig:dmlg} illustrates the virtue of longitudinal (rather than radial) scans and foretokens the difficulty of fitting any single radial profile to the data. A prominent feature is the Bragg peak which persists to quite large radii, becoming somewhat shallower.  We also see the distinct change in character between the halo and the aura. The semilog plot shows it best distal to the Bragg peak at small radii, while the linear plots show it as a broad background to the nonelastic bump at larger radii.

\section{Model-Dependent Fit to the Core/Halo}\label{sec:MD}
We have measured the core/halo dose at a finite number of points $D(r_j,z_{ij})$. For further work (e.g. estimation of field-size effects) it is convenient to reduce these data to a continuous function $D(r,z)$. We offer two such fits, one model-dependent (MD) and the other model-independent (MI). 

The MD fit, described in this section, is rather complicated, difficult to optimize, and certainly not unique. However, it does give some insight into the kinds of reactions that contribute and in roughly what proportion, and may be scalable to other energies.  The MI fit described in the next section is more straightforward, more accurate, and considerably more portable. It is probably more suitable for follow-on computations. 

As before, we reserve mathematical details for the appendices. Anticipating that core/halo measurements at other energies will scale with beam range, we have expressed most of the functions in the MD fit in terms of dimensionless normalized depth $z_\mathrm{n}\equiv z/R_0$ and radius $r_\mathrm{n}\equiv3\,r/R_0$. The arbitrary `3' is based on the considerations of Section\,\ref{sec:QE}. 

\subsection{Goodness-of-Fit Criterion}
The MD and MI functions to be fitted and the optimization technique are entirely different but the final figure of merit is the same. We seek a least-squares fit, but ordinary $\chi^2$ will not work because of the orders-of-magnitude variation of $D(r,z)$. Therefore we minimize $\chi^2$ as computed in $\log_{10}$ space. If the fit is relatively good, that is equivalent to minimizing the mean squared deviation from 1 of the {\em ratio} of measured $D_M$ to fitted $D_F$ since, provided $D_{M}/D_{F}$ is not too different from 1, it is easy to show that 
\begin{equation}\label{eqn:fom}
\log_e10\,\left(\sum_{i=1}^N\Big(\log_{10}D_{Mi}-\log_{10}D_{Fi}\Big)^2\right)^{1/2}\,\approx\,\left(\sum_{i=1}^N\Big(\frac{D_{Mi}}{D_{Fi}}-1\Big)^2\right)^{1/2}
\end{equation}

\subsection{Decomposition of the Dose}\label{sec:breakdown} 
Let $D$ be the physical absorbed dose per incident proton. Since physical doses add, the dose every\-where in an infinite water tank from a pencil beam can be written as a sum of electromagnetic, nuclear elastic, nuclear nonelastic (subscripted `qe' for `quasi-elastic') and aura doses:
\begin{equation}\label{eqn:D}
D(\vec x)\,=\,D(r,z)\,=\,D_\mathrm{em}(r,z)\,+\,D_\mathrm{el}(r,z)\,+\,D_\mathrm{qe}(r,z)\,+\,D_\mathrm{au}(r,z)
\end{equation}
We have assumed radial symmetry, which will obtain if the incident beam is radially symmetric.\footnote{~If not, ($r,z$) can easily be generalized to $(x,y,z)$.} The categories implied by Eq.\,\ref{eqn:D} are flexible. For instance, we think of $D_\mathrm{el}$ as including dose from the single scattering tail of the Moli\`ere distribution even though that is a Coulomb interaction. Similarly, we include elastic scattering from free H in $D_\mathrm{qe}$. $D_\mathrm{em}$ will be found to include a tiny dose from recoil nuclei. In other words, Eq.\,\ref{eqn:D} is meant to separate processes according to their pattern of dose deposition rather than any strict classification scheme.

Without losing generality we can write dose as fluence $\Phi$ (p/cm$^2$), the areal density of particles at a point in space, times mass stopping power $S/\rho$ (MeV/g)/(p/cm$^2$), their mass rate of energy deposition. Fluence in turn can be expressed as the product of a function $F$, normalized such that
\begin{equation}
\int_0^{2\pi}\int_0^\infty F(r,z)\,r\,dr\,d\phi\,=\,1\qquad\hbox{any $z$}
\end{equation}
 and a dimensionless multiplier $M$ (a normalizing constant). Thus
\begin{equation}\label{eqn:MFS}
D(r,z)\,=\,M_\mathrm{em}F_\mathrm{em}(S/\rho)_\mathrm{em}\,+\,M_\mathrm{el}F_\mathrm{el}(S/\rho)_\mathrm{el}\,+\,\ldots
\end{equation}
Such a decomposition is helpful only if we have independent models of the factors $F$ and $S/\rho$ for each process. That is completely true for `em', less so for `el' and so on. By the time we reach `au' the fit is purely empirical and we will simply fit $D_\mathrm{au}$ directly. 

\subsection{Electromagnetic Term (Core)}\label{sec:EM}
Not surprisingly, this is the term we know most about. It can be computed almost entirely from first principles. Its adjustable parameters are needed only because of our ignorance of parameters of the incident beam, numbers which in principle could be measured by ancillary experiments.

The only exception is $M_\mathrm{em}$\,, the number of primaries remaining at $z$. We will set it equal to $(1-\alpha z)$ where $\alpha$ itself depends linearly on $z$.\footnote{~For the mathematical definition and fitted numerical value of any parameter see Appendix\,\ref{app:MD}.} Our $\alpha$ differs from the similar parameter ($\approx0.01$/cm) employed by Bortfeld \cite{bortfeld}, which only accounts for {\em nonelastic} reactions \cite{janni82}.

$F_\mathrm{em}(r,z)$ is a 2D cylindrical Gaussian\,\footnote{~In our convention $\sigma_r^2\,\equiv\,<r^2>\,=\,2\,\sigma^2$. In other words our $\sigma$ is the {\em projected} transverse width parameter.}
\begin{equation}\label{eqn:G2D}
G_\mathrm{2D}(r,\sigma(z))\;\equiv\;\frac{1}{2\pi\,\sigma^2}\;e^{\displaystyle{-\frac{1}{2}}\left(\frac{r}{\sigma}\right)^2}
\end{equation}
$\sigma(z)=\sigma_\mathrm{em}(z)$ is found by means of Fermi-Eyges theory, which transports the parameters of the incident beam (transverse spread, angular spread and emittance) to any depth $z$  \cite{transport2012}. For details see Appendix\,\ref{app:SigEM}.

$S_\mathrm{em}$ is an electromagnetic stopping power based on the Bethe-Bloch $-dT/dz$ \cite{janni82,icru49} convolved with a 1D Gaussian whose $\sigma_\mathrm{sem}$ combines range straggling with the energy spread of the incident beam. Unlike Bortfeld \cite{bortfeld} we perform that convolution numerically. Since that is computationally expensive, we tabulate a family of curves at several values of $\sigma_\mathrm{sem}$ and interpolate as necessary during the search for a fit. Because $dT/dz$ is singular at end-of-range, a cutoff procedure is required. For details see Appendix\,\ref{app:SEM}. 

In summary, the electromagnetic term has a simple form which, if the incident beam parameters are known, is mostly derivable from first principles:
\begin{equation}\label{eqn:Dem}
D_\mathrm{em}\,=\,M_\mathrm{em}\,F_\mathrm{em}\,S_\mathrm{em}\,=\,(1-\alpha z)\;G_\mathrm{2D}(r,\sigma_\mathrm{em}(z))\;S_\mathrm{em}(z_n,\sigma_\mathrm{sem})
\end{equation} 
where $S_\mathrm{em}$ is an electromagnetic stopping power which includes the energy spread of the incident beam, $G_\mathrm{2D}$ is a Gaussian which, via $\sigma_\mathrm{em}$, describes the beam's transverse characteristics, and $\alpha$ is empirical. Because $\alpha$ is of order $\%/$cm the core retains most of the integrated dose (energy). We estimated the energy lost to nonelastic reactions in Appendix \ref{app:energyLost}.  

\subsection{Elastic Scattering}\label{sec:EL}
We have reviewed elastic scattering in Section\,\ref{sec:elastic}. The angular dependence is roughly exponential. Elastic reactions feeding the halo can occur anywhere along the primary track. To a considerable extent that does not matter. Because the angles are relatively small, the scattered protons will stop at more or less the same depth. For the transverse dependence $F$ we will use a 2D exponential with a variable $1/e$ width namely
\begin{equation}\label{eqn:E2D}
E_\mathrm{2D}(r,r_\mathrm{el}(r))\,\equiv\,\frac{1}{2\pi r_\mathrm{el}^2}\;e^{\displaystyle{-\frac{r}{r_\mathrm{el}}}}
\end{equation}
$E_\mathrm{2D}$ is independent of $z$. $E_\mathrm{2D}$ would be normalized with $<r^2>^{1/2}=\sqrt{6}\;r_\mathrm{el}$ if $r_\mathrm{el}$ were constant. We will allow $r_\mathrm{el}$ to be a polynomial in $r_\mathrm{n}$ 
but still keep the form of Eq.\,\ref{eqn:E2D}.

Since elastically scattered protons, as well as large-angle single scatters, are barely distinguishable from primaries, we can still use $S_\mathrm{em}(z)$ for the stopping power. The number of such protons crossing a plane at $z$ is proportional to the number of primaries that have interacted so a normalizing constant of the form $(\alpha \,z\,M_\mathrm{el})$ seems appropriate.

At larger radii we observe the Bragg peak to pull back and broaden somewhat. The pullback is presumably due to a combination of trigonometry and the increase of nuclear recoil energy with angle; the broadening, to an increased admixture of inelastic reactions at larger momentum transfers. We parameterize this effect by allowing both $z_\mathrm{n}$ and $\sigma_{sem}$ to depend on $r_n$. The electromagnetic stopping power then has the form
\begin{equation}
S_\mathrm{em}\,(z_\mathrm{n,adj}(r_\mathrm{n})\;,\sigma_\mathrm{sem}(r_\mathrm{n}))
\end{equation}
We use this same general form throughout since the modification has no effect at small $r_\mathrm{n}$. 

\subsection{Nonelastic Reactions}\label{sec:NES}
Of the many nonelastic channels, quasi-elastic (p,2p) and (p,pn) reactions contribute the most, have simple final states, and define the envelope of the halo at mid range (Section\,\ref{sec:QE}, Appendix\,\ref{app:QE}). To conform to our computer program we subscript nonelastic reactions `qe'.

Our parameterization of nonelastic reactions is largely empirical. The main ingredient is a dose `bump' at mid range characterized by an amplitude, mean $z$, half-width and exponential radial falloff, all functions of $r_n$ (Appendix\,\ref{app:MD}). We experimented with three forms for the bump, selectable through a jump index $j$:
\begin{equation}A_j/(1+z_j'^2)\qquad A_j/(1+z_j'^2)^2\qquad A_j\exp{-z_j'^2}
\end{equation}
with amplitudes $A_j$ and shifted scaled variables $z'$ adjusted so that each function had the same maximum value, mean $z_\mathrm{n}$ and half-width in $z_\mathrm{n}$ for easy interchangeability during optimization. We eventually chose the Gaussian, but all three worked reasonably well.

The complete $D_\mathrm{qe}$ term consists of the bump multiplied by 
\begin{equation}
\alpha\, z\, E_\mathrm{2D}(r_\mathrm{n},r_\mathrm{e}(r_\mathrm{n}))\,S_\mathrm{em}(z_\mathrm{n,adj},\sigma_\mathrm{sem}(r_\mathrm{n}))
\end{equation}
We retained $S_\mathrm{em}$ because, broadly speaking, all protons should have specific ionization increasing with $z$ per Bethe-Bloch theory. Alternatively, it could be absorbed into the normalization of the bump.

\subsection{Aura}
The increasing importance of the aura at large $r$ is already obvious in the raw data (Figs.\,\ref{fig:dmlg},\,\ref{fig:dmlin}). Our parameterization is entirely empirical, simply an adjustable multiple of $E_\mathrm{2D}(r_\mathrm{n},r_\mathrm{e}(r_\mathrm{n}))$ where the decay length $r_\mathrm{e}$ is itself linear in $r_\mathrm{n}$.

\subsection{Recoil Nuclei}\label{sec:RN}
So far we have not discussed recoil nuclei which are present in every reaction and deposit energy very near the reaction site. Let us calculate the energy deposited in an infinitesimal volume $dA\times dz$ at $(r,z)$. The number of primaries per incident proton is $(1-\alpha z)\,G_\mathrm{2D}(r,\sigma_\mathrm{em}(z))\,dA$. Of these a fraction $\alpha\, dz$ interact, each recoil nucleus depositing let us say an average energy $T_\mathrm{rn}$. Putting that together and dividing by the mass contained in the volume we find
\begin{equation}\label{eqn:Drn}
D_\mathrm{rn}\;=\;\frac{(1-\alpha z)\,G_\mathrm{2D}(r,\sigma_\mathrm{em}(z))\,dA\,\alpha\,dz\,T_\mathrm{rn}}{\rho\,dA\,dz}\;=\;
  (1-\alpha z)\,G_\mathrm{2D}(r,\sigma_\mathrm{em}(z))\;\,\frac{\alpha\,T_\mathrm{rn}}{\rho}
\end{equation}
In other words, we can include the effect by simply adding an equivalent $S_\mathrm{rn}\equiv\alpha T_\mathrm{rn}$ to $S_\mathrm{em}$ in Eq.\,\ref{eqn:Dem}. It remains to estimate its magnitude. 

In quasi-elastic reactions the recoil nucleus is a spectator. Its most probable momentum is $pc=75$\,MeV and its kinetic energy is $T=(pc)^2/(2(A-1)mc^2)=0.2$\,MeV. In elastic scattering on O we can use the data of Gerstein et al. at 96\,MeV on C \cite{Gerstein1957}. Ignoring small differences between C and O and between $\theta_\mathrm{CM}$ and $\theta_\mathrm{lab}$ and fitting the observed differential cross section with a simple exponential we find $\theta_{rms}=7$\degr\;\; corresponding to $T=0.1$\,MeV. Using the larger recoil energy, and our fitted value for $\alpha$, $S_\mathrm{rn}\approx$\;0.006\,MeV/cm, much smaller than $S_\mathrm{em}\approx3.5$\,MeV/cm, the lowest possible value for radiotherapy protons in water.

In summary, the energy loss by $Z>1$ recoil nuclei adds, in effect, a negligible amount to $S_\mathrm{em}$. The `recoil energy' in elastic scattering on H is huge, of course, but that reaction is effectively included in the $D_\mathrm{qe}$ term. Apart from that, what we have discussed here is identical to the `nuclear stopping power' considered in \cite{icru49}. Their much more complete calculations confirm that it is indeed very small, some 0.035\% of the electronic stopping power in liquid water at 175\,MeV.

\subsection{Optimization}
We used the Marquardt algorithm \cite{bevington,nr} for final adjustment of the 25 fit parameters. In a fit of this complexity, however, that and other automatic methods are hopeless unless good starting values are supplied for most of the fit parameters. Those were obtained by a rather haphazard semi-interactive process aided by the following software capabilities:
\begin{itemize}
\item Fitting any subset of the data;
\item Locking any subset of parameters;
\item Varying any parameter by hand and quickly displaying graphs and goodness-of-fit;
\item Writing final parameters of a session to a file to cut and paste to new starting values.
\end{itemize}
We grouped parameters according to the terms in Eq.\,\ref{eqn:D} (Appendix\,\ref{app:MD}) and adjusted them in the order (em), (au), (el) and (qe). Only at the very end was automatic optimization used. Even then $p_{10}$ and $p_{11}$ had to be locked to avoid exponential overflow.

\subsection{Results}
\begin{figure}[p]
\centering\includegraphics[width=4.72in,height=3.5in]{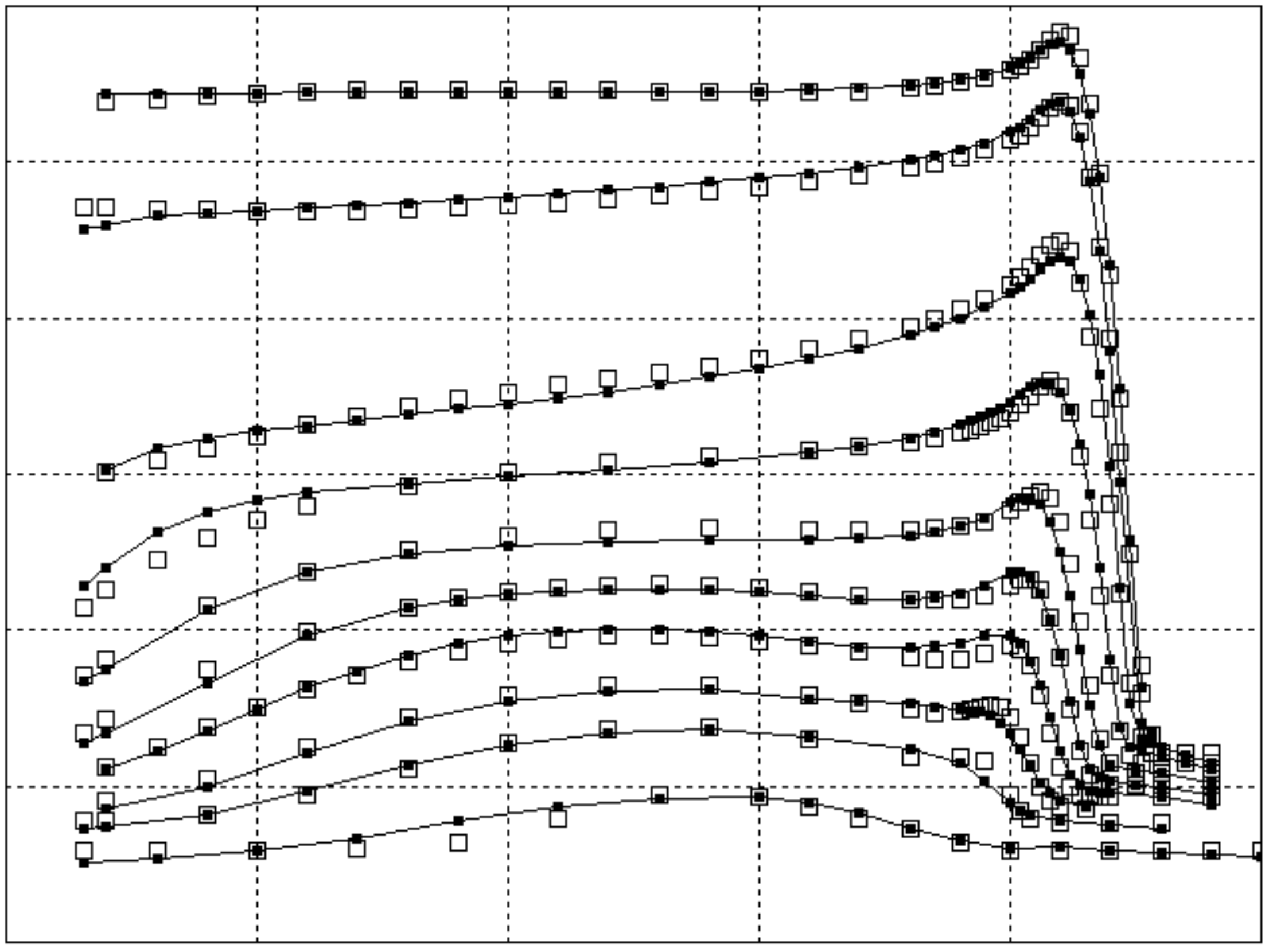}
\caption{Same as Fig.\,\ref{fig:dmlg} with model-dependent fit (open square) of each measured point.\label{fig:dflgMD}}
\end{figure}

\begin{figure}[p]
\centering\includegraphics[width=4.72in,height=3.5in]{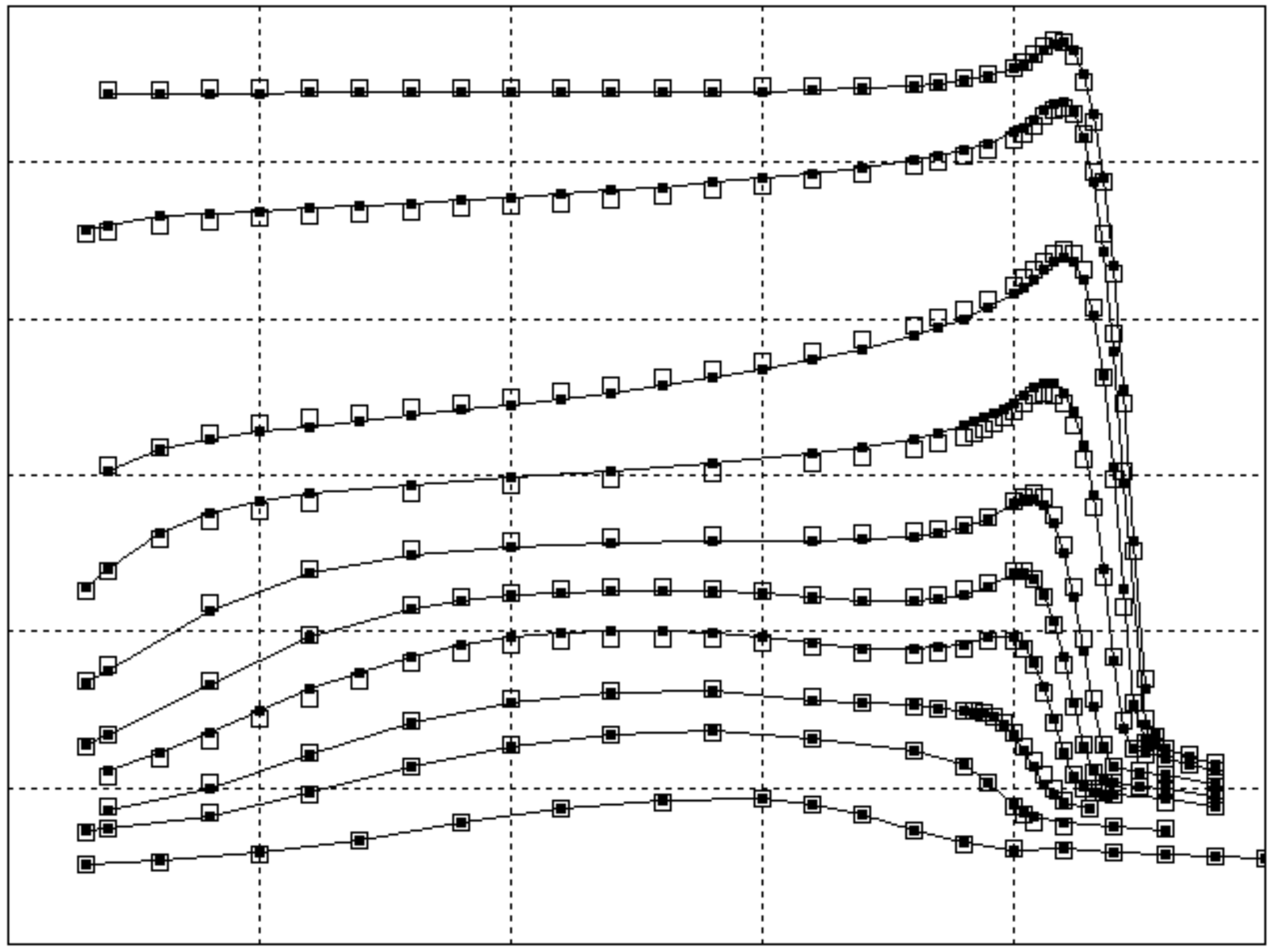}
\caption{Same as Fig.\,\ref{fig:dflgMD} with model-independent fit.\label{fig:dflgMI}}
\end{figure}

\begin{figure}[p]
\centering\includegraphics[width=4.72in,height=3.5in]{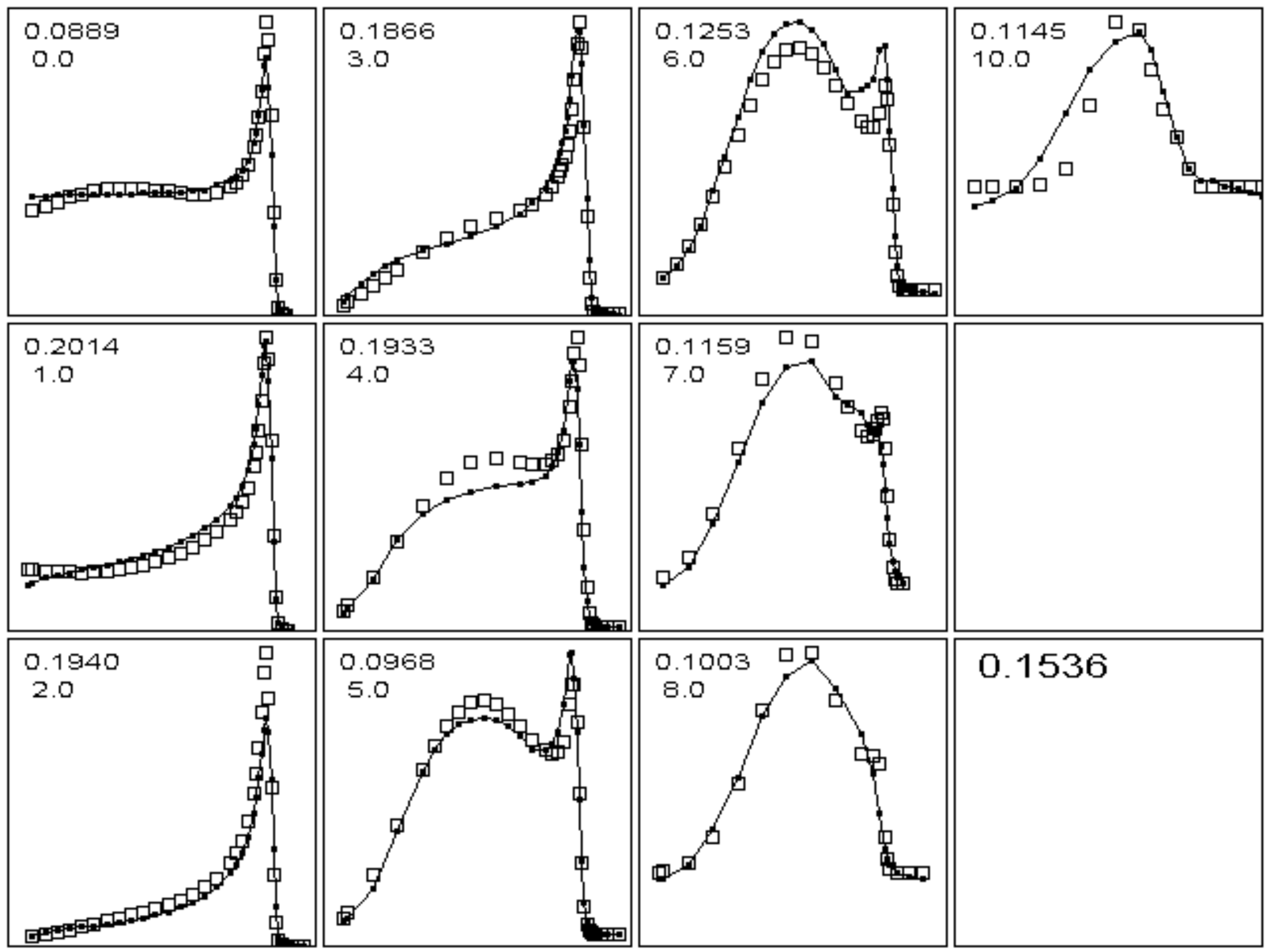}
\caption{Same as Fig.\,\ref{fig:dmlin} with model-dependent fit (hollow square) of each measured point. The upper number in each frame is the goodness of fit, the rms value of the (ratio\,$-$\,1) of measurement to fit for that scan (see text). The isolated number is the global value.\label{fig:dlinMD}}
\end{figure}

\begin{figure}[p]
\centering\includegraphics[width=4.72in,height=3.5in]{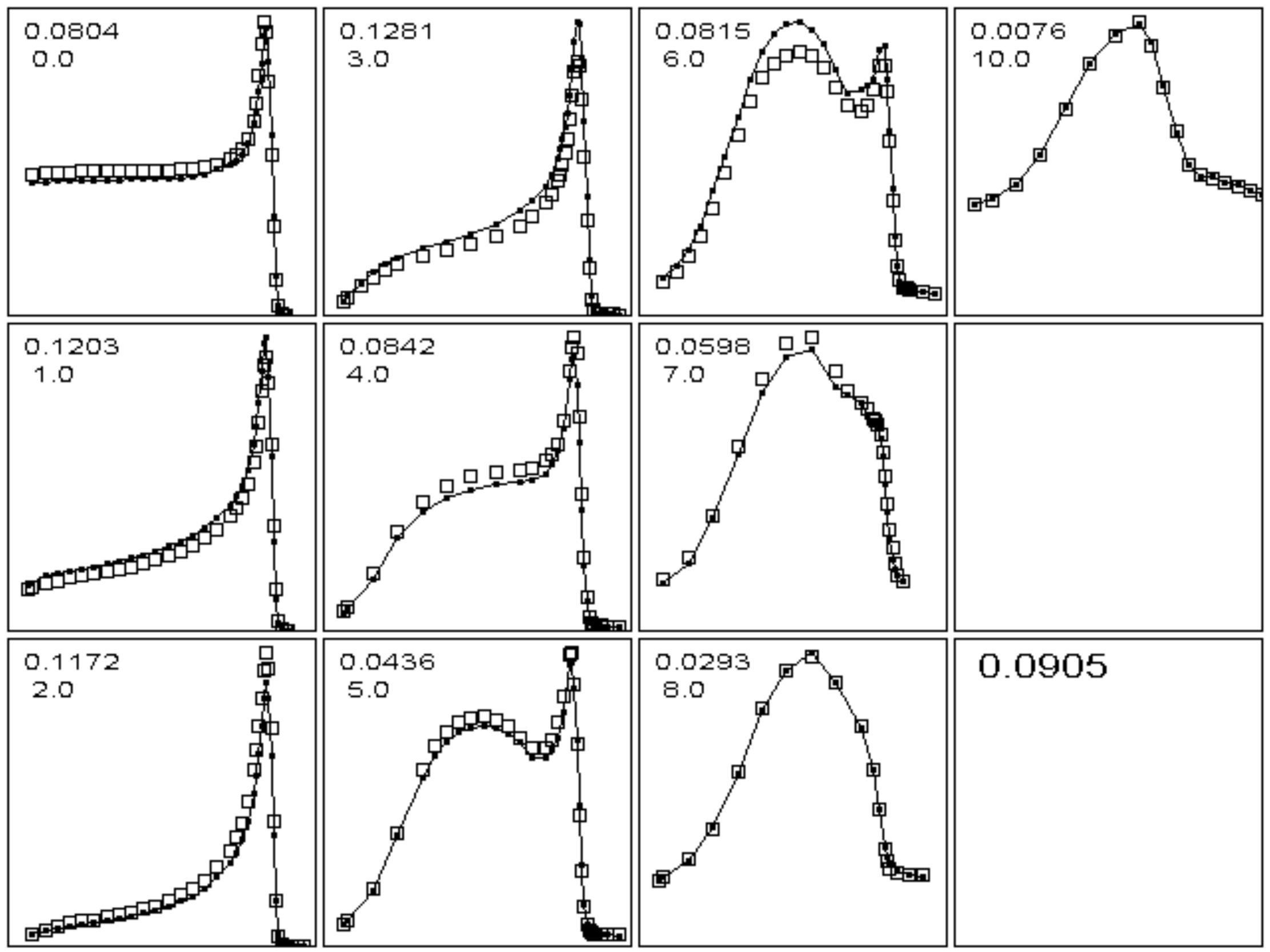}
\caption{Same as Fig.\,\ref{fig:dlinMD} with model-independent fit.\label{fig:dlinMI}}
\end{figure}

\begin{figure}[p]
\centering\includegraphics[width=4.57in,height=3.5in]{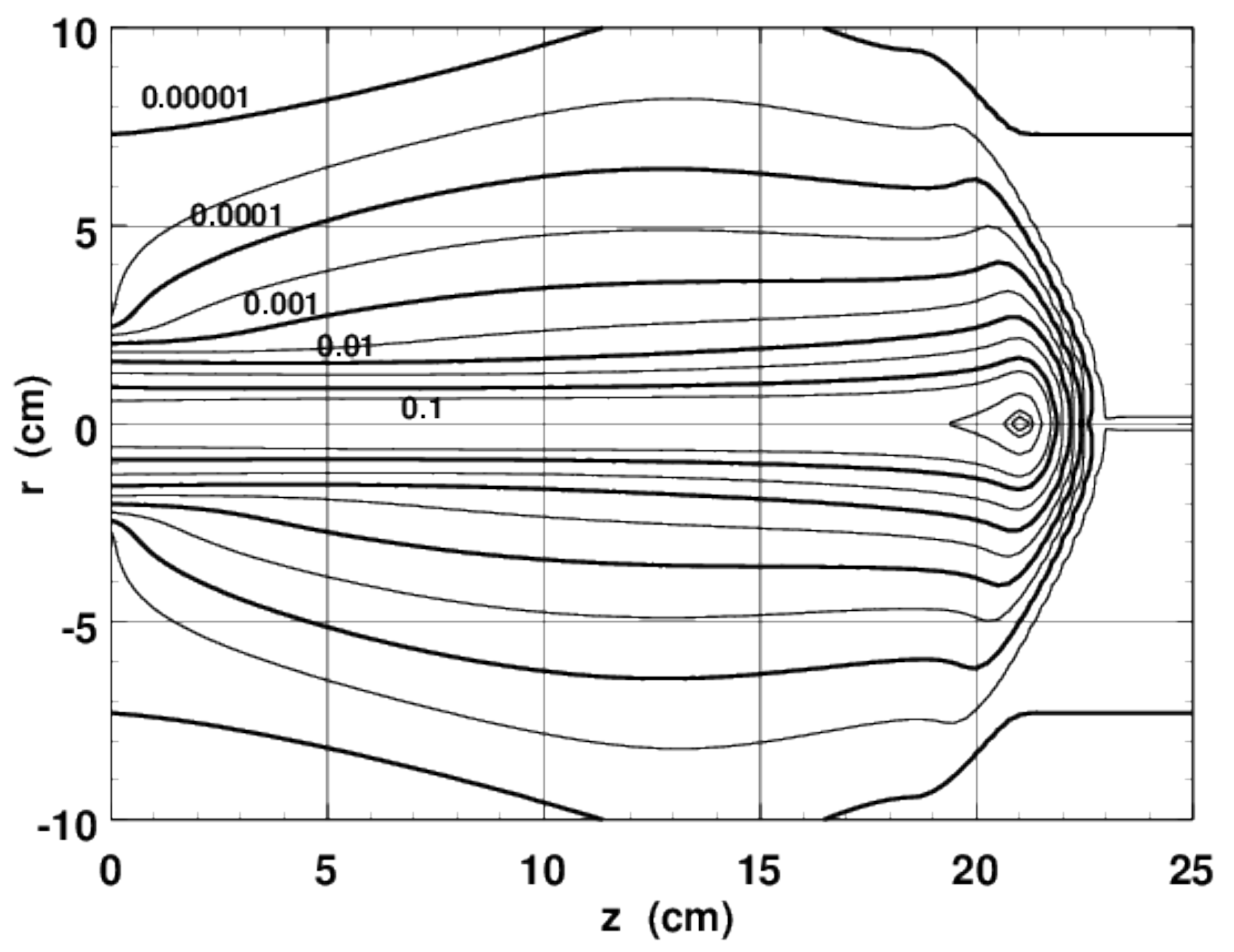}
\caption{Contours of MD fit at .9 .8 .5 .2 .1 .0316 .01 $\ldots$ .00001 of maximum dose.\label{fig:DMDcontours}}
\end{figure}

\begin{figure}[p]
\centering\includegraphics[width=4.57in,height=3.5in]{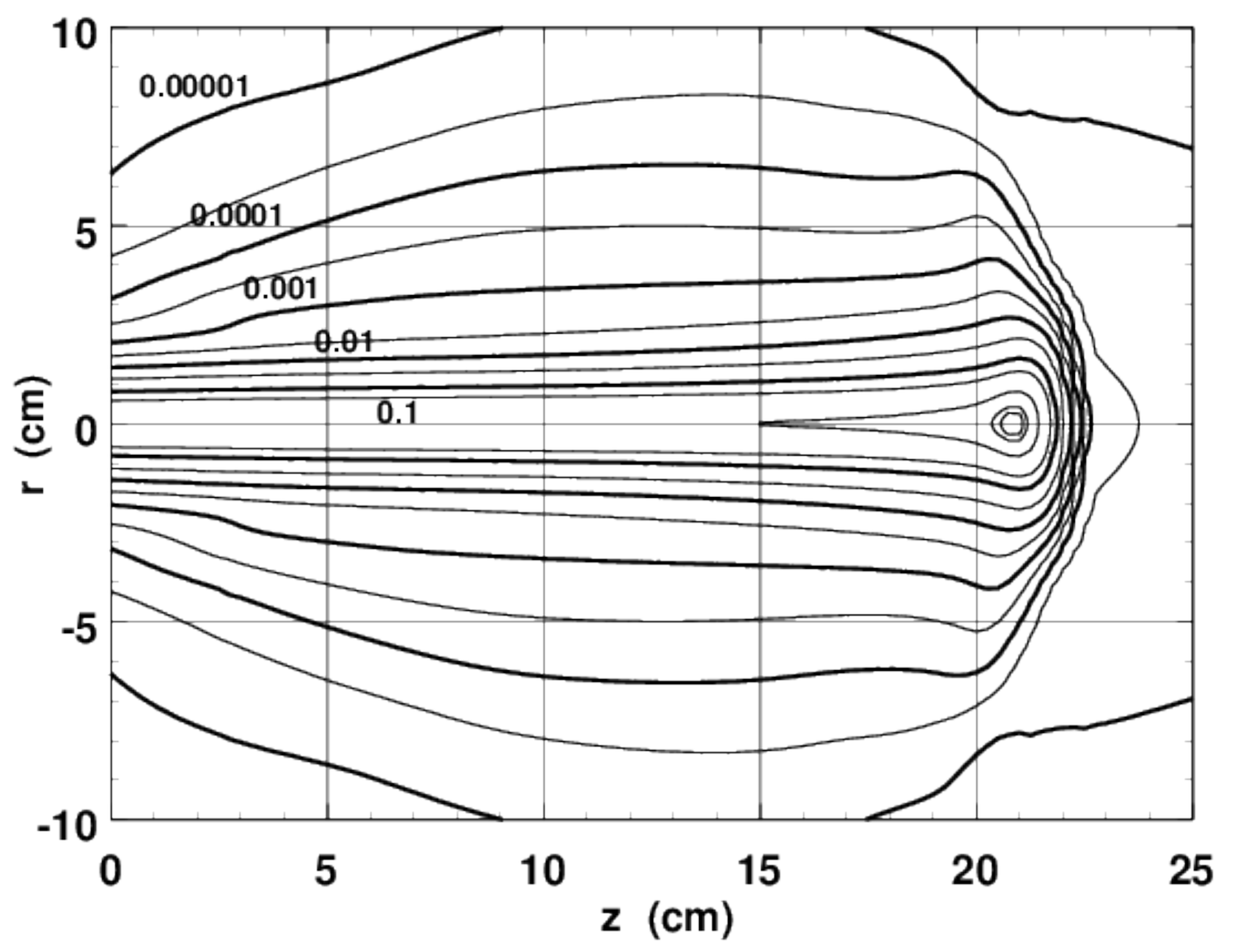}
\caption{Same as Fig.\,\ref{fig:DMDcontours} for MI fit.\label{fig:DMIcontours}}
\end{figure}

Anticipating Section\,\ref{sec:MI} we present plots in pairs to compare the MD and MI fits.
Figs.\,\ref{fig:dflgMD} and \ref{fig:dflgMI} are semilog plots of measured dose with the fits to each measured point superimposed as open squares. Figs.\,\ref{fig:dlinMD} and \ref{fig:dlinMI} are the same in autoscaled linear format. The goodness-of-fit is given at each radius. Finally, Figs.\,\ref{fig:DMDcontours} and \ref{fig:DMIcontours} are contour plots of $D(r,z)$ for each fit.

Although our measurements were scans in $z$, either fit allows us to reconstruct what scans in $r$ would look like, and the MD fit allows us to display each physical process (Eq.\,\ref{eqn:D}) separately. Fig.\,\ref{fig:trans12} shows a virtual radial scan at $z=12$\,cm in semilog presentation. The bold line is the MD fit and the full circles are measurements. The core, elastic, nonelastic and aura components are shown as light lines. The core is, of course, Gaussian in profile. The others are roughly exponential with varying degrees of curvature. Fig.\,\ref{fig:trans21} shows the same thing at the Bragg peak, $z=21$\,cm. The two transverse profiles are quite different. Neither would be easily described by a few Gaussians.\footnote{~With enough Gaussians one can, of course, fit anything.}

\subsection{Discussion}\label{sec:MDdiscussion}  
It is easy to be carried away by model dependent curve fitting, losing sight of its purpose and what can reasonably be achieved. The purpose is twofold: to produce a continuous function $D(r,z)$ that describes the discrete measurements reasonably well, and to obtain values for some of the parameters that may have simple physical interpetations. (Appendix\,\ref{app:parms} summarizes the functional form of our fit and the numerical values of all fitted parameters.)

The parameters best determined and most likely to be useful in further work are those associated with the electromagnetic or core term: the fitted range $R_0$, its spread $\sigma_\mathrm{sem}$ combining straggling and beam energy width, and the transverse size of the incident beam $\sigma_x=\sigma_\mathrm{em}$ as well as (much less well determined) its angular divergence and emittance.

The overall compliance of the MD fit to measurement, about 15\%, is unimpressive. On the other hand, we cannot reasonably expect a handful of simple functions to describe a phenomenon of this complexity. Only an accurate Monte Carlo program can hope to do that.

\section{Model-Independent Fit to the Core/Halo}\label{sec:MI}

\subsection{Method}\label{sec:MImethod}
Because of the complexity of the MD fit we sought an alternative that would be more portable and hopefully more accurate. 

\begin{figure}[p]
\centering\includegraphics[width=4.72in,height=3.5in]{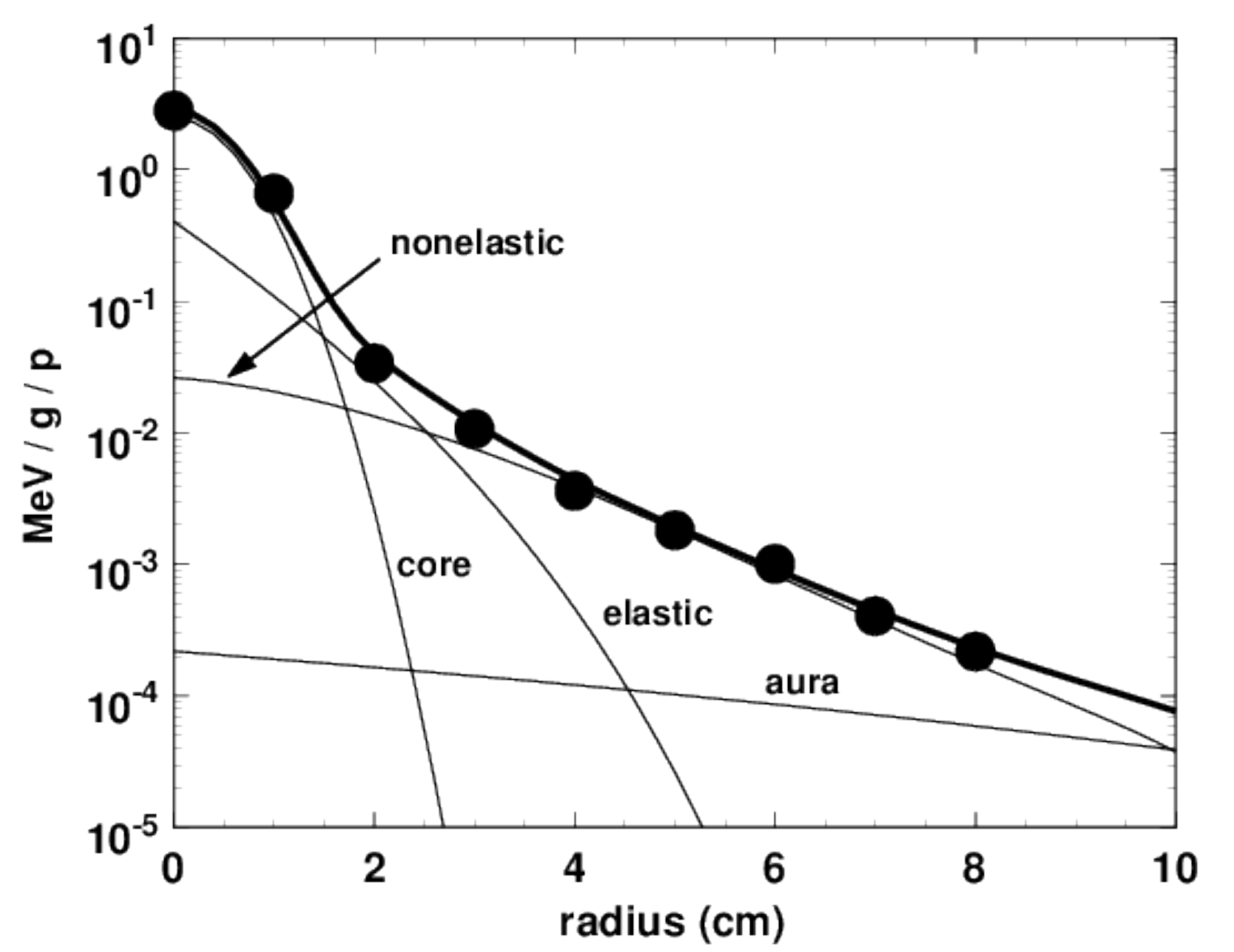}
\caption{Model dependent fit (bold line) to the transverse dose distribution at $z=12$\,cm with experimental points (full circles). The contribution of each term in the fit is also shown.\label{fig:trans12}}
\end{figure}

\begin{figure}[p]
\centering\includegraphics[width=4.72in,height=3.5in]{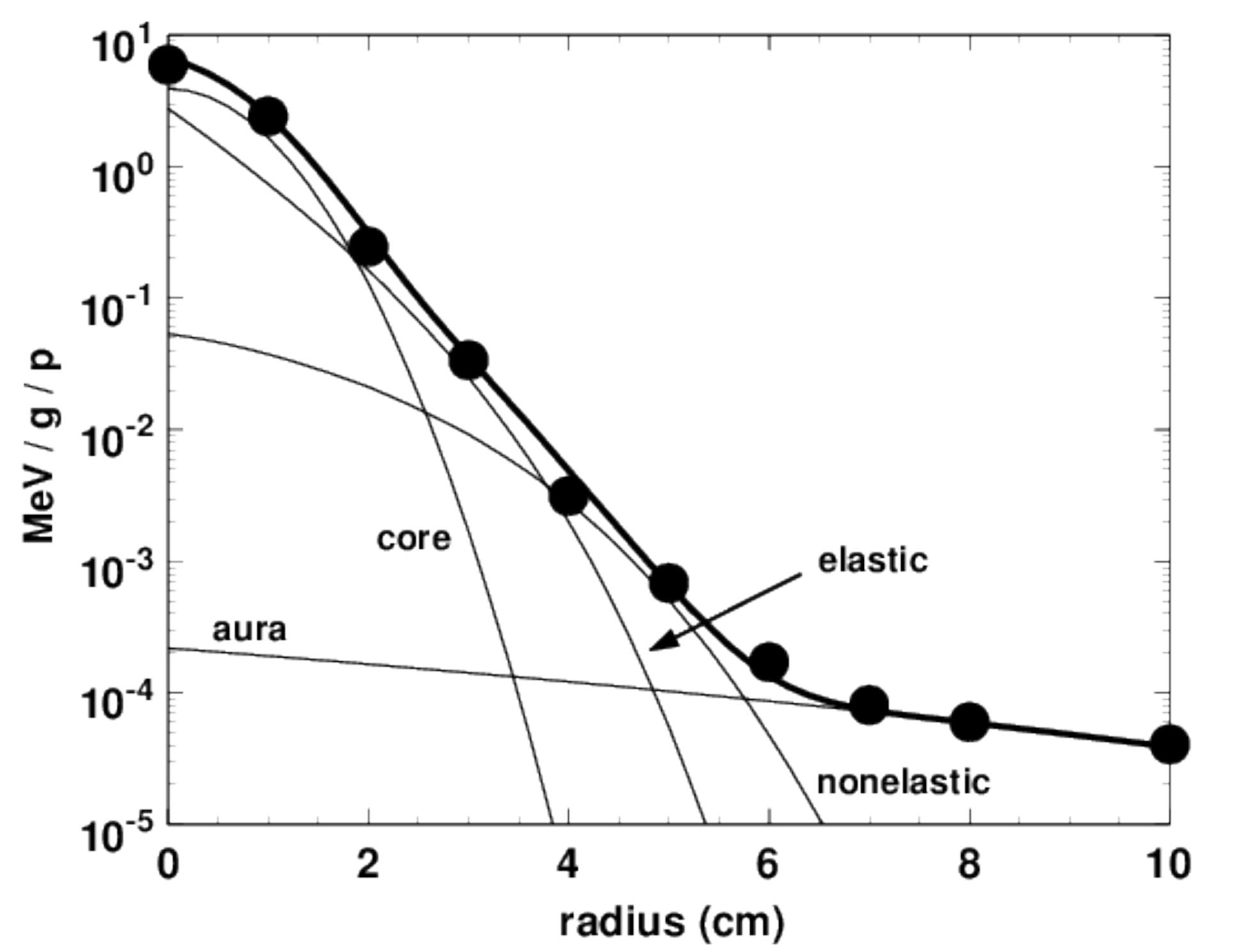}
\caption{The same as Fig.\,\ref{fig:trans12} but at $z=21$\,cm.\label{fig:trans21}}
\end{figure}

The first strategy that comes to mind is suggested by Figs.\,\ref{fig:dmlg} and \ref{fig:trans12}: Describe $D(r,z)$ by a polynomial in $z$ whose coefficients are polynomials in $r$ or vice versa. That is a straightforward and unique linear least-squares fit. However, it gives very poor results here, because both the $z$ and virtual $r$ scans (no matter in which order taken) must be separately fittable by polynomials. That is emphatically not true for the $z$ scans even in logarithmic form (Fig.\,\ref{fig:dmlg}). Straight regions of differing slopes separated by regions of high curvature require polynomials of very high degree. That creates its own problems of matrix inversion \cite{nr} and pathological behavior just outside the fitted region.

Fortunately, both the $z$ and virtual $r$ scans are well fit by cubic splines, leading to the following strategy:
\begin{itemize}
\item Fit each semilog $z$ scan by a cubic spline, yielding ten continuous functions of $z$.
\item At equally spaced $z$\,s, use these functions to generate virtual ten point $r$ scans.
\item Fit each virtual $r$ scan with a cubic spline, yielding a continuous function of $r$.
\item Evaluate those functions at equally spaced $r$\,s to yield finally a rectangular grid in $(r,z)$.
\item Write those values of $\log_{10}(D/$(MeV/g/p)) to a text file with an appropriate header.
\end{itemize}
We have now created and exported the fit. To use it later or on a different computer
\begin{itemize}
\item Initialize by reading the text file and repopulating the 2D array.
\item Use bilinear interpolation to find $\log_{10}(D/$(MeV/g/p)) at any $(r,z)$.
\item Raise 10 to that power to find the dose in MeV/g/p at any $(r,z)$.
\end{itemize}
We used a 1\,mm grid (overkill) in both $r$ and $z$ spanning $2\le Z\le25$\,cm and $0\le r\le10$\,cm yielding a 187\,KB text file. The entire fit takes a few seconds on a high-end laptop running Fortran. To simplify the program we needed to pad some of the measured data sets with made-up values to $z=25$\,cm. Only actual data points are shown in the figures.

The text file referred to is available as \bs BGware\bs fitHalo\bs DlgCS.TXT in BGware.ZIP at either \BGsite\ or \LPPCsite. 

An important advantage of cubic spline fits in the present context is that a boundary condition on the first or second derivative of the fit may be  (indeed, must be) imposed at each end of the data set. That allows us to require the physically necessary condition that the derivative of the dose with respect to $r$ be zero at $r=0$ for every radial scan (cf. Figs.\,\ref{fig:trans12} and \ref{fig:trans21}). This powerful constraint is in effect a free datum, and is the only model dependent feature of the MI fit.

A further advantage of cubic splines is that they extrapolate gracefully just outside the fit range, emphatically not true of high order polynomials. Fitted values in the unmeasured region $0\le z\le2$\,cm, though they may not be exactly correct, are not unreasonable.

Interpolation is by far the most common application of cubic splines. Least-squares fitting (or `regression') with cubic splines is not unknown, but usage is far from standard. For that reason, we give a brief description of our cubic spline fitting technique in Appendix\,\ref{app:CSfit} and for completeness, bilinear interpolation in Appendix\,\ref{app:bilinear}.

\subsection{Results}\label{sec:MIdiscussion}
We have already shown the MI results in Figs.\,\ref{fig:dflgMI}, \ref{fig:dlinMI} and \ref{fig:DMIcontours}. The MI procedure (once the program is written) is strikingly simpler and the goodness of fit considerably better.

It is probably not accidental that the $r=6$\,cm scan is an outlier in both the MD and MI fits. That, and certain irregularities in the variation of the Bragg peak distal edge with $r$, suggest that the experiment suffered positioning errors of order 1\,mm in both $z$ and $r$. We did not fit or otherwise try to correct them.

\section{Review of the Literature}\label{sec:review}

We now review the literature to establish early and current practice in measuring, simulating and parameterizing the core/halo. Table\,\ref{tbl:lit} summarizes, in order, the papers reviewed. The subdivision by topics is fairly arbitrary as most of the papers, Pedroni's in particular, cover more than one aspect of the problem. First, we discuss two topics that apply to all.

\subsection{Mixed Stopping Power}
The quantity
\begin{equation}
S_\mathrm{mixed}(z)\;\equiv\;\rho_\mathrm{water}\;\int_0^{2\pi}\int_0^{r_\mathrm{halo}}D(r,z)\;r\,dr\,d\phi\hbox{\qquad MeV/cm/p}
\label{eqn:Sav}\end{equation}
was introduced (in the PBS context) by  Pedroni et al. \cite{pedroniPencil} as the `integral dose' $T(w)$:
\begin{quote}
`$T(w)$ is the total dose (in Gy cm$^2$) integrated over the whole plane perpendicular to the beam at the depth $w$.'
\end{quote}
It plays a critical part in the literature. Dimensionally, $S_\mathrm{mixed}$ is a stopping power per proton whereas $T(w)$ is a {\em mass} stopping power per proton or some fixed number of protons.\footnote{~The $y$-axis label in Pedroni's Fig.\,1 should read `nGy cm$^2$/p' not `nGy cm$^2$/10$^6$\,p'.} We call them `mixed' because they mix electromagnetic and nuclear effects. That is clear from Pedroni's Section\;2.2.2, the definition and physical model of $T(w)$.

Eq.\,\ref{eqn:Sav} suggests applying a single pencil beam, adding up its contributions to annuli extending from zero to a sufficiently large radius, and dividing by the total number of protons. Mathematically, the same result could be obtained by filling those annuli uniformly with pencil beams, adding up their contributions on axis, and dividing by the total number of protons. Thus $S_\mathrm{mixed}$ is directly related to the central axis depth dose or `Bragg peak' in a sufficiently broad uniform beam of quasi-monoenergetic protons. 

More precisely, it is the (dose to water) per (fluence in air) on the axis of a monoenergetic beam broad enough that nuclear equilibrium prevails.\footnote{~If nuclear equilibrium prevails, Coulomb equilibrium will prevail also.} That has long been a quantity of interest in the description of scattered proton beams. $S_\mathrm{mixed}$ is basically Bortfeld's analytic Bragg peak \cite{bortfeld}, the object modeled by Berger's PTRAN Monte Carlo \cite{berger2}, and the topic of numerous other papers. By assuming Coulomb and nuclear equilibrium, so that small excursions from the axis make no difference, it collapses electromagnetic and nuclear processes into a single function of $z$ only.

\subsection{The Bragg Peak Chamber}
Assume we adopt Pedroni's approach, and therefore find it necessary to measure $S_\mathrm{mixed}$. Assume also that we have at our disposal all the hardware and software for PBS treatments. We could fill our water tank with pencils and measure the depth-dose on axis. However, with a single dosimeter that would be very time consuming because each point in the depth-dose requires a complete set of pencils. The preferred method, introduced by Pedroni at al. \cite{pedroniPencil} and suggested by Eq.\,\ref{eqn:Sav}, is to scan a single pencil on axis in a water tank using a plane-parallel IC (PPIC) of large radius. Two such `Bragg peak chambers' (BPCs) are currently available commercially: the PTW \cite{ptw} Model 34070 of 4\,cm radius and more recently, the IBA \cite{iba} `Stingray' of 6\,cm radius. The former is generally conceded to be too small. As we will see, several papers employ it but use Monte Carlo to correct for the missing dose.

In the present work we have measured and parameterized the core/halo to a large radius, so we can carry out the integral of Eq.\,\ref{eqn:Sav} numerically on real data. Fig.\,\ref{fig:Smixed} shows the result for $r_\mathrm{halo}=10$\,cm using both the MI and MD models (top frame) as well as the defect (missing dose) when the integral is only taken to 4 or 6\,cm. The missing dose, not surprisingly, depends on $z$ with a maximum near mid range of 2.8\% for the Model 34070 and $<1$\% for the Stingray. Those numbers will increase at higher energies.

\subsection{Measurements}

\subsubsection{Pedroni}
Pedroni et al. \cite{pedroniPencil} had a strong influence on subsequent papers, so a close reading of this paper is required. The PSI group was the first to use magnetic scanning in proton radiotherapy, to publish an end-to-end dose calculation model, and to point out the dosimetric importance of the halo.

PSI has been significantly upgraded in recent years and \cite{pedroniPencil} does not describe current practice. Then, they used seven predefined beams in conjunction with degraders just upstream of the patient. It was not a `pure' PBS system, in that it used a combination of magnetic scanning, couch motion, upstream energy variation and local degraders. We will ignore complications from the degraders. For brevity, we refer to Pedroni's Fig.\,1 (for instance) as `Fig.\,1P'.

Eq.\,(7P), Pedroni's parameterization of the core/halo, reads verbatim
\begin{equation}\label{eqn:PedEqn7}
D(x,y,w)\;=\;T(w)\times\left[\,(1-f_\mathrm{NI}(w))\times G_2^p(x,y,\sigma_\mathrm{P}(w))\;+\;f_\mathrm{NI}(w)\times G_2^\mathrm{NI}(x,y,\sigma_\mathrm{NI}(w))\,\right]\nonumber
\end{equation}
In our notation, and assuming radial symmetry (which \cite{pedroniPencil} also assumes later on)
\begin{equation}\label{eqn:ourPedEqn7}
D(r,z)\;=\;\frac{N_\mathrm{p}\;S_\mathrm{mixed}}{\rho_\mathrm{water}}\times\bigl[\,(1-f_\mathrm{NI}(z))\,G_\mathrm{2D}(r,\sigma_\mathrm{P}(z))\;+\;
f_\mathrm{NI}(z)\,G_\mathrm{2D}(r,\sigma_\mathrm{NI}(z))\,\bigr]
\end{equation}
the familiar `dose = mass stopping power $\times$ fluence'. Pedroni's numerical model of $T(w)$ combines
\begin{itemize}
\item The Bethe-Bloch stopping power convolved with a Gaussian that combines range straggling with beam energy spread (our $S_\mathrm{em}$).
\item A function describing the attrition of primaries by `protons undergoing nuclear interactions'. An early sign of trouble is that a similar, but explicit, attrition function multiplies the first term in square brackets. Therefore it is used twice.
\item Energy deposition by secondaries: one-third local, one-third longer range and one-third missing (neutrals).
\end{itemize}
$T(w)$ at seven energies including 177\,MeV is presented in Fig.\,1P along with confirming relative dose measurements taken with a 4\,cm radius Bragg peak chamber. (At the time, 4\,cm was considered large enough.)

Returning to Eq.\,\ref{eqn:ourPedEqn7}, the first term in square brackets represents the core, the second the halo. $f_\mathrm{NI}(z)$ is an arbitrary function given at several energies in Fig.\,6P. $G_\mathrm{2D}$ is a fluence per proton, here assumed to be a normalized 2D Gaussian, Eq.\,\ref{eqn:G2D}. $\sigma_\mathrm{P}(z)$ is the projected rms transverse spread of the core, found as a function of $z$ by applying Fermi-Eyges theory (albeit in Pedroni's notation) to the phase space parameters of the incident beam. It is shown at several energies in Fig.\,3P. Finally, $\sigma_\mathrm{NI}(z)$ is the width parameter of the second Gaussian, supposed to describe nuclear interactions. It is shown at several energies in Fig.\,7P.

Unlike $T(w)$ and $\sigma_\mathrm{P}(z)$ which derive from first principles plus a few beam measurements,  $f_\mathrm{NI}(z)$ and $\sigma_\mathrm{NI}(z)$ are entirely arbitrary and fitted to measurements of the halo. Unlike our measurement, Pedroni used a `frame' technique which was widely copied later with variants. Data were taken with an IC on the central axis, while monenergetic pencil beams were scanned in a full square $3\times3$\,cm and hollow squares 3--5, 5--7 and 7--10\,cm. Fig.\,5P shows these frame measurements (multiplied by 10 for the hollow frames) as well as the resulting fit using Eq.\,\ref{eqn:ourPedEqn7}.

We retrieved Pedroni's fit at 177\,MeV by reading data from Figs.\,1P, 3P, 6P and 7P and checked our work by reproducing Fig.\,5P.  (That step required some reasonable assumptions about exactly how the frames were populated with pencil beams.) 

\begin{figure}[p]
\centering\includegraphics[width=4.70in,height=3.5in]{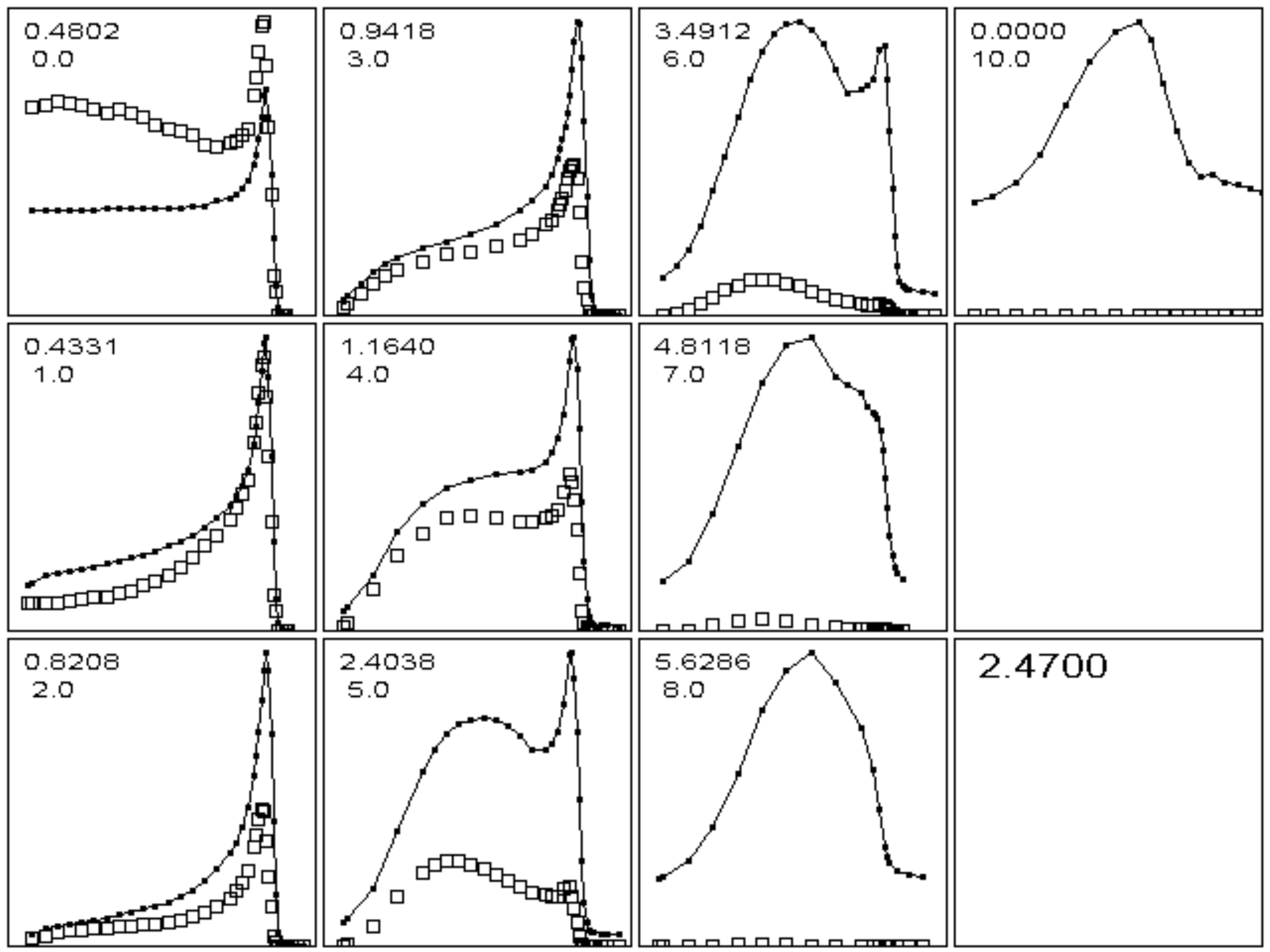}
\caption{Our data with Pedroni's fit (absolute comparison). The upper number in each frame is our usual figure of merit (Eq.\,\ref{eqn:fom}), the rms {\em ratio} of measurement to fit, which may have contributions hard to see in these linear plots. The large fit error at small $r$ is due to the smaller cross section of Pedroni's beam.\label{fig:PedLinFit}} 
\end{figure}

Fig.\,\ref{fig:PedLinFit} shows our data with Pedroni's fit (absolute comparison). The gross discrepancy at small radii is due to Pedroni's beam having a smaller $\sigma_x$ than ours. We have corrected that in Fig.\,\ref{fig:PedLinFitEM} by substituting our $\sigma_\mathrm{em}$ (Appendix \ref{app:SigEM}) for his $\sigma_\mathrm{P}$, with no other changes. 

The adjusted Pedroni fit at $r=0$ has a significant dose excess at mid range. That is because $T(w)$ (Eq.\,7P) includes, around mid range, dose deposited by secondaries which, in reality, is almost all deposited at larger radii. Stated more formally, the dose {\em on axis} of a single pencil beam is not proportional to $\Phi_\mathrm{em}\,S_\mathrm{mixed}$; it is very nearly proportional to $\Phi_\mathrm{em}\,S_\mathrm{em}$. The difference near $r=0$, being in the high-dose region, is potentially of clinical significance. That is our fundamental critique of the Pedroni parameterization, and has far reaching consequences discussed in Sec.\,\ref{sec:discussion}.

The second point of interest is more obvious, and has been widely noted: the Gaussian transverse dose distribution from nuclear secondaries falls far too quickly with $r$.

The third point is that the Bragg peak seems suppressed compared to the nonelastic bump at larger radii. That may be because the peak region of $T(w)$, which factors into the entire fit (cf. Eq.\,7P) is {\em reduced} by nuclear interactions whereas, in fact, the opposite is true. The Bragg peak at larger $r$ is {\em caused by} nuclear interactions, in this case elastic and inelastic. 

\subsubsection{Sawakuchi}
Sawakuchi et al. \cite{Sawakuchi2010} measure lateral dose profiles of a single beam on axis with a small IC. They use a reference IC, so their dose measurements are relative.\footnote{~They also use optically stimulated luminescent detectors (OSLDs) which generally confirm the IC data, and Gafchromic EBT film, which is of no value at low doses.} To accommodate very low doses they take 10 measurements of 4\;sec each. They study beams of 72.5, 148.8 and 221.8\,MeV, and take measurements in air and in water. In addition to the single beam measurements they perform direct field size factor (FSF) measurements at two depths on the axis of square fields of various sizes, uniformly populated with pencil beams. 

Their Figs.\,1$-$4 show results in air and are strongly influenced by beam contamination (a second Gaussian) from the tungsten wires of an upstream beam profile monitor. Their Figs.\;5a$-$c show FWHM, FW0.01M and FW0.001M as a function of depth in water $d$ at the three energies, essentially low-resolution contour plots. These too are heavily influenced by beam contamination (large widths at $d=0$) but the highest energy and lowest dose (221.8\,MeV, FW0.001M) curve has features not inconsistent with our data: a tendency to large radius at mid range, and the hint of a rapidly disappearing Bragg peak at large $d$. From their FSF measurements they conclude `$\ldots$ the low-dose envelope can be influential even for fields as large as 20 $\times$ 20\,cm.', not inconsistent with our findings.

\subsubsection{Clasie}
{\bf Clasie et al.} \cite{Clasie2012} compute, by Monte Carlo, `Golden Bragg peak' functions at 0.5\,cm range intervals over the clinical regime. Apart from normalization, these equal our $S_\mathrm{mixed}(z)/\rho$ or Pedroni's $T(w)$, namely the dose in water per fluence in air (i.e. mass stopping power) on the axis of an infinitely broad proton beam, except that they assume incident $\Delta E/E=0$ and are therefore universal, not facility-specific. 

The normalization of these fields is not taken from the Monte Carlo but determined (presumably to comply with absolute proton dosimetry protocol) by spot measurements at three depths insensitive to $\Delta E/E$ using a calibrated Markus chamber for the dose and a Faraday cup for the number of incident protons. The $\approx7$\% correction to Monte Carlo seems rather large, but is not discussed.

From the facility-independent Golden peaks a facility-dependent data set is constructed, by convolution with Gaussians of $\Delta E/E$ appropriate to each range at MGH. The relation between $\Delta E/E$ and range $R_{80}$ (their Fig.\,3, right) is found by determining, at a number of ranges, the convolution required to make the Golden peak agree with the MGH$-$specific peak as measured by a Bragg peak chamber ($R=4.08$\,cm) corrected (by Monte Carlo) for missing dose. The missing dose, shown at two ranges in their Fig.\,2, peaks (like the LTDCFs of Anand et al. \cite{Anand2012}) near full range, rather than mid range as we find (Fig.\,\ref{fig:Smixed}).

Clasie et al. go on to measure and parameterize `secondary contributions'. The measurement is a variant of Pedroni's square frames, concentric circles of increasing radii populated uniformly with monoenergetic pencils. The parameterization is essentially Pedroni's with a change in notation. Pedroni's $T(w)$ becomes $D_\infty(d,R_{80},\Delta E/E)$ with $D$ (the MGH$-$specific function just described) characterized as a `depth-dose'. Finally, the pencil beam model is used to predict the field size dependence of dose on axis for square fields of increasing size. Agreement is good for $R_{80}=25$\,cm, less good for 20\,cm.

\subsubsection{Anand}
Anand et al. \cite{Anand2012} are concerned with obtaining PISDs (Planar Integral Spot Dose, essentially $S_\mathrm{mixed}$), specifically with correcting them for the fact that the PTW Bragg peak chamber (BPC) is too small (radius = 4.08\,cm). The correction relies on measurement, but is compared with Monte Carlo .

Relative radial scans are measured with a small IC at selected depths at five beam energies, to a radius at which the dose is 0.001$\times$ central dose. (Azimuthal asymmetry of the spot is considered and dealt with.) The radial scans are integrated, once to the BPC radius and again to the full radius, and each integral is normalized to the dose $D_0$ measured on axis with an IC. (Detector size effect on $D_0$ is considered and dealt with.) That yields two computed numbers at the selected depths, PISD$_\mathrm{FULL}$ and PISD$_\mathrm{RBPC}$ which can be compared with the actual measurement PISD$_\mathrm{BPC}$. A `Long Tail Dose Correction Factor' (LTDCF\;$\equiv$\;PISD$_\mathrm{FULL}$/PISD$_\mathrm{RBPC}$), meant to be applied to PISDs as measured with the too-small Bragg peak chamber, is computed. 

\begin{figure}[p]
\centering\includegraphics[width=4.57in,height=3.5in]{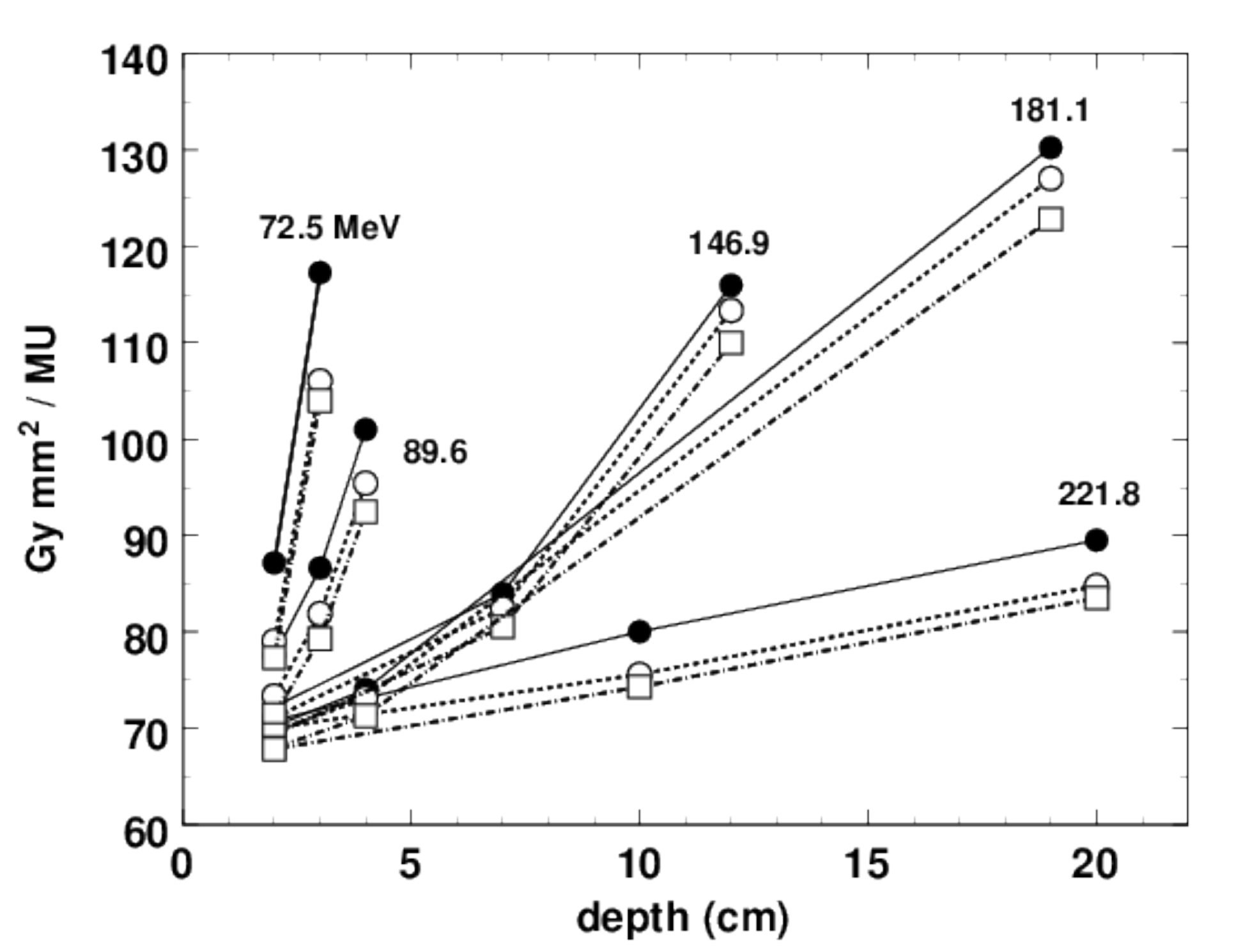}
\caption{From Table\,3 of Anand et al. \cite{Anand2012}. Full circles PISD$_\mathrm{FULL}$, open circles PISD$_\mathrm{RBPC}$ from radial integrals normalized to $D(0)$. Hollow squares, measured PISD$_\mathrm{BPC}$.\label{fig:AnandT3}}
\end{figure}

\begin{figure}[p]
\centering\includegraphics[width=4.57in,height=3.5in]{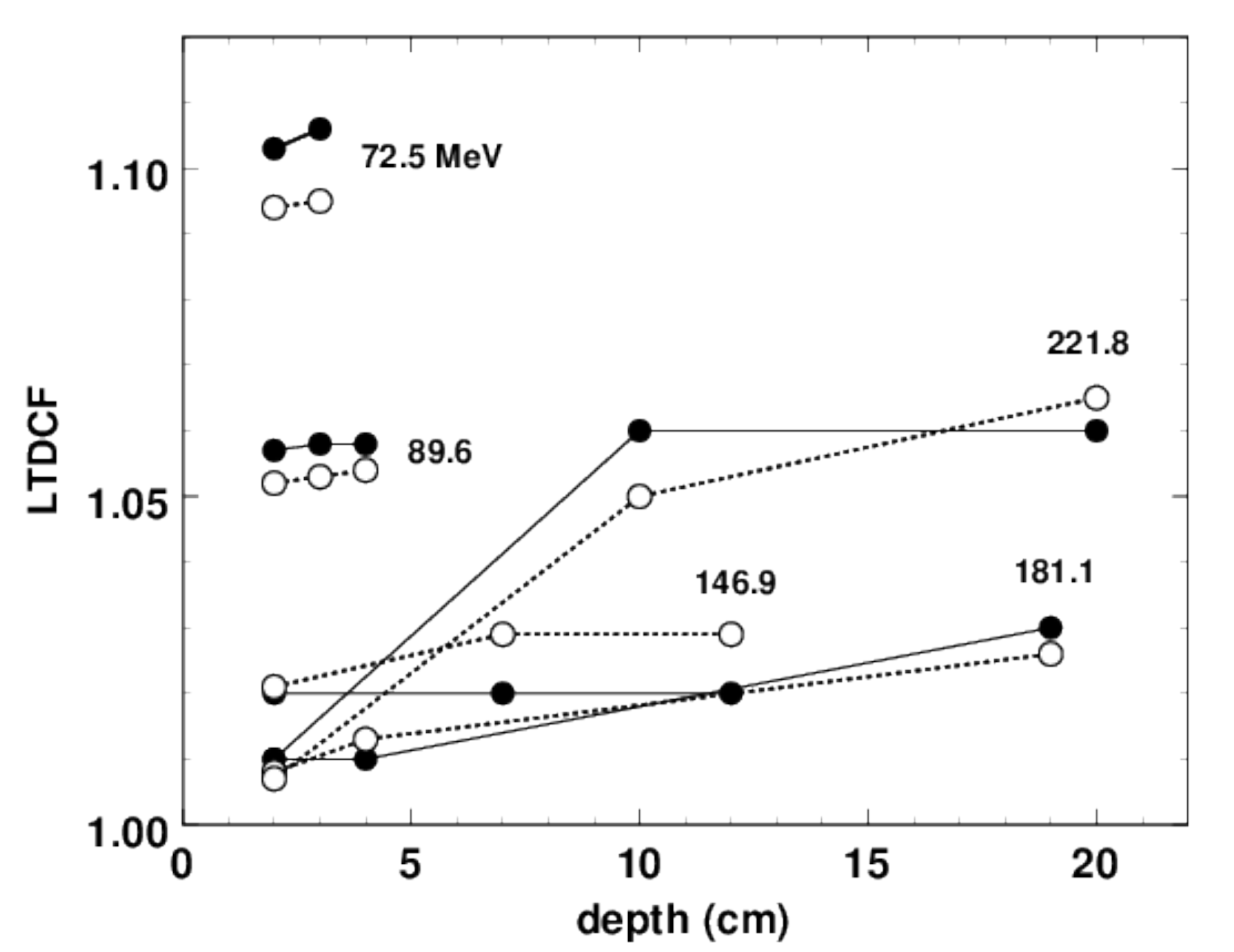}
\caption{From Table\,4 of Anand et al. \cite{Anand2012}. LTDCFs (Long Tail Dose Correction Factors) from the experimental method (full circles) and from Monte Carlo (open circles). (LTDCF - 1) is the fractional dose defect in the PTW BPC, and at 181.1\,MeV, $d=19$\,cm is nearly twice what we find at 177\,Mev, $d=19$\,cm (Fig.\,\ref{fig:Smixed}).\label{fig:AnandT4}}
\end{figure}

\begin{figure}[p]
\centering\includegraphics[width=4.72in,height=3.5in]{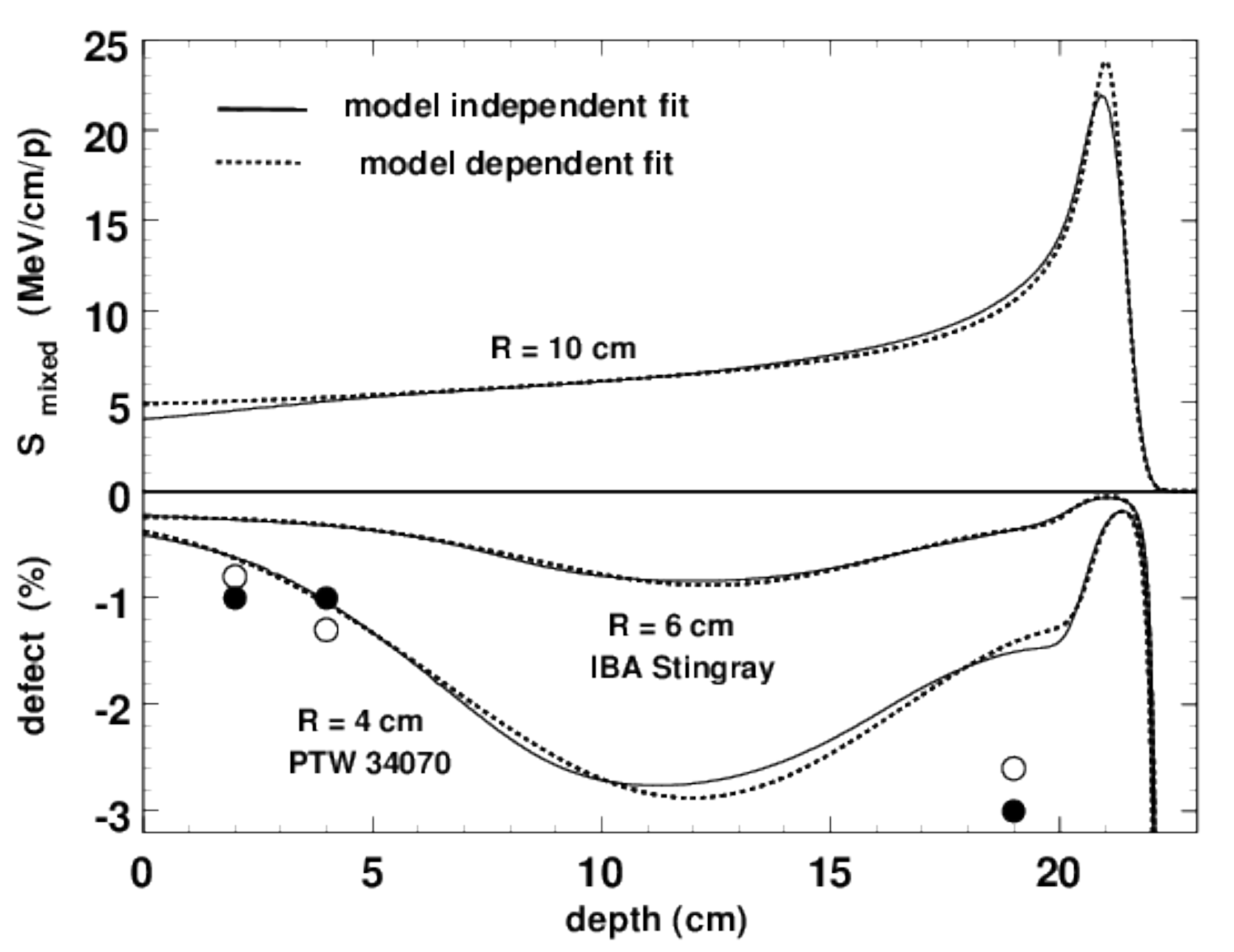}
\caption{Top: mixed stopping power $S_\mathrm{mixed}(z)$, the 2D integral over $r$ to $r_\mathrm{halo}=10$\,cm of $D_\mathrm{MI}(r,z)$ (solid line) and $D_\mathrm{MD}(r,z)$ (dashed line). Bottom: \% defect, or signal missed, if the integral is only taken to 4\,cm or 6\,cm, radii of the commercial Bragg peak chambers indicated. Filled circles are LTDCF$_\mathrm{meas}$, empty circles are LTDCF$_\mathrm{MC}$ at 181\,MeV from Anand et al. \cite{Anand2012}.\label{fig:Smixed}}
\end{figure}

Fig.\,\ref{fig:AnandT3} represents Table\,3 of \cite{Anand2012} giving the three PISDs at the energies and depths considered. Overall, the agreement between PISD$_\mathrm{RBPC}$ and PISD$_\mathrm{BPC}$ is only fair. Fig.\,\ref{fig:AnandT4} represents Table\,4, LTDCFs from measurement (Anand) compared with Monte-Carlo \cite{SawakuchiMCNPX}. The very large values at low energy cannot be from nuclear effects (which should be smaller, not larger) and reflect beam contamination. The results for 221.8\,MeV are of a reasonable order of magnitude, and hint at the expected peaking (cf. Fig.\,\ref{fig:Smixed}) of the correction at mid range. (Full range at 221.8\,MeV is 31\,cm.) At 181\,MeV, the shallow points agree with our computation (Fig.\,\ref{fig:Smixed}). At 19\,cm, just proximal to the Bragg peak, the LTDCFs are too large by a factor 2.

\subsection{Monte Carlo Studies}
\subsubsection{Sawakuchi MC\,\caproman1}
Sawakuchi et al. \cite{SawakuchiMC2010} compare Monte Carlo (MCNPX) with measurements described earlier \cite{Sawakuchi2010}. Beams of (72.5, 148.8, 221.8)\;MeV are studied, in air and at depths in water of (2, 10, 20)\;cm for (depth/range) = (0.46, 0.64, 0.64). 

MCNPX is used in all the M.D. Anderson (MDACC) papers and its limitations, in tallies and in the distinction between primaries and secondaries, are important. Sawakuchi writes
\begin{quote}
`Thus, because of limitations in the available MCNPX tallies, two simulations were performed in order to discriminate between
(1) primary protons, which included elastically scattered protons, and (2) secondary particles, which included secondary protons (inelastically scattered protons), deuterons, tritons, alphas, neutrons, etc.'
\end{quote}
The distinction between primary and secondary protons is inadequate. In elastic scattering on H, are both of the (indistinguishable) final state protons primaries? See Sec.\,\ref{sec:primaries}.
 
Returning to \cite{SawakuchiMC2010}, `secondaries' are displayed by subtracting a simulation of `primaries' from one for `total' events. Similarly, tungsten scattered events are displayed by subtracting `directly' (protons passing between the wires) from `total'. The attendant statistical errors are ameliorated by a variance reduction technique. Finally, the standard MCNPX MCS algorithm is apparently incorrect for large-angle scatters and is replaced, except at 72.5\,MeV.

The following lateral profile comparisons with data are made: in air at isocenter (primaries v. secondaries); in water at depth/range given above, primaries v. secondaries; in air at isocenter, directly v. scattered. FSF measurements and simulations are also shown, at (72.5, 221.8)\,MeV and (depth/range) = (0.46, 0.64). Generally speaking, simulations reproduce the data fairly well and show that the `low-dose envelope' comes from two distinct effects: beam protons passing through the tungsten wires of the upstream profile monitor (dominates at low energy), and nuclear reactions in water (dominates at high energy). 

\subsubsection{Sawakuchi MC\,\caproman2}
A second paper by this group \cite{SawakuchiMCNPX} overlaps the previous paper. The improved MCS model is spelled out in more detail, as is the modeling of the proton source. Beam divergence is ignored (scattering in the vacuum window is said to dominate) and $\sigma_x, \sigma_y$ of the beam just upstream of the vacuum window are adjusted, as functions of energy, so as to fit measured transverse profiles at isocenter. The fitted values are not compared with measurements from the profile monitor.

Measured in-air lateral profiles at (72.5, 148.8, 221.8)\;MeV compare well with the MC except for the tail of the 72.5\,MeV distribution (Fig.\,4Sa). Overall agreement of in-air FWHM, FW0.01M and FW0.001M as a function of energy seems about as good for the standard MCS model as for the new model, though details differ (Fig.\,4Sb). Measured in-air FW(HM, 0.01M, 0.001) levels as a function $z$ near isocenter agree better with the standard MCS model especially at 72.5\,MeV (Fig.\,5S). Since the source was adjusted to fit at isocenter, Fig.\,5S mainly proves that the virtual source of scanned beams is indeed halfway between the $x$ and $y$ magnets to a sufficiently good approximation.

In-water measurements as a function of depth at the same energies and relative dose levels agree reasonably well though again, the standard MCS model seems a bit better. The 221.8\,MeV result at the lowest level shows a hint of the expected maximum size around 2/3 range, and is well predicted by both MCS models (Fig.\,6Sc).

3D dose distributions with a $10\times10$\,cm field size and flat SOBPs at four energies are compared with depth-dose measurements using a Markus chamber and lateral measurements using either a pinpoint chamber or EBT film. The comparisons with MC are relative, not absolute. The SOBPs are predicted well, with no difference between standard and new MCS, as one expects because MCS is a second-order effect in depth-dose situations (Fig.\,7S). However, there is also very little difference between the two models in the lateral MCs (Fig.\,8S).

Finally, in Fig.\,9S, the authors compare MC predictions of the integral depth-dose of the 221.8\,MeV beam for various radii (4.08\,cm, 10\,cm and very large) with measurements using a 4.08\,cm radius Bragg peak chamber. The maximum predicted dose defect for that chamber and energy is 7.8\% at about 2/3 range. (Compare with our Fig.\,\ref{fig:Smixed}.)

Overall, the authors conclude that the standard MCS algorithm is better at the lower energies and worse at the highest energy. From the figures presented, it seems to us that standard is better at low energy and comparable at high energy.

\subsubsection{Grevillot}
L. Grevillot's PhD. thesis \cite{GrevillotPhD2011} concerns tuning and evaluation of the Gate/Geant4 Monte Carlo (MC) package for PBS and consists largely of three publications in their entirety.

The first paper G1 \cite{Grevillot2010} is concerned with optimizing parameters in Geant4. Our present paper is not about MC, but MC will undoubtedly play a large role eventually, in transporting the core among other things. Therefore it is troubling  (because of the implications for water) that G1 finds that Geant4 underpredicts beam spreading in PMMA by as much as 20\%.  

We have no new measurements of $\sigma_x$ at our disposal so we will compare Grevillot's PMMA and water simulations with Fermi-Eyges (FE) theory, which is reliable to a few percent if the correct scattering power is used. The results we need were found, and verified experimentally in water and aluminum, by Preston and Koehler \cite{preston}. They are discussed in \cite{transport2012}. If $R_1$ (cm) is the range of the beam then
\begin{enumerate}
\item $\sigma_x(z=R_1)\propto R_1$. Ref. \cite{transport2012} Table\,1 lists recommended coefficients for common materials.
\item $\sigma_x(z<R_1)$ obeys a universal law
\end{enumerate}
\begin{equation}\label{eqn:PKuniversal}
f_\mathrm{PK}\,\equiv\;\frac{\sigma_x(z)}{\sigma_x(R_1)}\;=\;\left[2\,(1-t)^2\ln\left(\frac{1}{1-t}\right)
  +\,3\,t^2-2\,t\right]^{1/2}\hbox{\quad,\quad}t\;\equiv\; z/R_1
\end{equation}
Therefore $\sigma_x$ at any depth in any homogeneous slab, for any incident charged particle at any energy, can be predicted quite accurately.\footnote{~The small-angle approximation must hold, which is always true for protons.} In Table\,\ref{tbl:G4oFE} we compare Grevillot's Geant4 results, corrected for $\sigma_x(0)$ by subtraction in quadrature (G4C), with the FE value.\footnote{~Subtraction in quadrature is an approximation. Without all three incident beam parameters we cannot do better.}

The smallest $\sigma$\,s for both materials should be given far less weight because of their sensitivity to $\sigma_x(0)$. Thus it would seem that G4 underpredicts both PMMA and water by some 10\%, the latter in qualitative agreement with Grevillot's Table\,5.

(We have found that radiochromic film is unsuitable for absolute dosimetry and for accurate beam profile measurements. Its nonlinearity with dose varies with readout technique and probably batch to batch. A small diode on a 1D scanner is the best way to measure beam profiles. The $\approx8\%$ over-response at low energies observed for some diodes \cite{amkdiode} does not matter in this application. Reference \cite{mcsbg} contains MCS measurements, with errors, for many materials and thicknesses, and it, or Moli\`ere theory itself, would be the best starting point for validating Monte Carlo MCS algorithms.) 

The second paper G2 \cite{Grevillot2011} is concerned with simulating the pencil beam and checking Gate/Geant4 against various measurements: pristine Bragg peaks, an SOBP in a 10$\times$10$\times$10\,cm volume, field size factors (FSFs), and a test pattern in air. The strategy is not to simulate the nozzle each time, but to use a library of beam parameters at nozzle exit, at 27 energies from 100 to 226.7\,MeV. 

\begin{figure}[p]
\centering\includegraphics[width=5.50in,height=3.5in]{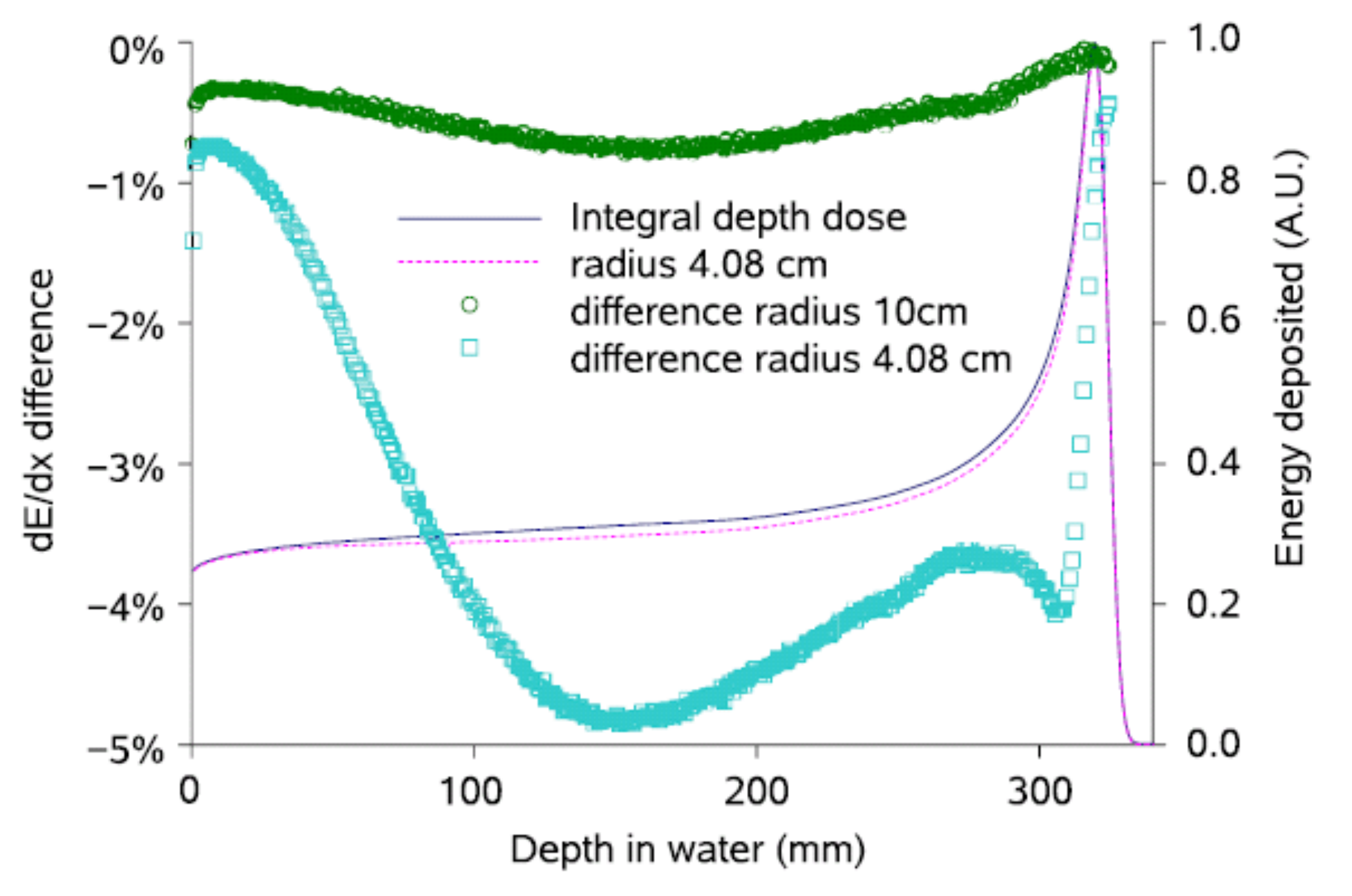}
\caption{Fig.\,7 from Grevillot et al. \cite{Grevillot2011}: Gate/Geant4 simulation of a 226.7\,MeV proton pencil beam in water. Empty squares: dose defect in a 4.08\,cm radius Bragg peak chamber. Peak dose defect occurs near mid range. The ratio of peak defect to Fig.\,\ref{fig:Smixed} is $4.8/2.8=1.71$ while the ratio of ranges is $32.4/21.2=1.53$.\label{fig:GrevillotFig7}}
\end{figure}

A set of pristine Bragg peaks is measured with a 4.08\,cm radius Bragg peak chamber in water. They are well simulated using the site-specific $\Delta E/E$ and (for consistency) {\em no} correction for dose defect in the Bragg peak chamber. Neverthless, G2 does consider the dose defect problem (dose inside 4.08\,cm compared to a very large field) and presents a result of interest to us.  Fig.\,\ref{fig:GrevillotFig7} is a copy of Fig.\,7G2, the dose defect in a 4.08\,cm radius chamber as a function of depth at 226.7\,MeV (simulation). The resemblance to our  Fig.\,\ref{fig:Smixed} (experimental) is striking. 

The two results are at different energies, 177 and 226.7\,MeV. The ratio of dose defects at mid range is $4.8\%/2.8\%=1.71$. For reference, the ratio of ranges is $32.4/21.2=1.53$. The increase with energy is certainly in the right direction, but it is hard to say at this stage whether the magnitude of the increase is reasonable.

\begin{figure}[p]
\centering\includegraphics[width=4.51in,height=4.in]{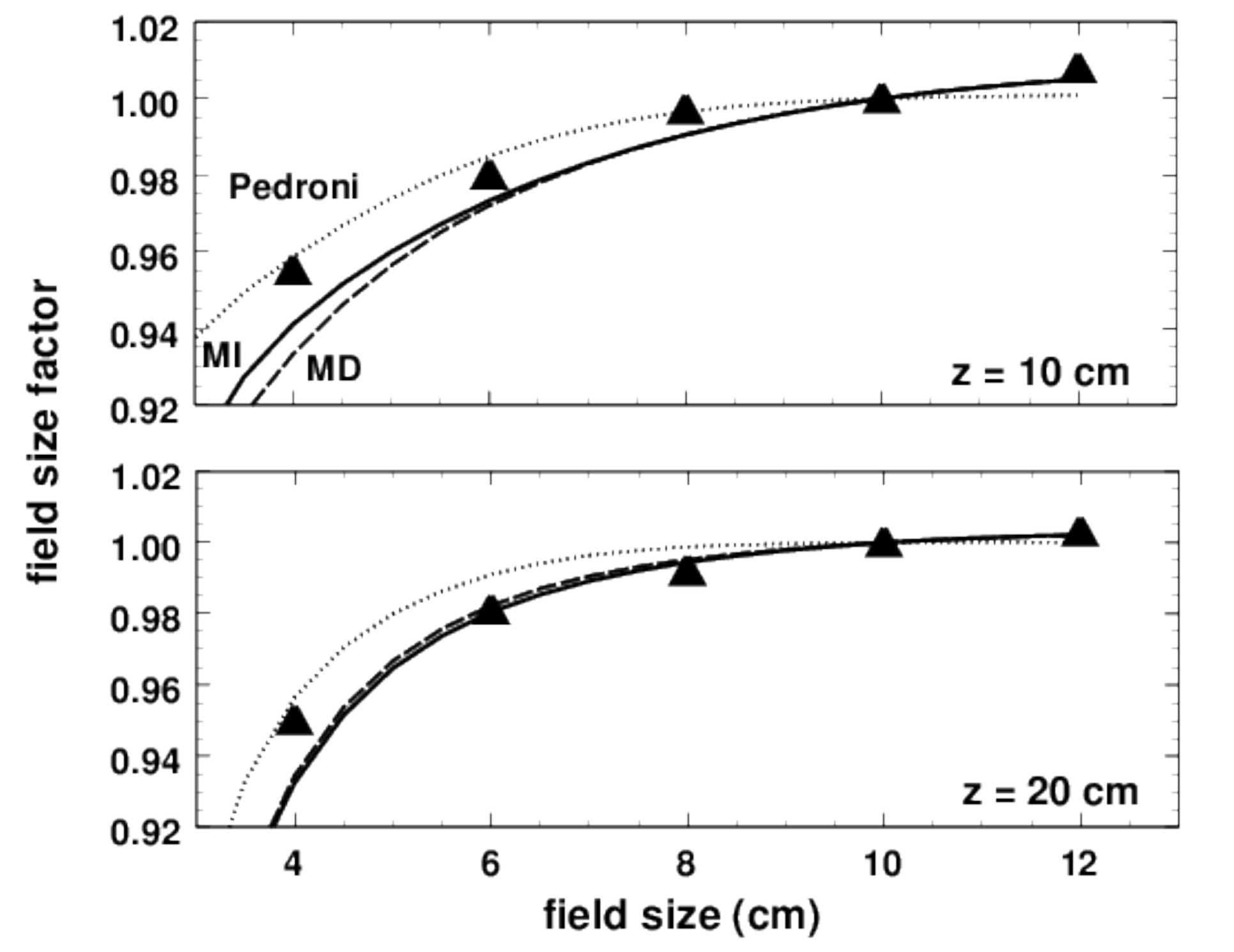}
\caption{Measured field size factors from Grevillot et al. \cite{Grevillot2011} for a monoenergetic 180\,MeV beam (triangles), compared with our calculation using three parameterizations of the core/halo at 177\,MeV. `Pedroni' is taken from \cite{pedroniPencil} adjusted to our incident beam size. $z=10$\,cm is at midrange, $z=20$\,cm is at the Bragg peak. The calculation used squares uniformly filled with pencils spaced at 0.25\,cm.\label{fig:FSF}}
\end{figure}

G2 also measures and simulates FSFs at several energies and depths in water. The usual trends are observed, and the simulation is good, generally within 2\%. Not surprisingly, the largest difference (3.2\%) is near mid range at the highest energy. Figure\,\ref{fig:FSF} shows Grevillot's measurements compared with FSFs we computed for our MD and MI parameterizations and Pedroni's adjusted to our incident beam size, for the same pencil to pencil spacing (0.25\,cm) as Grevillot's. At mid range the measured FSFs split the difference between us and Pedroni. At the Bragg peak, ours is favored. Absent experimental errors, it is hard to say more.

In Sec.\,4.4 of the thesis Grevillot compares the precompound and binary cascade (Bertini \cite{bertini}) nuclear models in their contributions to pristine and spread-out Bragg peaks. The fundamental problem is that the electromagnetic stopping power dominates both cases, and nuclear reactions are a small effect. The direct measurement of the core/halo we have presented will provide a far cleaner test of the nuclear model. We can already say the halo has Bertini-like features (evidence of quasi-elastic reactions).

The remainder of the thesis is devoted to a comparison, in a variety of configurations, of the XiO treatment planning system with GATE/Geant4 and with measurement. The most striking differences between XiO and MC occur when the protons travel some distance parallel to an interface, a well known problem for any pencil beam algorithm. In general these comparisons are dominated by electromagnetic effects (slowing down and scattering of the core) and insensitive to the nuclear model in either XiO or Geant4.

\subsubsection{Peeler}
Peeler and Titt \cite{Peeler2012} perform a Monte Carlo study to measure the effect, on the axis of a broad field exposed to monenergetic protons, of the transverse size of the initial core, comparing beams with $\sigma_x=0$ at the beam source with larger beams yielding $\sigma_x=0.73$\,cm (148.8\,MeV) or $\sigma_x=0.50$\,cm (221.8\,MeV) at the surface of the water tank. They find  a very small effect, at most 0.43\% for a 2$\times$2\,cm field at 221.8\,MeV, concluding that it is not necessary to consider the initial size of the beam in field size effect studies.

Of greater interest, they show (their Fig.\,1) depth dose distributions (integrated over radius) at the two energies. These show a maximum effect at mid range, and the onset of the aura. There is no Bragg peak in the halo because MCNPX counts elastically scattered protons as primaries \cite{SawakuchiMC2010}. Also, the transverse distribution of separated primaries and secondaries (their Fig.\,2) shows that the radial dependence is exponential, not Gaussian, except for the core. A break in slope at $r\approx5$\,cm in the 148.8\,MeV data strongly suggests the transition to the aura. The `primary' dose carries on to large radii, eventually reaching approximately the same slope as secondaries because elastically scattered protons are treated as primaries.

The main result of this paper is unsurprising, but it reveals other features in qualitative agreement with our experiment.

\subsection{Parameterizations}

\subsubsection{Soukup}
These authors describe \cite{soukup} a pencil beam algorithm designed for PBS with the possibility of beam modifying devices such as a collimator and ridge filter. Evidently their algorithm is used in the XiO treatment planning system \cite{GrevillotPhD2011}. They call the halo the `nuclear pencil beam'. 

They use Pedroni's parametrization of the core/halo (their Eq.\,11) but the nuclear interaction parameters (weight $W_\mathrm{nuc}$ and Gaussian $\sigma_\mathrm{nuc}$) are derived from Monte Carlo (Geant4/binary cascade) rather than experiment. They provide approximate empirical formulas for $W_\mathrm{nuc}$ and $\sigma_\mathrm{nuc}$ and claim that Geant4/precompound gives similar results. Their Figs.\,4\,--\,6 really only demonstrate mathematical consistency because they use essentially the same Monte Carlo for verification as was used to obtain $W_\mathrm{nuc}$ and $\sigma_\mathrm{nuc}$.

\subsubsection{Zhang}
Zhang et al. \cite{Zhang2011} focus on broad-field Bragg peaks and point out the desirability of an analytic formula such as Bortfeld's \cite{bortfeld}. They seek to remedy the small disagreement with experiment of Bortfeld's analytic formula (BAF), near 3/4 range, by generalizing four of Bortfeld's quantities ($\Phi_0$, $R$, $\sigma$ and $\epsilon$) to be functions of nominal range $R_0$ and introducing four new parameters $a\cdots d$, also functions of $R_0$. 

Each function is a quadratic polynomial (3 terms) in $R_0$. The required 24 fit parameters are simultaneously optimized over broad-field Bragg peaks at 7 energies, measured with a 4.08\,cm radius BPC. Those data are corrected for missing dose (up to 6\%) by what seems to be the method of Anand et al. \cite{Anand2012}, derived from measured lateral profiles.

The modified BAFs are compared with 3 Bragg peaks not used in the optimization and indeed found to agree better than the original BAF. When the modified BAFs are integrated into the Eclipse TPS, good agreement is also found with two 10\,cm SOBPs, at 20.6 and 30.6\,cm range. The transverse field size is not given directly but other numbers suggest 10$\times$10\,cm.

In summary, this paper shows that flat SOBPs can be obtained on axis in large fields by using accurate integral depth dose functions. The fact that these are analytic is a bonus.

\subsubsection{Zhu}
Zhu et al. \cite{Zhu2013} optimize the parameters of a commercial treatment planning system, Eclipse\;V8.917. We'll use suffix Z to reference their figures, equations and citations.

A great deal of complexity, as in all the MDACC papers, comes from the second Gaussian (what we have called the `second core') caused by a nonuniform scatterer (wire chamber) located upstream. This paper in fact treats it as a second core, transporting it by a partial Fermi-Eyges theory.\footnote{~Details of their treatment are insufficient. The in-text formula (following Eq.\,3Z) for $\sigma_i(E_k,z)$ is correct for a slab of thickness $z$ when initial phase space parameters $A_i$, $B_i$ and $C_i$ are known and scattering in the slab is ignored. For mixed slabs with scattering the more general form of Fermi-Eyges theory must be used \cite{transport2012}. Thus the second core could be correctly transported even through the water, if attrition of primary protons were put in.}

We are, however, more interested in the halo. Eq.\,2Z is promising, since it divides the `beamlet' dose into two terms, a core and a halo, with {\em separate} stopping powers. That raises two questions. First, what processes do the core and halo comprise? Second, how are the two stopping powers implemented in what follows? 

The first question is essentially: what is a primary and what is a secondary? Everyone agrees that charged particles from nonelastic nuclear interactions are secondaries. We argue that protons scattered coherently from nuclei (either elastically or inelastically, either Coulomb or nuclear force) are also secondaries (belong in the halo) because they stem from nuclear interactions and, more important, because their pattern of transverse dose deposition is dramatically different from the Gaussian core. MDACC papers, however, classify them as primaries. Thus \cite{Zhu2013}: `Primary protons include protons that only undergo elastic interactions with electrons and elastic proton-nucleus scatterings in the medium.' That definition gets into trouble with elastic scattering from H, where the outgoing protons look very like secondaries from (p,2p) reactions in O.

As to the second question, after Eq.\,2Z there is no further mention of two stopping powers. Eclipse requires IDDs (Integral Depth Dose, the same as PISD in some other MDACC papers) as input. In \cite{Zhu2013} these are Monte Carlo generated (presumably with realistic $\Delta E/E$) and checked by Bragg peak chamber measurements whose dose defect is corrected by Monte Carlo. One can only conclude that Eclipse eventually implements Pedroni's parametrization\footnote{~Reference\,[4Z] is the wrong Pedroni paper; they meant our \cite{pedroniPencil}.} using what we have called $S_\mathrm{mixed}$ for both the core and halo.

Fig.\,2Z(b) shows measured and Monte-Carlo corrected IDDs at all energies. The impression given by these curves that the correction vanishes at high energy is misleading, because the IDD is evaluated at $d=2$\,cm at all energies. At low energies, the IDD is undoubtedly dominated by the second Gaussian in the incident fluence. At high energies, 2\,cm is nowhere near mid range, where the correction is greatest.

Fig.\,5Z(a-g) shows measured FSFs (reference field is 10$\times$10\,cm) for monenergetic beams of three energies, along with predictions and errors for three models: single Gaussian incident fluence (SG), double Gaussian with empirical parameters (DG), and double Gaussian as implemented in Eclipse (DG-EFP). DG is best, and the second Gaussian is clearly needed. Of greater interest to us is Fig.\,5Z(g), a field size factor measurement at 221.8\,MeV and $d=23.2$\,cm (0.74\,$\times$\,range) where the error even in the DG model is greatest, $\approx\pm3\%$. The other graphs are at $d=2$\,cm or near the Bragg peak, where the effect of the nuclear halo is expected to be small.

\subsection{Alternative Transverse Profiles}

\subsubsection{Fuchs}
Fuchs et al. \cite{Fuchs2012} deal with He ions rather than protons. They develop a dose algorithm using Monte Carlo generated (GATE/Geant4) data, and find that a Voigt function (convolution of a Gaussian and a Lorentzian) best fits the transverse distribution in the halo.

\subsubsection{Knutson}
In his Master's thesis \cite{Knutson} Knutson evaluates a pencil beam algorithm in a 2D geometry (semi-infinite slabs) against the MCNPX Monte Carlo. He uses Pedroni's parametrization \cite{pedroniPencil} but, to better fit the transverse form of the halo, adds a Cauchy-Lorentz term proportional to $\sigma_\mathrm{N2}/(x^2+\sigma_\mathrm{N2}^2)$, with its own depth dependent weight factor $W_\mathrm{N2}(z)$ and width parameter $\sigma_\mathrm{N2}(z)$. The fit to the MC generated halo is considerably improved. Knutson notes that the eventual sharp falloff of the MC halo with $x$ is not reproduced, but that does not matter because the dose there is extremely low. 

\subsubsection{Li}
Li et al. \cite{Li2012}, evidently working with the data of Anand et al. \cite{Anand2012}, also find that an improved fit to the transverse distribution, particularly at mid range,  can be obtained by adding a Cauchy-Lorentz function to the traditional Gaussian. FSFs at 221.8\,MeV and mid range are then better, though still not perfectly, predicted.

\subsection{Summary}
Every paper dealing with parameterization of the core/halo uses Pedroni's form Eq.\,7P with $T(w)$ (alias IDD, PISD, improved BAF, golden Bragg Peak, $S_\mathrm{mixed}$) multiplying both the core and the halo. A Bragg peak chamber, corrected for missing dose either by inference from measured transverse profiles or by Monte Carlo, is used to measure or confirm $T(w)$. 

Only Zhu et al. \cite{Zhu2013} Eq.\,2Z identifies the IDD as a stopping power and hints that $S_\mathrm{mixed}$ might not be appropriate, indeed that different stopping powers might govern the core and halo. That insight does not appear to affect what follows in the same paper.

By contrast, several authors substitute an improved form, falling more slowly with $r$, for Pedroni's Gaussian halo. 

In defense of Pedroni \cite{pedroniPencil}, he writes
\begin{quote}
`These are first preliminary estimates of the `first moment' of the spatial lateral dose
distribution due to nuclear interaction products and large angle Coulomb scattering of a proton
pencil beam in water. Unfortunately the dosimetric method (the smallness of the effect) is
not sufficiently precise for extracting information on the higher moments (the shape of the
beam halo).'
\end{quote}
Too often, Pedroni's `first preliminary estimate' is treated as received truth.

\section{Discussion}\label{sec:discussion}

\subsection{Theory}
Often the domain of nuclear secondaries is viewed as a {\em terra incognita}, with Monte Carlo as the only hope of sorting things out. We have shown that, in fact, the qualitative features of the halo can be understood in terms of relatively simple two- and three-body processes which were well studied in the 1950s and 60s. They are the Moli\`ere single scattering tail, scattering on H, coherent elastic/inelastic scattering on O, and incoherent quasi-elastic scattering on O. We have also shown that the envelope of the halo can be accounted for by kinematics. More complicated processes exist, of course, and presumably populate the middle region of the halo.

\subsection{Measurement}
Measurements of the core/halo in the literature fall into two categories. There are depth-dose studies on axis using variants of the `frame' technique (e.g. \cite{pedroniPencil,Sawakuchi2010,Clasie2012}) and there are transverse studies of a single pencil at selected depths (\cite{Sawakuchi2010} and other MDACC papers). In the present paper we did depth-dose studies at various transverse distances from a single pencil. That has several advantages. First, the equipment was designed for depth-dose measurements and therefore lends itself most naturally to that technique. Second, in each pass the signal only varies by one or two orders of magnitude and is therefore easier to measure. Most important, in hindsight the results show that there is no single depth where the transverse distribution is characteristic of the whole. (If one has to chose a few key depths, one should measure near zero, at mid range, and near the Bragg peak.)

We have also shown that, using a fluence calibrated plane parallel ionization chamber instead of a reference chamber, it is as easy to obtain absolute doses as relative ones. Stating results in MeV/g/p rather than Gy/MU makes it easier to compare measurements from various sites.

Our positioning accuracy, dosimeter size and absolute accuracy could certainly be improved. Measurements at other energies are needed. All that said, we may claim to have obtained the first absolute and comprehensive measurement of the dose distribution in the core/halo.

\subsection{Disputing the Role of Integral Depth-Dose}\label{sec:IDD}
We come now to a topic which may be controversial. Because of the influence of Pedroni et al. \cite{pedroniPencil}, the integral of the depth-dose distribution over radius (our $S_\mathrm{mixed}$) plays a key role in parameterizing the pencil beam.  We argue that it should play no role whatsoever.

We do {\em not} argue that $S_\mathrm{mixed}(z)$ is unimportant. It is an essential concept when designing a range modulator for a flat SOBP in a broad field. Therefore it will also play a role in determining the relative weights of pencil beams if we pursue the same goal with PBS. We only argue that it is not useful, and even conceptually incorrect, in describing a {\em single pencil}. It mixes the very processes we are trying to tease apart.

Pedroni's Eq.\,1P
\[D(x,y,w)\;=\;T(w)\times G(x,w,\sigma_x(w))\times G(y,w,\sigma_y(w))\]
is {\em incorrect if applied to any small beam}. As $T(w)$ is defined, Eq.\,1P holds only near the axis when both $\sigma$s are so large that nuclear equilibrium prevails there. As a consequence, Eq.\,7P, where Pedroni splits the fluence into a core and halo, but leaves $T(w)$ multiplying both, is also incorrect.

To argue the matter more positively, we essentially repeat our derivation of the model dependent fit, specifically Eq.\,\ref{eqn:Dem}. Consider a monoenergetic beam of Gaussian cross section entering a water tank. At $z=0$ it is all core. As it propagates, it sheds primaries to elastic, inelastic and nonelastic nuclear interactions, spinning off a halo. At finite depth, the core still consists of quasi-monoenergetic protons, but fewer, more spread out, and of lower energy. Its dose is
\begin{equation*}
D_\mathrm{core}(r,z)\,=\,(1-\alpha z)\;G_\mathrm{2D}(r,z)\;
S_\mathrm{em}(z)/\rho_\mathrm{water}
\end{equation*}
What primaries remain are {\em still} governed by Bethe-Bloch theory ($S_\mathrm{em}$). The only manifestation of nuclear physics is in the `primary reduction coefficient' $\alpha$ which should not, logically speaking, be incorporated in $S$ but in the fluence multiplier, as shown.

Of course, some secondaries {\em do} deposit energy in the central region dominated by the core. But that is taken care of by a proper transverse model of the halo, which indeed overlaps the central region cf. Figs\,\ref{fig:trans12} and \ref{fig:trans21}.

Unlike \cite{Zhu2013} Eq.\,2Z we do not, in our model-dependent fit, use different stopping powers in the core and halo. We use $S_\mathrm{em}$ for both. $S_\mathrm{em}$ is exactly correct for the core and very nearly correct for the elastic/inelastic component of the halo. It is obviously not correct for the nonelastic component, or at the very least depends upon $r$ as well as $z$. But in that component we merely use it to set the overall absolute scale, so to speak. The more complicated behavior of the bump is absorbed into other parameters, and we could obtain an equally good fit without explicitly using $S_\mathrm{em}$.

\begin{figure}[p]
\centering\includegraphics[width=4.70in,height=3.5in]{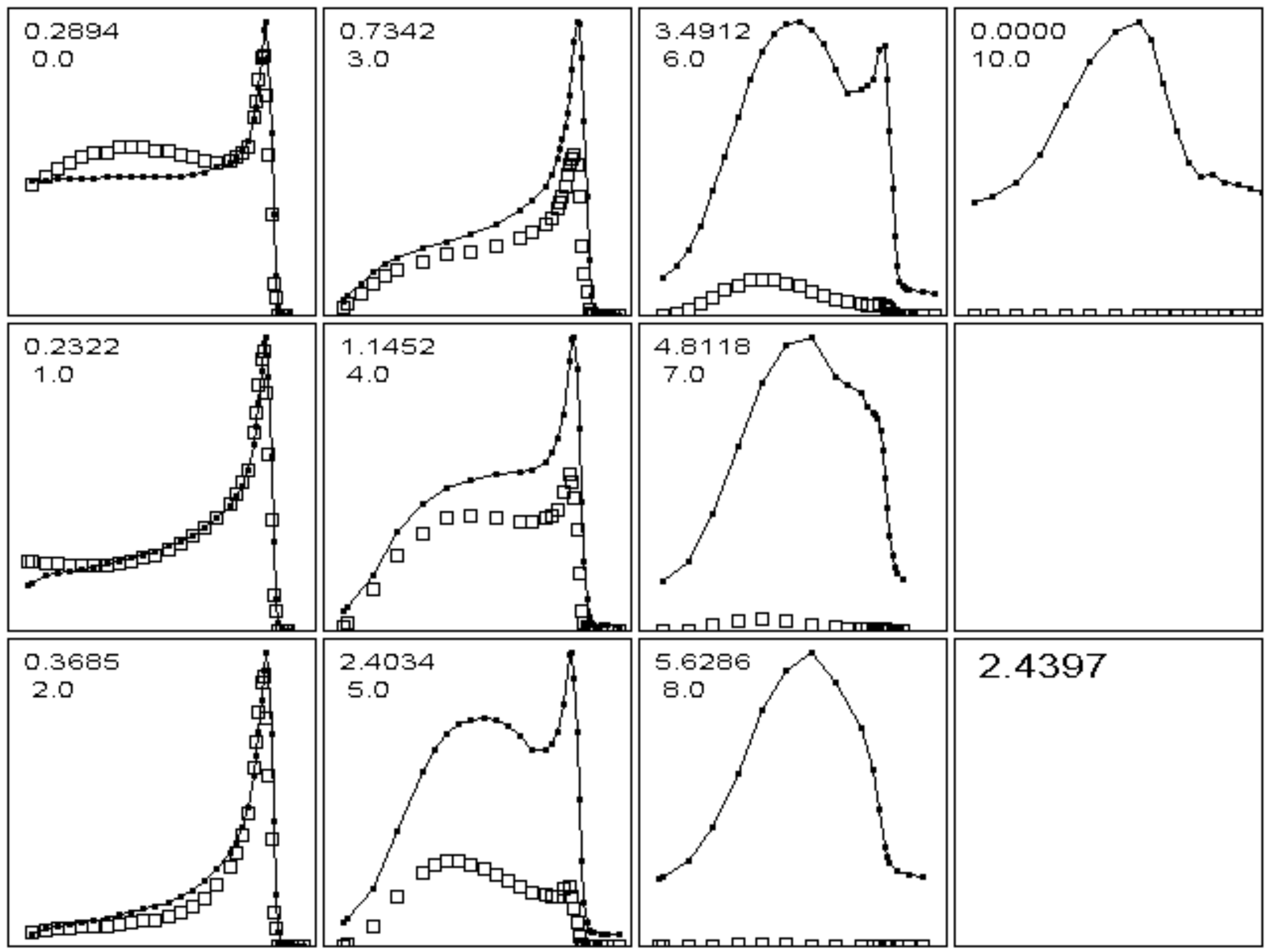}
\caption{Our data with Pedroni's fit (absolute comparison) using our beam size $\sigma_\mathrm{em}$. Agreement is much improved in the core, but largely unaffected for $r>3$\,cm.\label{fig:PedLinFitEM}}
\end{figure}

It remains to consider the quantitative effect of erroneously using $S_\mathrm{mixed}$ for both core and halo. The most serious error may be in the core around mid range (cf. Fig\,\ref{fig:PedLinFitEM}, $r=0$). The spurious dose enhancement is in the high-dose region and therefore may result in an incorrect target dose especially for small fields requiring few pencils. In the halo it may contribute, as argued previously, to the de-emphasis of the Bragg peak noted in Fig.\,\ref{fig:PedLinFitEM}.

\subsection{Disputing the Role of the Bragg Peak Chamber}
If our previous section is correct, the Bragg peak chamber is a red herring. Measurements with a Bragg peak chamber and Monte Carlo corrections for its being too small are not necessary in parameterizing the pencil beam. The proof is this very paper. We have displayed two independent fits to the core/halo, never mentioning integral depth-dose until we came to discuss the literature.

$S_\mathrm{mixed}$ is still needed in computing weight factors if we wish to use PBS to make a large uniform SOBP with a flat depth-dose on axis. But that  (apart from commissioning studies) is precisely the clinical case where we are least likely to use PBS. Large uniform fields are more efficiently delivered by scattered beams, unless we need magnetic deflection to maintain penetration. 

If $S_\mathrm{mixed}$ is required it can be obtained, as we did, by numerically integrating point measurements of the core/halo. Alternatively, one can use a small dosimeter in a sufficiently large water tank, or a sufficiently large multi-layer ionization chamber (MLIC), on the axis of a large uniform field. Except for time structure, the dosimeter does not know whether that field was obtained by scattering or magnetic scanning.

\subsection{The Full Integral}
We have integrated the dose over $r$ from 0 to 10\,cm to obtain $S_\mathrm{mixed}(z)$ (MeV/cm). It is instructive to further integrate that over $z$. That should equal the beam energy except for the energy deposited by neutrals (the aura) outside a cylinder of length 23\,cm and diameter 20\,cm. We performed that integral four ways. First, recognizing that we did not really measure the halo for $z<2$\,cm, we integrated from 2 to 23\,cm and compared the answer with the residual energy at 2\,cm. We also took the integral from 0 to 23 compared with the full energy, and each of those using the MD and MI model. The four answers are very similar, and averaging them we obtain $8.2\%\pm0.3\%$ (1$\sigma$) for the energy defect $\equiv$ (beam energy/full integral) $-$ 1.

At first glance that seems rather large. The energy lost to nonelastics is about 12\% (Appendix \ref{app:energyLost}) and one-third of that or 4\% is thought to go to neutrals \cite{pedroniPencil}. However, the integral is directly affected by the absolute normalization of our measurements. If our numbers were 4\% low, that would account for the difference. At present all we can say is that the full integral is in the right ballpark, and shows that our absolute numbers are not wildly off.

\subsection{Implications for Nuclear Buildup}
Our investigation of the core/halo may have inadvertently solved an old problem. When a broad proton beam is sent into a water tank from air or vacuum, a dose buildup of a few percent with a characteristic length of the order of 1\,$-$\,2\,cm is observed. This comes from charged secondaries, mostly protons, and was already measured in 1977 by Carlsson and Carlsson \cite{carlsson}. They noted that the magnitude of the buildup (at 185\,MeV) was smaller than expected.

However, `expected' refers to conditions of transverse nuclear equilibrium, with the beam radius comparable to one-third the range, or $\approx7$\,cm. Unfortunately Carlsson and Carlsson did not report their beam size but assuming (as is very likely) that it was smaller, they would have observed a smaller buildup. A consequence of the core/halo model is that nuclear buildup as measured on axis by any detector of finite (but not very large) transverse size is a fairly complex matter, and its magnitude will be smaller than at equilibrium. 

\subsection{Primaries and Secondaries}\label{sec:primaries}
At the risk of repeating ourselves we revisit the definition of `primaries' and `secondaries'. Obviously any outgoing non-proton is a secondary. Recoil nuclei (except from scattering on H) are a special case and, as we have shown, can be ignored to a good approximation. Protons are the only contributor to the core and by far the largest contributor to the halo. They arise from
\begin{enumerate}
\item Multiple Coulomb scattering, nearly Gaussian transverse distribution.
\item Single Coulomb scattering, tends to $1/\theta^4$ distribution (Moli\`ere tail).
\item Coherent nuclear scattering, elastic, roughly exponential, quantum interference with 2.
\item Coherent nuclear scattering, inelastic, resembles 3 with peak shifted.
\item Incoherent nuclear scattering, nonelastic, peaks around mid range.
\end{enumerate}
We have proposed that all final state protons except the first category be termed `secondaries'. The strongest argument for that definition is that it includes final state protons from elastic scattering on H, which strongly resemble the last category kinematically. It also means that anything non-Gaussian is a secondary.

It appears that different Monte Carlos handle this issue differently.

\subsection{The Core/Halo as a Monte Carlo Benchmark}
The measurements reported here (Table\;\ref{tbl:dmlg}) are by far the best data set yet available with which to benchmark the nuclear model of a Monte Carlo in the energy region appropriate to proton radiotherapy. The region dominated by EM effects (core) and the region dominated by nuclear effects (halo) are separated nearly as well as can be (a smaller beam would be better), so that any defects in the nuclear model should be separable from those in the MCS model and therefore comparatively easy to diagnose.

An advantage of our data set is that division into primaries and secondaries is unnecessary. They are separated by their spatial characteristics, and a single run even with MCNPX should reproduce Table\;\ref{tbl:dmlg}. At large $r$, statistics will obviously be a problem, and the dose should be scored in annuli, perhaps with $\Delta r$ increasing with $r$ and $\Delta z$ decreasing around the Bragg peak if that is feasible. The comparison with MC should be absolute, and the units should be MeV/g/p. 

\section{Grand Summary}
We have measured the core/halo at 177\,MeV using a test beam, depth scans at fixed distances from the beam in a water tank, a fluence calibrated beam monitor, and a small dose calibrated thimble chamber. Though our measurements could be improved in absolute accuracy, dosimeter size, and positioning accuracy, they are the most comprehensive to date, and presented in full (Table\,\ref{tbl:dmlg}).

We also present model-dependent and model-independent fits to out data. The former breaks the dose down into component nuclear processes, and therefore gives some measure of the contribution of each. However it is complicated and therefore not very portable. The latter is quite simple and a better fit, but takes no advantage at all of the physics. A hybrid approach using calculation from first principles for the EM part (core) but model independent for the rest (halo) may ultimately prove more convenient for computing dose distributions in PBS. 

We reviewed the literature and found the parametrization of Pedroni et al. used universally, sometimes with the Gaussian transverse dependence of the halo improved. In our opinion there is a more fundamental problem. A stopping power that mixes electromagnetic and nuclear processes is used where a purely electromagnetic one is indicated. If that is true, the considerable fraction of the literature given over to Bragg peak chamber measurements, corrected by Monte Carlo or empirical means, is irrelevant.

Our data set should provide an incisive test of Monte Carlo nuclear models. One need simply send a beam, with the initial parameters found in our mode-dependent fit, into a water tank, and look at the absolute dose to water (MeV/g/p) everywhere. Outside the core (which, by the way, tests the electromagnetic model) the dose is of purely nuclear origin and not masked (as in Bragg peak measurements) by much larger electromagnetic effects.

\section{Acknowledgements}
BG thanks Harvard University, the Physics Department, and the Laboratory for Particle Physics and Cosmology for their continuing support. We thank Drs. Grevillot, Pedroni, Sawakuchi and Zhu for communication regarding their work.

\appendix
\section{Fraction of Energy Lost to Nonelastic Reactions}\label{app:energyLost}
Let $R$ (cm) be a range, $T$ (MeV) a kinetic energy and $z$ the depth in a water tank. Janni \cite{janni82} has tabulated the total fraction of nonelastic nuclear reactions experienced by a proton beam of initial kinetic energy $T_0$ stopping in water. The rate of loss of primaries to nonelastic events is nearly constant at $\delta\approx0.012/$cm \cite{bortfeld}. (In this calculation, unlike the rest of the paper, we treat elastically and inelastically scattered protons as primaries. The only important difference is their angle.) To compute the total energy lost we must take into account how much energy each primary still had when it interacted. To a fair approximation the range-energy relation of protons in water is \cite{bortfeld}
\[R\,=\,a\,T^b\,=\,0.0022\,T^{1.77}\]
leading to
\[T(z)\,=\,(1-z/R_0)^{0.565}\,T_0\]
In $dz$ at $z$, $N_0\,(1-\delta z)$ primaries remain, a fraction $\delta\,dz$ have nonelastic nuclear reactions, and each time $T(z)$ is lost to the primary beam. Thus the total energy lost is
\begin{eqnarray*}
\Delta T_\mathrm{beam}&=&N_0\int_0^{R_0}(1-\delta z)\,T(z)\,\delta\,dz\\
&=&\delta\,N_0\,T_0\int_0^{R_0}(1-\delta z)(1-z/R_0)^{0.565}\,dz\\
&=&(\delta R_0)\,N_0\,T_0\int_0^1(1-(\delta R_0)\,x)(1-x)^{0.565}\,dx
\end{eqnarray*}
$x$ is the normalized depth, $N_0T_0$ is the total incident beam energy and $\delta R_0$ is the fraction of primaries lost while going the full range. Therefore the fraction of energy lost is given by two integrals which depend only on the power-law exponent $b$ : 
\[(\delta R_0)\times\left(\int_0^1(1-x)^{0.565}\,dx\,-\,(\delta R_0)\int_0^1x(1-x)^{0.565}\,dx\right)\,=\,0.639\,(\delta R_0)-0.249\,(\delta R_0)^2\]
For instance, if $(\delta R_0)=20\%$ of the primaries are lost, as at 175\,MeV in water \cite{janni82}, then 12\% of the energy is lost to nonelastic reactions.

\section{Center-of-Mass Angle in Elastic Scattering}\label{app:CM}
To estimate the typical kinetic energy of the recoil nucleus (a function of scattering angle) in $^1$H(p,p)$^1$H and $^{16}$O(p,p)$^{16}$O reactions, we must consult the literature on elastic scattering which invariably presents results in terms of the center-of-mass (CM) angle rather than the lab angle. Assume we have solved the kinematic problem relativistically in the lab frame (Appendix\,\ref{app:QE}) so we know the recoil energy as a function of lab angles. At a given lab angle, what is the CM angle?\footnote{~The first thing to recall is that there is only one CM angle, because the particles are back-to-back in that frame even when their lab angles differ.}

Let $1,2,3,4$ denote respectively the incident proton of rest energy $m_pc^2$ entering along the $z$ axis, the target nucleus of rest energy $A\,m_pc^2$, the scattered proton, and the recoil nucleus. Let the $x$ axis correspond to some transverse direction. For any particle of rest energy $mc^2$, velocity $\vec{v}$, momentum $\vec{p}$\,, kinetic energy $T$ and total energy $E$ the following equations apply: 
\begin{eqnarray}
E&\equiv&T\,+\,mc^2\\
\vec{\beta}&\equiv& \vec{v}/c\;=\;\vec{p}\,c/E\\
\gamma&\equiv&1/\sqrt{1-\beta^2}\;=\;E/mc^2\\
(pc)^2&=&((T/mc^2)+2)\,mc^2\,T\label{eqn:pcsq} 
\end{eqnarray}
The velocity of the CM frame as seen from the lab is \cite{hagedorn}
\begin{eqnarray}
\vec{\beta}_\mathrm{CM}&=&\vec{p}c/E=(\vec{p}_1c+\vec{p}_2c)/(E_1+E_2)\\
\gamma_\mathrm{CM}&=&1/\sqrt{1-\beta_\mathrm{CM}^2}
\end{eqnarray}
In our special case the CM moves along the $z$ axis (beam direction) and we find
\begin{equation}
\beta_\mathrm{CM}=(pc)_1/(m_pc^2(\gamma_1+A))
\end{equation}
Having already solved the lab problem we know $(pc)_{3x}\,,(pc)_{3z}$ and $T_3$ and a Lorentz trans\-formation \cite{hagedorn} to the CM yields
\begin{eqnarray}
(pc)_{3x\mathrm{CM}}&=&(pc)_{3x}\\
(pc)_{3z\mathrm{CM}}&=&\gamma_\mathrm{CM}\,\left((pc)_{3z}-\beta_\mathrm{CM}(T_3+mc^2)\right)\\
\theta_\mathrm{CM}&=&\tan^{-1}\big((pc)_{3x\mathrm{CM}}/(pc)_{3z\mathrm{CM}}\big)
\end{eqnarray}

\section{Kinematics of Elastic and Quasi-Elastic Reactions}\label{app:QE}
\begin{figure}[p]
\centering\includegraphics[width=4.57in,height=3.5in]{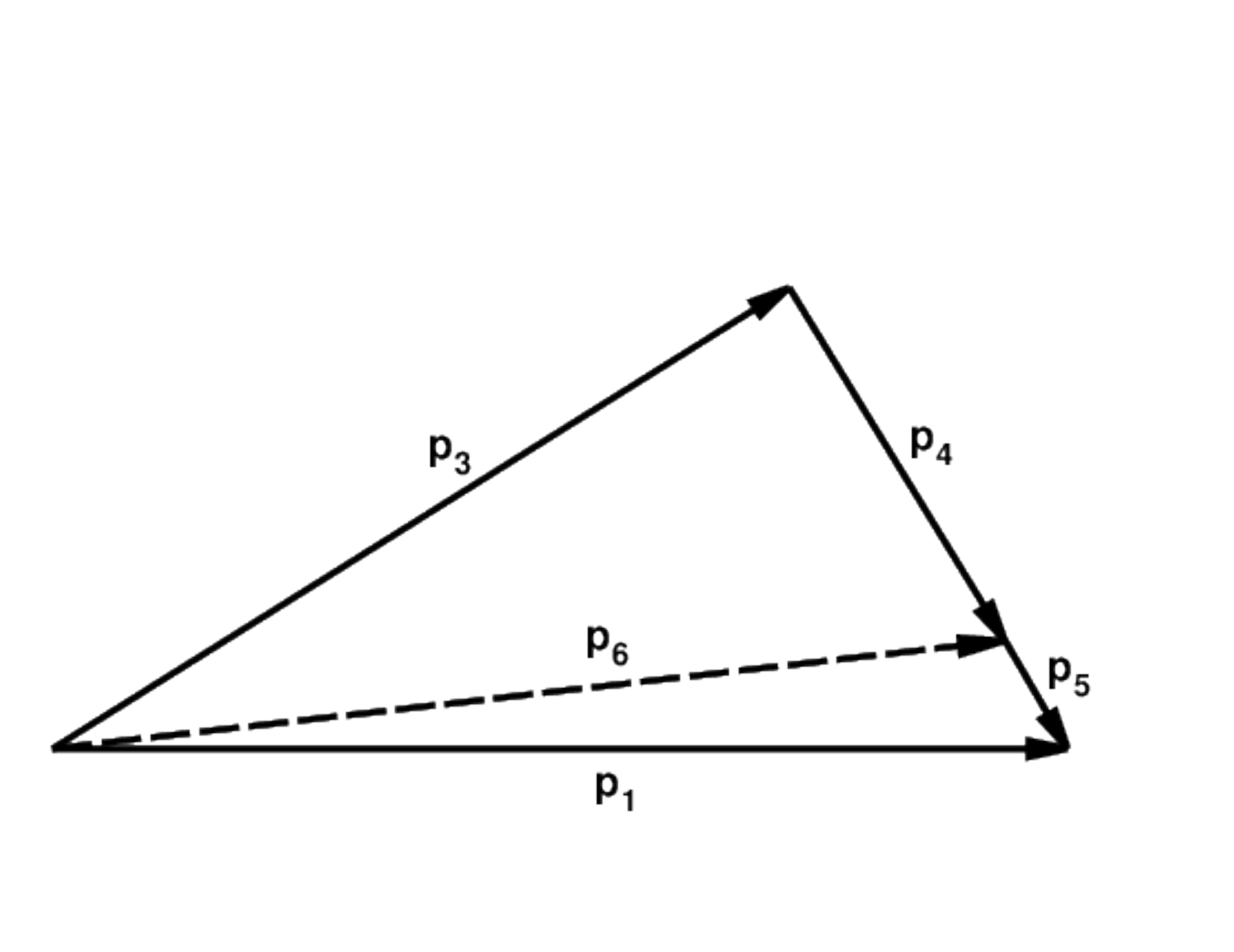}
\caption{Conservation of momentum corresponding to the largest radius in Fig.\,\ref{fig:haloBump}. $\vec p_5$ is oriented along $\vec p_4$ to maximize the transverse component of $\vec p_3$.\label{fig:haloDiagram}}
\end{figure}

We use the notation of Appendix\,\ref{app:CM} except that particles $3,4,5$ are the outgoing proton of interest, the other outgoing proton, and the recoiling residual nucleus of rest energy $(A-1)m_pc^2$. $\theta$ is a polar angle. If we set $T_2=p_2=0$ (the target nucleus is at rest in the lab) and put all the unknowns on the LHS, conservation of energy and momentum read 
\begin{eqnarray}
T_3+T_4&=&T_1-T_5-E_\mathrm{B}\label{eqn:consT}\\
\vec p_3+\vec p_4&=&\vec p_1-\vec p_5\;\equiv\;\vec p_6\label{eqn:consp}
\end{eqnarray}
where $\vec p_6$ is a known vector (Fig.\,\ref{fig:haloDiagram}). If the residual nucleus is left in its ground state, the binding energy $E_\mathrm{B}$ equals $-Q_\mathrm{M}$. Otherwise it is greater. The momentum of any particle is related to its rest energy and kinetic energy by Eq.\,\ref{eqn:pcsq}.\footnote{~Using this equation rather than its nonrelativistic limit is the only place relativity appears in our calculations.}

To begin, assume coplanar scattering in the $z,\,x$ plane; then Eqs.\,(\ref{eqn:consT},\,\ref{eqn:consp}) are three equations in four unknowns. Assuming the value of any one unknown final-state quantity allows us to find the three others. Let us assume some small value for $T_3$. We can immediately find $T_4$ (Eq.\,\ref{eqn:consT}) and $p_3$ and $p_4$ (Eq.\,\ref{eqn:pcsq}). We then have a triangle of known sides at a known orientation to the axes (Fig.\,\ref{fig:haloDiagram}) and can find the angles of $\vec p_3$ and $\vec p_4$ with respect to the beam by trigonometry. With $\theta_3$ and a proton range-energy table we can find the endpoint of the trajectory of proton 3. Next, we assume a larger value of $T_3$, and repeat until $T_4$ goes negative.

In the noncoplanar case, the plane containing $\vec p_1,\,\vec p_5$ and that containing $\vec  p_3,\,\vec p_4$ intersect in the line defined by $\vec p_6$ (Fig.\,\ref{fig:haloDiagram}). Any possible noncoplanar reaction can be obtained by rotating the $\vec  p_3,\,\vec p_4$ plane about that line, which has no effect on the procedure just outlined. The space angle between $\vec p_3$ and $\vec p_1$ changes, but it (and therefore the transverse range) are maximal in the coplanar case already considered.

We are free to choose $\theta_5$ by trial and error so as to maximize the transverse range of proton 3. Not surprisingly, $\vec p_5$ wants to be collinear with $\vec p_4$ (Fig.\,\ref{fig:haloDiagram}).

The same equations and procedure cover elastic scattering on H (which is always coplanar) if we simply set $A$ to 1 and $T_5,\,p_5$ and $E_\mathrm{B}$ to 0. 

\section{Model-Dependent Fit}\label{app:MD}

\subsection{Computation of $\sigma_\mathrm{em}$}\label{app:SigEM}
$\sigma_\mathrm{em}$ stands for $\sigma_x(z)$ (cm), the rms transverse spread of the electromagnetic core. This has some initial value at $z=0$ (the rms spread of the incident beam) and increases thereafter due to multiple Coulomb scattering. However, the incident beam is characterized by not one but three quantities and all three must be transported through $z$ to get an accurate picture of any one of them. The procedure is called Fermi-Eyges theory. This is a bare-bones account giving only the key equations with a few comments. Far more detailed accounts are available online \cite{transport2012,scatPower2010}.

The three quantities in question are moments of the scattering power $T_\mathrm{dM}$ of which more below.\footnote{~In this section $T_\mathrm{dM}$ is a scattering power (radian$^2$/cm) whereas $T$ (unsubscripted) is kinetic energy.} Their definitions and physical interpretations are
\begin{eqnarray}
A_0(z)&\equiv&\int_0^z\,T_\mathrm{dM}(z')\,dz'\;=\;<\theta^2>\;=\;\sigma_\theta^2\label{eqn:A0}\\
A_1(z)&\equiv&\int_0^z(z-z')\,T_\mathrm{dM}(z')\,dz'\;=\;<x\,\theta>\label{eqn:A1}\\
A_2(z)&\equiv&\int_0^z(z-z')^2\,T_\mathrm{dM}(z')\,dz'\;=\;<x^2>\;=\;\sigma_x^2\label{eqn:A2}
\end{eqnarray}
where $\theta$ in this section stands for $\theta_x$, the projected angle in the $x$ direction. The related quantity
\begin{equation}\label{eqn:B}
B(z)\;\equiv\;A_0A_2\;-\;A_1^2
\end{equation}
is also important. The $A$\,s can be viewed geometrically as giving the bounding box and shape of a centered ellipse, the {\em beam ellipse}. The area enclosed by that ellipse is $\pi\sqrt{B}$\,, the {\em emittance}.

Eqs.\,\ref{eqn:A0}$-$\ref{eqn:A2} assume that the $A$\,s are initially zero (ideal pencil beam incident). In reality, they usually have finite initial values, and there are standard equations for propagating or {\em transporting} these through any stack of (transversely) uniform slabs \cite{transport2012}. In our special case we have but a single slab, water, with a set of trial $A$\,s incident, and we require the transported $A$\,s at many depths since we wish to compute dose at many depths. The differential form of the transport equations \cite{kanematsu08} is most convenient:
\begin{eqnarray}
\Delta A_0&=&\tilde{T}_\mathrm{dM}\Delta z\label{eqn:a0k}\\
\Delta A_1&=&(A_0+(\tilde{T}_\mathrm{dM}/2)\Delta z)\Delta z\label{eqn:a1k}\\
\Delta A_2&=&(2A_1+(A_0+(\tilde{T}_\mathrm{dM}/3)\Delta z)\Delta z)\Delta z\label{eqn:a2k}
\end{eqnarray}
the tilde meaning `value at step center'. Eqs.\,\ref{eqn:a0k}$-$\ref{eqn:a2k} are equivalent to integration by the midpoint rule. The residual range $R$ needed to compute kinematic quantities is transported simply by
\begin{equation}
\Delta R\;=\;-\Delta z
\end{equation}

Let us return to scattering power and supply some formulas. Multiple scattering theories such as Moli\`ere's \cite{mcsbg} apply to {\em finite} slabs whereas any transport theory, deterministic or Monte Carlo, implies a {\em differential} approach. We divide the target into many small steps, and hope that the answer will converge (approach a limiting value independent of step size). Moli\`ere theory and its equivalents do not converge when used that way. Mathematically, we require a well defined derivative which we will call the {\em scattering power} namely 
\begin{equation}
T_\mathrm{xy}\;\equiv\;d<\theta^2>/dz
\end{equation} 
$T_\mathrm{xy}$ (the subscripts are placeholders) is best thought of as a differential approximation to Moli\`ere theory concocted to fill our need: some formula which, integrated over a finite slab, agrees with Moli\`ere theory and converges with step size. Of a number of such formulas \cite{scatPower2010} we recommend
\begin{equation}\label{eqn:TdM}
T_\mathrm{dM}(z)\;=\;f_\mathrm{dM}(pv,p_1v_1)\times\left(\frac{15.0\,\mathrm{MeV}}{pv(z)}\right)^2\;\frac{1}{X_S}
\end{equation}
Unlike stopping power, whose instantaneous value depends only on local variables (particle energy stopping material), scattering power must have some nonlocal dependence on the `history' of the proton (material already traversed) to give an accurate result over a wide range of slab thicknesses.\footnote{~That seems counterintuitive, but {\em multiple} Coulomb scattering, unlike stopping, is not a primitive process between one proton and one atom. It can be viewed as the statistical behavior of a cohort of protons. There is no fundamental reason their mean squared angle should be exactly proportional to target thickness, and indeed it is not.} That is supplied by $f_\mathrm{dM}$ which depends on $pv$ (local) {\em and} $p_1v_1$ (initial):
\begin{eqnarray}\label{eqn:fdM}
f_\mathrm{dM}&\equiv&0.5244+0.1975\log_{10}(1-(pv/p_1v_1)^2)+0.2320\log_{10}(pv/\mathrm{MeV})\nonumber\\
&&-\;0.0098\log_{10}(pv/\mathrm{MeV})\log_{10}(1-(pv/p_1v_1)^2)
\end{eqnarray}
$pv$ of any particle of rest mass $mc^2$ at any kinetic energy $T$ is
\begin{equation}
pv\;=\;\frac{(T/m_pc^2)+2}{(T/m_pc^2)+1}\;T
\end{equation}
The quantity $X_S$ in Eq.\,\ref{eqn:TdM} is the {\em scattering length} of water, a material property similar to radiation length $X_0$ but numerically larger for light materials. For an elementary material
\begin{equation}\label{eqn:XS}
\frac{1}{\rho X_S}\;\equiv\;\alpha N r_e^2\;\frac{Z^2}{A}\left\{2\log(33219\,(AZ)^{-1/3})-1\right\}
\end{equation}
For a compound
\begin{equation}\label{eqn:XSBragg}
\frac{1}{\rho X_S}\;=\;\sum_iw_i\left(\frac{1}{\rho X_S}\right)_i
\end{equation}
where $w_i$ is the fraction by weight of the $i^\mathrm{th}$ constituent. Other quantities are defined in \cite{transport2012} which also gives values for many elements and compounds.

We now have the machinery, once we initialize a beam ellipse, to transport it through water and find $\sigma_x$ at any depth $z$. Rather than the somewhat abstract $A$\,s used internally, we wish our initial parameters to be $\sigma_x$ (mm), $\sigma_\theta$ (mrad) and emittance (mm\;mrad). The conversion from practical units to $A$\,s is given in Eqs.\,\ref{eqn:p5}$-$\ref{eqn:fsq} Appendix\,\ref{app:SigEM}. 

The first two are straightforward but emittance (assigned parameter $p_7$) poses two problems. First, it has a maximum possible value namely $\sigma_x\,\sigma_\theta$ and in a realistic PBS beam line will probably be near that value. Second, specifying the emittance (an area) does not tell us whether the beam is converging ($A_1<0$) or diverging ($A_1>0$). Let us define a signed function $f(p_7)$ controlling the amount to which $A_1$ approaches its smallest or largest values (0, maximum emittance or $\sigma_x\,\sigma_\theta$, zero emittance). $f$ is large when $p_7$ is small, so realistic values of $p_7$ will be small and negative (fairly large emittance, converging). Sign information passes from $p_7$ to $f$ to $A_1$.

\begin{figure}[p]
\centering\includegraphics[width=4.51in,height=3.5in]{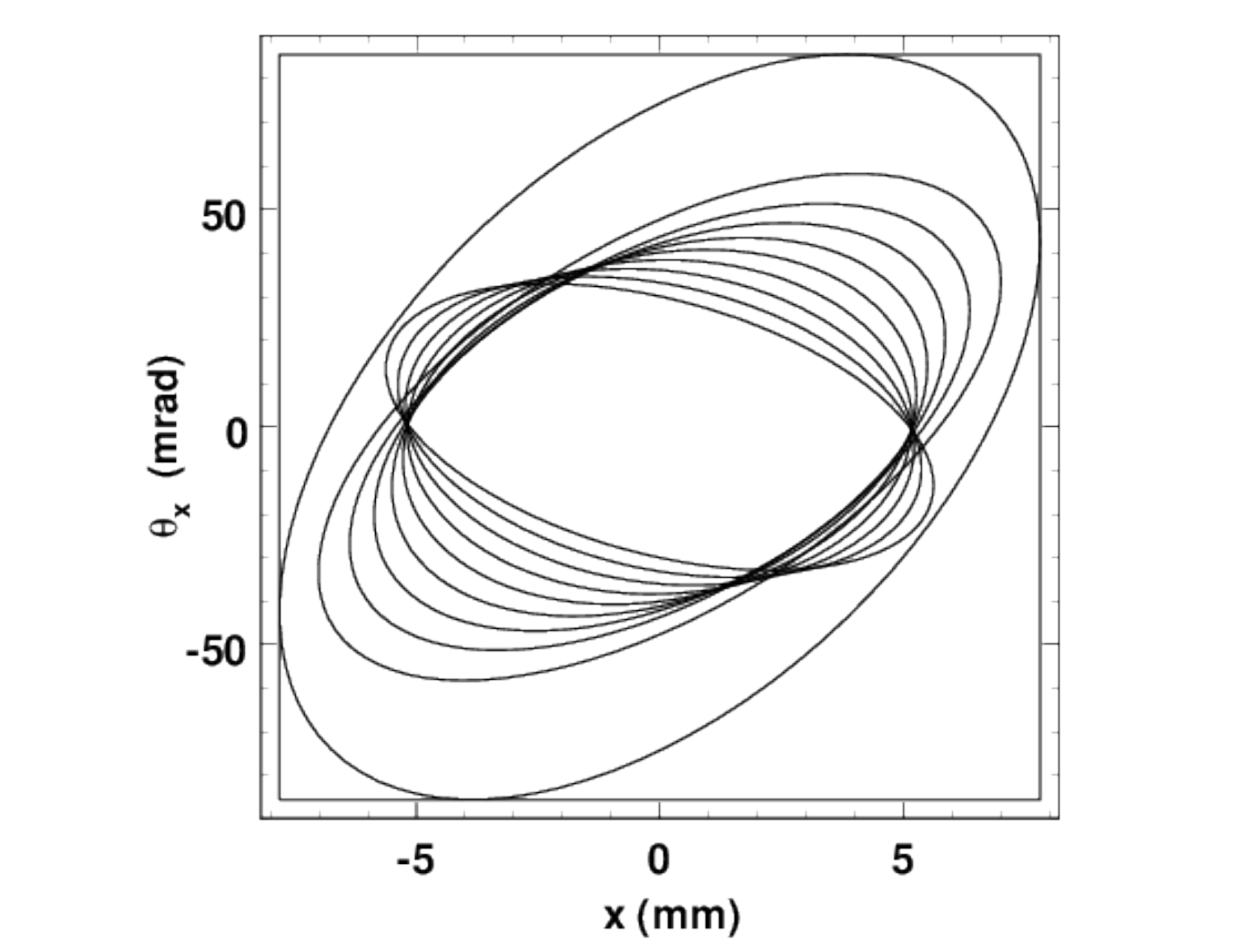}
\caption{Evolution of the beam ellipse in the water tank.\label{fig:ellipses}}
\end{figure}

Fig.\,\ref{fig:ellipses} shows the evolution of the beam ellipse in 21\,cm of water starting from $A$\,s defined by the best-fit values of $p_5-p_7$.

\subsection{Computation of $S_\mathrm{em}$}\label{app:SEM}
$S_\mathrm{em}$ is the electromagnetic stopping power (MeV/cm/p) of protons of initial range $R_0$ (cm) in water. If divided by the density $\rho$ of water, it equals the dose (MeV/g) per fluence (p/cm$^2$) that would obtain on the central axis of a very broad field exposed to a constant proton fluence if all nuclear reactions were turned off. It can be computed as the convolution of tabulated $-(dT/dz)$\footnote{~For consistent notation we use $dT/dz$ instead of the more conventional $dE/dx$.} with a Gaussian whose $\sigma$ combines range straggling with the energy spread of the incident beam.\footnote{~The correct `Gaussian' is also skewed towards smaller $z$ because projected range is, statistically, always less than pathlength, never greater. We will ignore this `detour factor' effect which is very small \cite{icru49}.} Thus
\begin{equation}\label{eqn:convo}
S_\mathrm{em}(z)\,=\,\int_{z-5\sigma}^{z+5\sigma}-\frac{dT}{dz}(z)\;G(z-z',\sigma_\mathrm{sem})\,dz'
\end{equation}
where $G$ is a conventional 1D Gaussian, normalized to unit area, and $\sigma_\mathrm{sem}$ is adjustable because we do not know the beam energy spread {\em a priori}. Eq.\,\ref{eqn:convo} assumes the Gaussian is sufficiently near zero at $\pm\,5\sigma$ (adjustable in the program). We perform the integral numerically using Simpson's Rule \cite{nr} but we must use a cutoff procedure because $-dT/dz$ is singular at end-of-range. We break the integral, at $z=z_c$\,, into two parts, one nonsingular and the other susceptible to approximation: 
\begin{equation}\label{eqn:cutoff}
S_\mathrm{em}(z)\,=\,\int_{z-5\sigma}^{z_c}-\frac{dT}{dz}(z)\;G(z-z',\sigma_\mathrm{sem})\,dz'\,+\,
  \int_{z_c}^{z+5\sigma}-\frac{dT}{dz}(z)\;G(z-z',\sigma_\mathrm{sem})\,dz'
\end{equation}
The second integrand is dominated by the singularity of $-dT/dz$ at $R_0\approx z_c$. We approximate that integral by setting $z'=z_c$ and find
\begin{equation}\label{eqn:Sem}
S_\mathrm{em}(z)\,\approx\,\int_{z-5\sigma}^{z_c}-\frac{dT}{dz}(z)\;G(z-z',\sigma_\mathrm{sem})\,dz'\,+\,
  T(R_0-z_c)\,G(z-z_c,\sigma_\mathrm{sem})
\end{equation}
$T(R_0-z_c)$ is the residual kinetic energy, which we obtain from a range-energy table, at $z_c$. The overall result is insensitive to $z_c$.

\begin{figure}[p]
\centering\includegraphics[width=4.51in,height=3.5in]{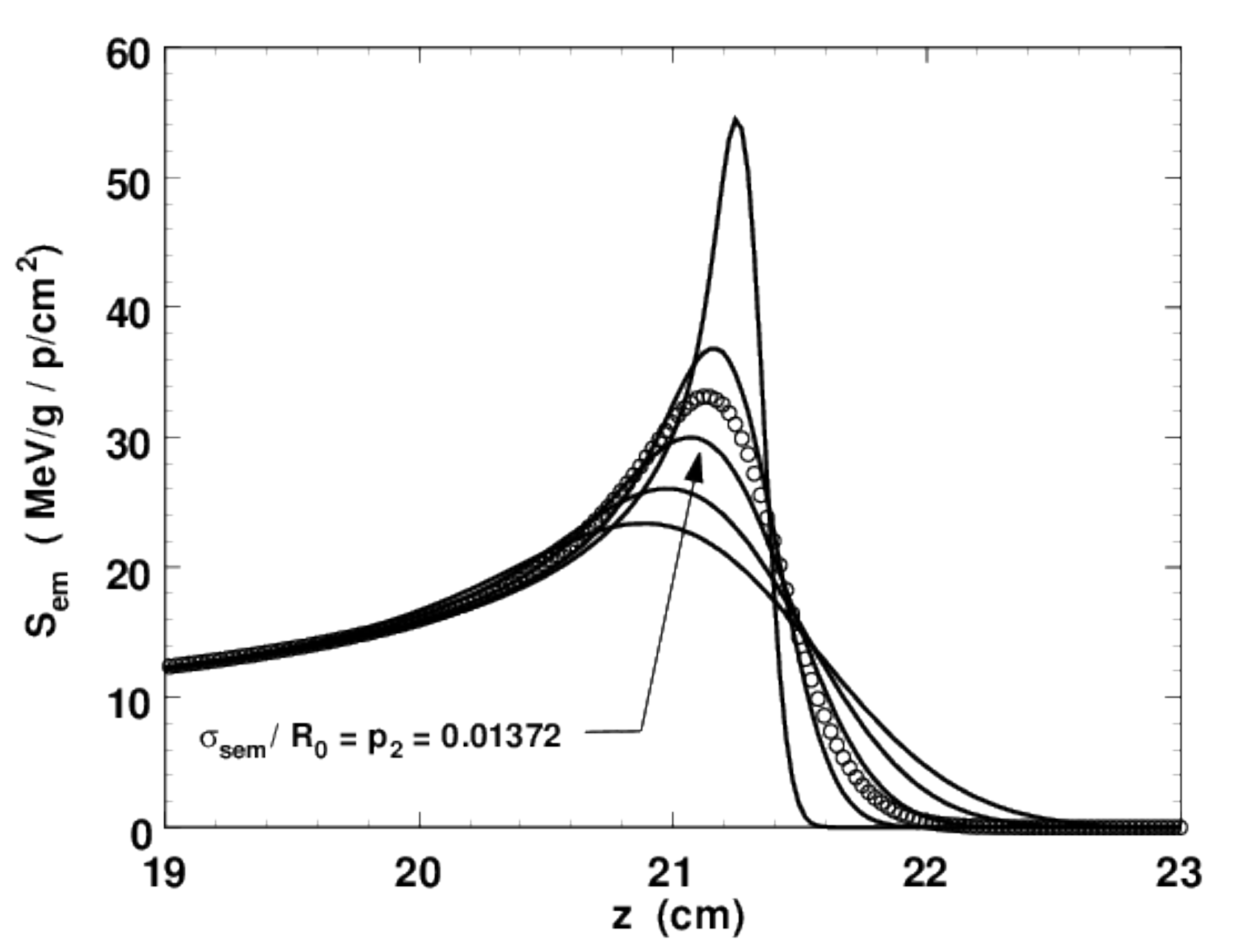}
\caption{Un-normalized $S_\mathrm{em}$ peaks of varying $\sigma_\mathrm{sem}$ (lines) and an interpolated peak (circles). The middle peak (arrow) corresponds to the best-fit value of $\sigma_\mathrm{sem}/R_0$. The full range of values is $\pm0.01$. The value of $S_\mathrm{em}$ at $z=0$ (off scale) is $4.82$\;(MeV/g)/(p/cm$^2$).\label{fig:SEMpeaksFull}}
\end{figure}

\begin{figure}[p]
\centering\includegraphics[width=4.51in,height=3.5in]{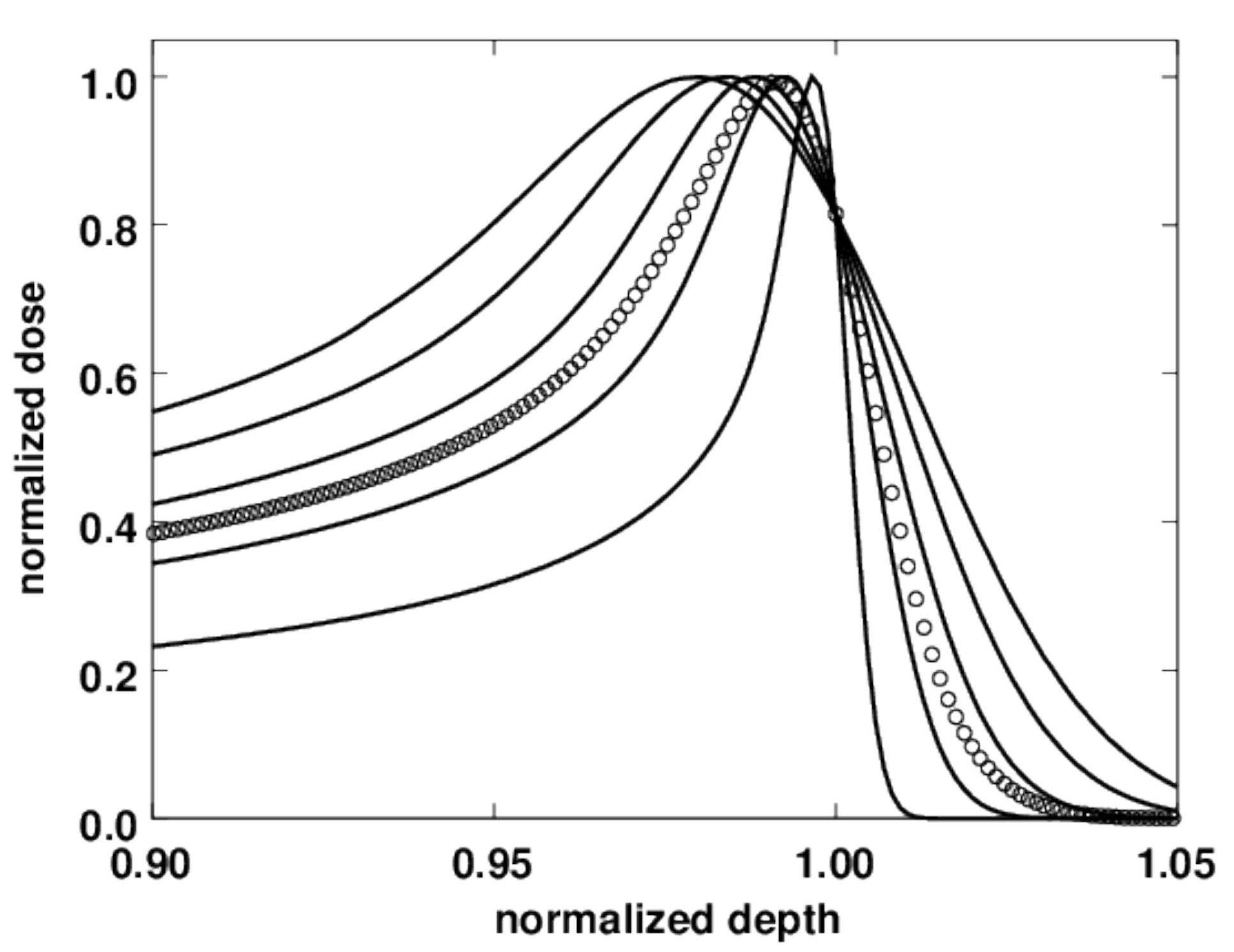}
\caption{Normalized $S_\mathrm{em}$ peaks corresponding to Fig.\,\ref{fig:SEMpeaksFull}.\label{fig:SEMpeaks}}
\end{figure}

The convolution is far too slow for a minimization procedure so we construct, at initiali\-zation time, a family of $S_\mathrm{em}$ curves which we will later interpolate. The raw peaks (Fig.\,\ref{fig:SEMpeaksFull}) are unsuitable for interpolation so we convert them to a family of curves $y_i(x_i\,;\sigma_j)$ where $y$ is normalized to 1 and $x=z/R_0$ (Fig.\,\ref{fig:SEMpeaks}). To find the value of $S_\mathrm{em}$ at a particular $x,\sigma$ we use
\begin{eqnarray}
y_a&=&y_{i,\,j}\,+\,(y_{i,\,j+1}-y_{i,\,j})\;(\sigma-\sigma_j)/\Delta\sigma\\
y_b&=&y_{i+1,\,j}\,+\,(y_{i+1,\,j+1}-y_{i+1,\,j})\;(\sigma-\sigma_j)/\Delta\sigma\\
y&=&y_a\,+\,(y_b\,-\,y_a)\;(x-x_i)/\Delta x
\end{eqnarray}
and multiply $y$ by $S_\mathrm{em,\,peak}(\sigma)$ (itself interpolated) to undo the normalization. $i,\,j$ are indices of the table entries just below $x,\sigma$ and $\Delta x,\,\Delta\sigma$ are the corresponding table increments.

Ultimately we find the best-fit value of $\sigma_\mathrm{sem}/R_0$ to be $1.37\,\%\pm0.1$\%. Taken at face value this means that the beam energy spread, though small, is not completely negligible, since tabulated range straggling \cite{janni82} corresponds to $\sigma/R_0=1.11$\%. The quadratic difference is $0.8$\%. 

\subsection{Equations and Parameter Values}\label{app:parms}

\begin{eqnarray}
D(r,z)&=&D_\mathrm{em}(r,z)\,+\,D_\mathrm{el}(r,z)\,+\,D_\mathrm{qe}(r,z)\,+\,D_\mathrm{au}(r,z)\\
\noalign{\vspace{15pt}}
D_\mathrm{em}(r,z)&=&(1-\alpha z)\,G_\mathrm{2D}(r,\,\sigma_\mathrm{em}(z))\,
  S_\mathrm{em}(z_\mathrm{n,adj},\sigma_\mathrm{sem}(r_\mathrm{n}))\\
G_\mathrm{2D}(r,\sigma)&\equiv&(2\pi\,\sigma^2)^{-1}\;\exp{(\displaystyle{-.5}\left(r/\sigma\right)^2)}\\
\sigma_\mathrm{em}(z)&=&\hbox{Fermi-Eyges theory}\,(A_0,\,A_1,\,A_2,\,z)\qquad\,\hbox{(Appendix\;\ref{app:SigEM})}\\
S_\mathrm{em}(z_\mathrm{n},\,\sigma_\mathrm{sem})&=&\hbox{convolution     integral}\;(z_\mathrm{n},\,\sigma_\mathrm{sem})\;\qquad\qquad\hbox{(Appendix\;\ref{app:SEM})}\\
R_0&=&p_1\\
z_\mathrm{n}&=&z/R_0\\
r_\mathrm{n}&=&r/(R_0/3)\\
\sigma_\mathrm{sem}&=&p_2\,(1+p_{13}\,r_\mathrm{n})\\
\alpha&=&p_3\,(1+p_4\,z_\mathrm{n})\\
A_2&=&(0.1\,p_5)^2\label{eqn:p5}\\
A_0&=&(0.001\,p_6)^2\label{eqn:p6}\\
f&\equiv&-(1+p_7)\;,\quad -1\le p_7\le0\label{eqn:p7}\\
&\equiv&+(1-p_7)\;,\qquad 0<p_7\le1\\
A_1&=&\pm((1-f^2)A_0A_2)^{1/2}\,\hspace{.9in}\hbox{(transfer sign from $f$)}\label{eqn:fsq}\\
\noalign{\vspace{15pt}}
D_\mathrm{el}&=&\alpha\,z\,p_8\,E_\mathrm{2D}(r_\mathrm{n},r_\mathrm{el})\,
  S_\mathrm{em}(z_\mathrm{n,adj},\sigma_\mathrm{sem}(r_\mathrm{n}))\\
E_\mathrm{2D}(r,r_e)&\equiv&(2\pi\,r_\mathrm{e}^2\,)^{-1}\,\exp\,(\displaystyle{-r/r_\mathrm{e}})\\
r_\mathrm{el}&=&p_9(1+p_{10}\,r_\mathrm{n}+p_{11}\,r_\mathrm{n}^2\,)\\
z_\mathrm{n,adj}(r_\mathrm{n})&=&(1+p_{12}\,r_\mathrm{n}^2\,)\,z_\mathrm{n}\\
\noalign{\vspace{15pt}}
D_\mathrm{qe}&=&\alpha\,z\,E_\mathrm{2D}(r_\mathrm{n},r_\mathrm{qe})\,G_\mathrm{1D}(z_\mathrm{n},c,z_0,h)\,
  S_\mathrm{em}(z_\mathrm{n,adj},\sigma_\mathrm{sem}(r_\mathrm{n}))\\
G_{1D}(z,c,z_0,h)&\equiv&c\,\exp(-z_t^2)\\
z_t&=&0.83255\,(z-z_0)/h\\
c&=&p_{14}\,(1+p_{15}\,r_\mathrm{n})\\
z_0&=&p_{16}\,(1+p_{17}\,r_\mathrm{n})\\
h&=&p_{18}\,(1+p_{19}\,r_\mathrm{n})\\
r_\mathrm{qe}&=&p_{20}\,(1+p_{21}\,r_\mathrm{n}+p_{22}\,r_\mathrm{n}^2\,)\\
\noalign{\vspace{15pt}}
D_\mathrm{au}&=&p_{23}\;E_\mathrm{2D}(r_\mathrm{n},r_\mathrm{au})\\
r_\mathrm{au}&=&p_{24}\,(1+p_{25}\,r_\mathrm{n})
\end{eqnarray}

\section{Model-Independent Fit}\label{app:MI}

\subsection{Cubic Spline Fit}\label{app:CSfit}
Assume we have a data set $y_i(x_i), i=1\ldots M$ with $x_i$ distinct and monotonically increasing.\footnote{~Most fits don't care about the order of the $x_i$\,, or whether they repeat. This one does, because we will need to take the numerical second derivative.} A function that passes through all the given points is called an `interpolant' and that is by far the most common application of cubic splines. 

Qualitatively, think of the cubic spline as an all-purpose smooth function which, in any single interval, has the considerable flexibility of a cubic polynomial: anything from a straight line to a skewed parabola. That gives it enormous flexibility as a fitting function.

More specifically, a cubic spline A) passes through each of a given set of points, B) is a cubic polynomial with (in general) four different coefficients in each successive interval between data points and C) has a continuous first derivative at each $x_i$ (the `smoothness' condition: no corners). Given the $y_i(x_i)$, the `passes through all points' and `continuous first derivative' conditions are not quite sufficient to determine all $4(M-1)$ polynomial coefficients, basically because we don't know the derivatives at $x_1^-$ and $x_M^+$ making the `continuity' condition powerless there.

The problem is usually resolved in one of two ways. Either stipulate the first derivative at the endpoints, or require that the  second derivative there be zero (the so-called `natural' boundary condition). We are free to use one method at one end and the other method at the other end. With that additional information, all the polynomial coefficients can be found \cite{nr}, and we have constructed a smooth interpolant of the original data set.

\begin{figure}[p]
\centering\includegraphics[width=4.66in,height=3.5in]{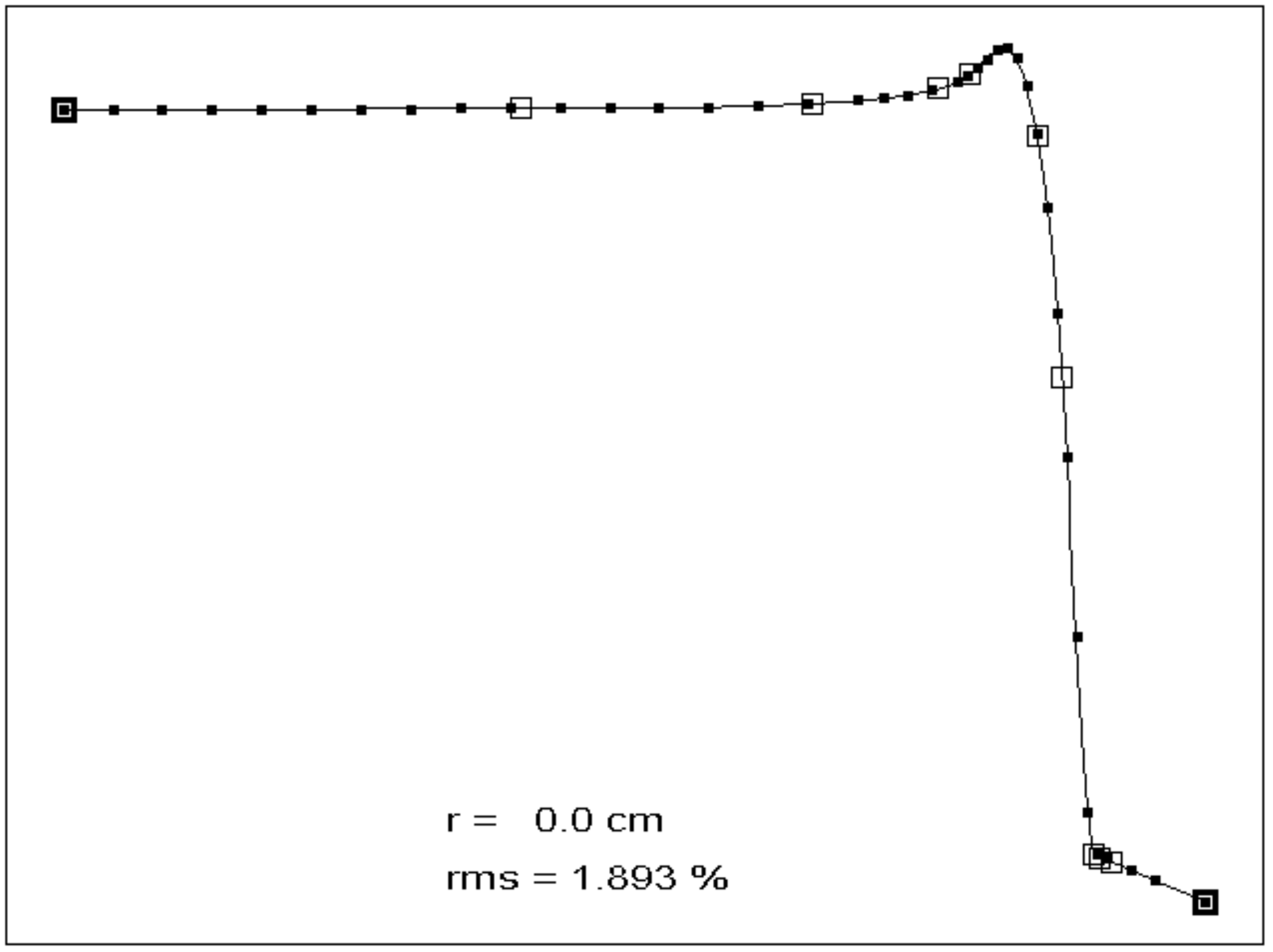}
\caption{Cubic spline fit to data of Fig.\,\ref{fig:CSparms}. Small squares: data points $y_i(x_i)$; light open squares: fitted spline points $v_i(u_i)$; bold open squares: spline end points, which only move vertically.\label{fig:CSfit}}
\end{figure}

Though that reduces the data set to a continuous function, that is not all we want to do. If the data set is noisy (random measurement errors in the $y_i$) we want to average out that noise. (When used on a noisy data set a cubic spline interpolant swings wildly in its effort to remain smooth, and is often a less reasonable choice than the common straight-line interpolant.) Consider for example Fig.\,\ref{fig:CSfit} where we have 39 data points. We wish to create a much smaller number (in this case 11) of `spline points' $v_i(u_i), i=1\ldots N$ such that {\em their} interpolant will constitute a least-squares fit to the original data set.\footnote{~Our spline points are often called `knots'.} The $v_i(u_i)$ are the hollow squares in Fig.\,\ref{fig:CSfit}.

Once we have picked starting values for the spline points, we can consider them (their horizontal and vertical coordinates) as the adjustable parameters of a nonlinear least-squares fit and optimize them with any standard technique \cite{nr}. As with any such fit, too few points will yield a poor fit while too many will yield a fit that chases random fluctuations in the data. A cubic spline fit converges quickly because A) if the initial points are chosen as described next, their starting values are already very good and B) the parameters are rather orthogonal. Moving one point usually has little effect beyond its nearest neighbors.

Unlike many fits (such as our own MD fit) where parameter definitions are buried deep in the mathematics, the adjustable parameters of a cubic spline fit are very like instances of the original data set. Fig.\,\ref{fig:CSfit} is instructive in another way: the fit to the peak is excellent even though there are no spline points at or near the peak.

In a full-fledged fit we allow any of the interior points to move in either the $x$ or $y$ directions. The end points, however, can only be allowed to move vertically, or the fit will instantly collapse to a single point. It will sometimes happen that points cross each other horizontally during optimization. In that case an `out of order' error is detected and the parameters reset to their previous value, which is usually a perfectly good fit. `Out of order' can be avoided completely by only allowing points to move vertically. Usually, that also yields a perfectly good fit. 

\subsection{Choosing Initial Spline Points}
To get the job done with the fewest points, we need to concentrate them near regions of greatest curvature. They can, of course, be positioned manually, but that introduces a subjective element and is in any case impractical for large jobs such as the present one. 

\begin{figure}[p]
\centering\includegraphics[width=3.33in,height=3.5in]{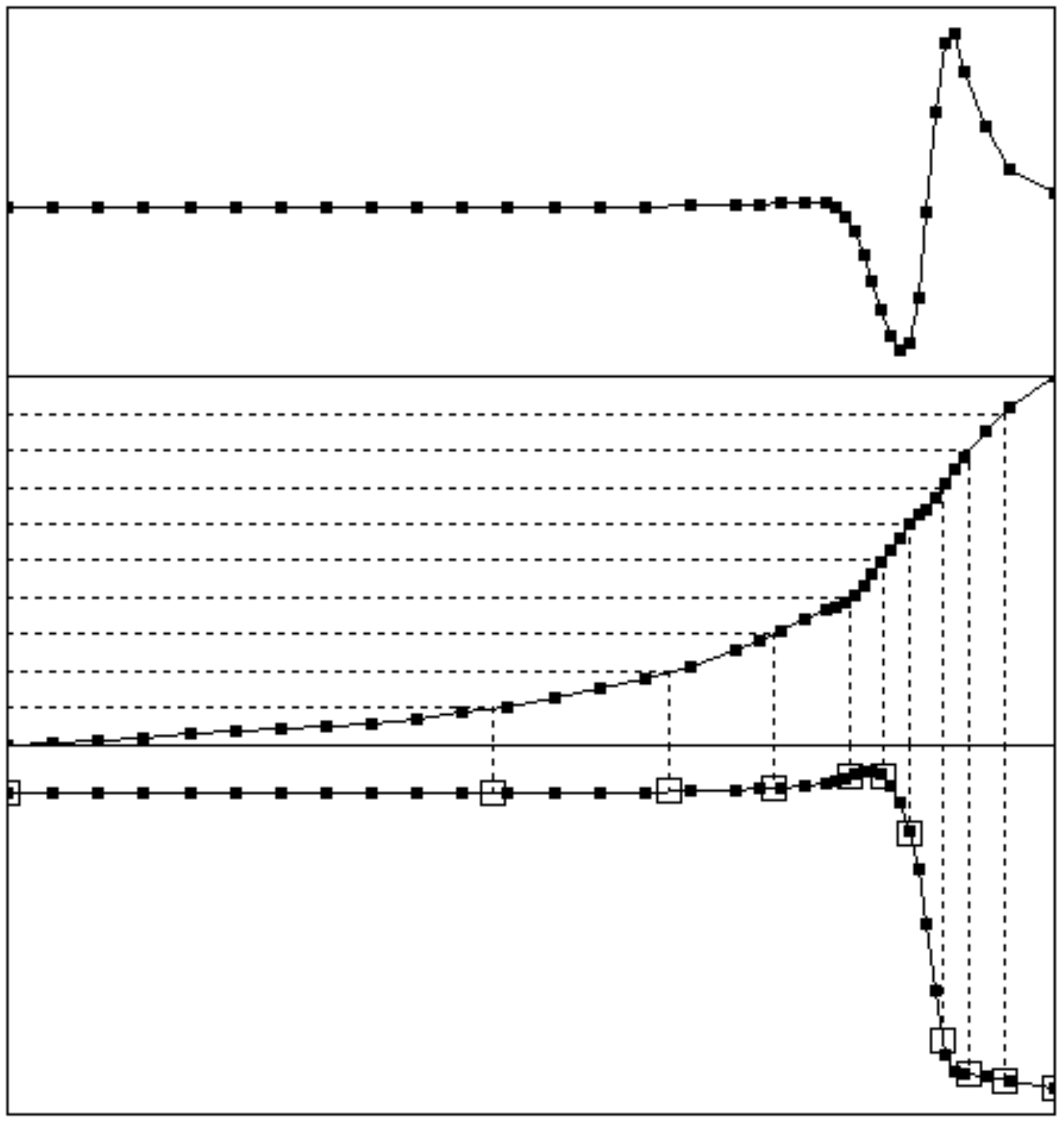}
\caption{How initial points are chosen. The smoothed second derivative of the function to be fit (top frame) is incorporated into a normalized density function (middle frame). The projection of equally spaced values onto the density function yields more initial points in regions of greater curvature (bottom frame).\label{fig:CSparms}}
\end{figure}

Our strategy is to create a normalized `density function' $F_i$ computed from the $y_i(x_i)$, onto which we can project points equally spaced between 0 and 1 to produce optimally spaced points (middle panel, Fig.\,\ref{fig:CSparms}). The numerical second derivative plays a key role since we seek regions of greatest curvature. It is
\begin{equation*}
s_i\;=\;2\,(\Delta_2y/\Delta_2x-\Delta_1y/\Delta_1x)/(\Delta_2x+\Delta_1x)
\end{equation*}
\begin{eqnarray*}
\Delta_1x\;=\; x_i-x_{i-1}\quad&,&\qquad\Delta_2x\;=\; x_{i+1}-x_i\\
\Delta_1y\;=\; y_i-y_{i-1}\quad&,&\qquad\Delta_2y\;=\; y_{i+1}-y_i
\end{eqnarray*}
Since numerical second derivatives are noisy, we smooth the $s_i$ by convolution with a suitable Gaussian to create an array $S_i$. The normalized density function $F$ is an integral approximated by a sum and is defined by
\begin{eqnarray*}
f_1&=&0\\
f_i&=&f_{i-1}\;+\;(x_i-x_{i-1})\;\bigl|S_i\bigr|^P\quad,\qquad i\,=\,2\ldots N
\end{eqnarray*}
followed by
\begin{equation*}
F_i\;=\;f_i/f_N\quad,\qquad i\,=\,1\ldots N
\end{equation*}
$P$ is an arbitrary power. Reducing it spaces the spline points more evenly. In Fig.\,\ref{fig:CSparms}, $P=0.3$\,.

\subsection{Bilinear Interpolation}\label{app:bilinear}
This standard technique is best explained by assuming that arbitrary values of $z$ are assigned to the four corners of a unit square in the $x,y$ plane spanning (0,0) to (1,1).\footnote{~Notation local to this subsection.} What is the best value of $z$ to assign to an arbitrary point $(x,y)$ on or inside the square?

The first thing to observe is that the answer is arbitrary. Interpolants are never unique, but represent a best (or simplest) guess. Suppose we seek a procedure that reduces to linear interpolation on any edge. The following formulas can be derived several ways. Perhaps the simplest is to seek a form A) with four free parameters, B) symmetric between $x$ and $y$ because there is no reason to prefer one over the other and C) of lowest possible order. That form is
\begin{equation*}
z\;=\;a+b\,x+c\,y+d\,x\,y
\end{equation*}
We now find $a\ldots d$ which force the $z$\,s at the corners to come out right:
\begin{eqnarray*}
a&=&z_{00}\\
b&=&z_{10}-z_{00}\\
c&=&z_{01}-z_{00}\\
d&=&z_{00}-z_{10}-z_{01}+z_{11}
\end{eqnarray*}
Generalization to an arbitrary element of a rectangular grid is straightforward: simply translate and scale $x$ and $y$.

\listoftables

\begin{table}[h]
\setlength{\tabcolsep}{5pt}
\begin{center}
\begin{tabular}{rrrrrrrrrrr}
\multicolumn{1}{c}{z/r}&           
\multicolumn{1}{c}{0.0}&           
\multicolumn{1}{c}{1.0}&           
\multicolumn{1}{c}{2.0}&           
\multicolumn{1}{c}{3.0}&           
\multicolumn{1}{c}{4.0}&           
\multicolumn{1}{c}{5.0}&           
\multicolumn{1}{c}{6.0}&           
\multicolumn{1}{c}{7.0}&           
\multicolumn{1}{c}{8.0}&           
\multicolumn{1}{c}{10.0}\\           
 1.533&       M& -0.4274&       M& -2.7179& -3.3276& -3.7252&       M&       M& -4.2708& -4.4897\\
 2.000&  0.4398& -0.3994& -1.9724& -2.5946& -3.2508& -3.6650& -3.8861& -4.1493& -4.2596&       M\\
 3.000&  0.4406& -0.3460& -1.8342& -2.3751&       M&       M& -3.7725&       M&       M& -4.4656\\
 4.000&  0.4438& -0.3267& -1.7684& -2.2437& -2.8720& -3.3366& -3.6435& -4.0060& -4.1789&       M\\
 5.000&  0.4444& -0.3116& -1.7244& -2.1716&       M&       M& -3.5098&       M&       M& -4.4177\\
 6.000&  0.4465& -0.2945& -1.6886& -2.1196& -2.6220& -3.0377& -3.3674& -3.7823& -4.0284&       M\\
 7.000&  0.4471& -0.2796& -1.6509&       M&       M&       M& -3.2603&       M&       M& -4.3311\\
 8.000&  0.4477& -0.2603& -1.6184& -2.0575& -2.5106& -2.8524& -3.1636& -3.5861& -3.8656&       M\\
 9.000&  0.4476& -0.2427& -1.5831&       M&       M& -2.8031& -3.0898&       M&       M& -4.2184\\
10.000&  0.4490& -0.2215& -1.5479& -2.0126& -2.4641& -2.7707& -3.0398& -3.4558& -3.7332&       M\\
11.000&  0.4506& -0.2025& -1.5114&       M&       M& -2.7500& -3.0099&       M&       M& -4.1351\\
12.000&  0.4504& -0.1787& -1.4714& -1.9694& -2.4401& -2.7402& -2.9951& -3.3940& -3.6628&       M\\
13.000&  0.4519& -0.1558& -1.4286&       M&       M& -2.7382& -2.9944&       M&       M& -4.0866\\
14.000&  0.4534& -0.1282& -1.3798& -1.9216& -2.4234& -2.7417& -3.0054& -3.3838& -3.6381&       M\\
15.000&  0.4566& -0.1000& -1.3279&       M&       M& -2.7546& -3.0280&       M&       M& -4.0727\\
16.000&  0.4618& -0.0690& -1.2657& -1.8632& -2.4153& -2.7726& -3.0690& -3.4454& -3.6837& -4.1002\\
17.000&  0.4801& -0.0300& -1.1926& -1.8213& -2.4084& -2.8035& -3.1125& -3.4600&       M& -4.1740\\
18.000&  0.4915&  0.0169& -1.1083& -1.7703& -2.3906& -2.8041& -3.1075& -3.4741& -3.7672& -4.2738\\
18.500&  0.5046&  0.0459& -1.0593& -1.7321& -2.3659& -2.7881& -3.1002& -3.4997&       M&       M\\
19.000&  0.5232&  0.0802& -1.0022& -1.6839& -2.3317& -2.7638& -3.0866& -3.5121& -3.8568& -4.3552\\
19.200&       M&       M&       M& -1.6602&       M&       M&       M& -3.5142&       M&       M\\
19.400&       M&       M&       M& -1.6348&       M&       M&       M& -3.5226&       M&       M\\
19.500&  0.5510&  0.1253& -0.9328&       M& -2.2810& -2.7112& -3.0350&       M& -3.9701&       M\\
19.600&       M&       M&       M& -1.6076&       M&       M&       M& -3.5468&       M&       M\\
19.800&       M&       M&       M& -1.5757&       M&       M&       M& -3.5921&       M&       M\\
20.000&  0.5993&  0.1898& -0.8433& -1.5368& -2.1845& -2.6303& -3.0308& -3.6658& -4.1091& -4.3940\\
20.200&  0.6261&  0.2266& -0.7987& -1.4900& -2.1520& -2.6292& -3.0882& -3.7601& -4.1539&       M\\
20.400&  0.6662&  0.2741& -0.7438& -1.4443& -2.1500& -2.6695& -3.1958& -3.8699& -4.1808&       M\\
20.600&  0.7169&  0.3306& -0.6859& -1.4146& -2.1974& -2.7669& -3.3555& -3.9737&       M&       M\\
20.800&  0.7625&  0.3765& -0.6353& -1.4176& -2.3110& -2.9298& -3.5583& -4.0469&       M&       M\\
21.000&  0.7747&  0.3835& -0.6096& -1.4712& -2.5038& -3.1666& -3.7683& -4.0937& -4.2238& -4.3924\\
21.200&  0.7208&  0.3199& -0.6364& -1.5919& -2.7751& -3.4568& -3.9238&       M&       M&       M\\
21.400&  0.5714&  0.1605& -0.7491& -1.8058& -3.1200& -3.7335& -3.9970&       M&       M&       M\\
21.500&       M&       M&       M&       M&       M&       M&       M& -4.1368&       M&       M\\
21.600&  0.3094& -0.1285& -0.9806& -2.1316& -3.4818& -3.8935& -4.0253&       M&       M&       M\\
21.800& -0.0945& -0.5719& -1.3676& -2.6019& -3.7381& -3.9465& -4.0363&       M&       M&       M\\
22.000& -0.6658& -1.2025& -1.9513& -3.1835& -3.8593& -3.9629& -4.0462&       M& -4.2504& -4.4152\\
22.200& -1.4468& -2.0547& -2.7337& -3.6274&       M&       M&       M&       M&       M&       M\\
22.400& -2.4194&       M& -3.4703& -3.7514&       M&       M&       M&       M&       M&       M\\
22.500&       M&       M&       M&       M& -3.8977& -3.9856&       M&       M&       M&       M\\
22.600& -3.3715& -3.5933& -3.6934& -3.7737&       M&       M&       M&       M&       M&       M\\
22.800&       M& -3.6613& -3.7298&       M&       M&       M&       M&       M&       M&       M\\
23.000&       M&       M& -3.7546& -3.8084& -3.9206& -4.0035& -4.0735&       M& -4.2678& -4.4219\\
23.500&       M&       M& -3.8008& -3.8496&       M&       M&       M&       M&       M&       M\\
24.000&       M&       M& -3.8456& -3.8863& -3.9752& -4.0544& -4.1141&       M&       M& -4.4374\\
25.000&       M&       M&       M&       M&       M&       M&       M&       M&       M& -4.4517\\
\multicolumn{1}{c}{z/r}&           
\multicolumn{1}{c}{0.0}&           
\multicolumn{1}{c}{1.0}&           
\multicolumn{1}{c}{2.0}&           
\multicolumn{1}{c}{3.0}&           
\multicolumn{1}{c}{4.0}&           
\multicolumn{1}{c}{5.0}&           
\multicolumn{1}{c}{6.0}&           
\multicolumn{1}{c}{7.0}&           
\multicolumn{1}{c}{8.0}&           
\multicolumn{1}{c}{10.0}\\           
\end{tabular}
\end{center}
\caption{Measured log$_{10}$(dose/(MeV/g/p)) at various depths $z$ (cm) and 10 distances $r$ (cm) from the beam centerline.\label{tbl:dmlg}}
\end{table}
\clearpage

\begin{table}[h]
\setlength{\tabcolsep}{5pt}
\begin{center}
\begin{tabular}{lrll}
\multicolumn{1}{c}{param}&           
\multicolumn{1}{c}{value}&           
\multicolumn{1}{c}{units}&           
\multicolumn{1}{c}{description}\\           
\noalign{\vspace{8pt}}
$p_1$&21.321&cm&R$_0$ (incident beam range or $d_{80}$)\\
$p_2$&0.01372&cm$^{-1}$&$\sigma_\mathrm{sem}=p_2\,R_0$ (range straggling + beam energy spread)\\ 
$p_3$&0.0312&cm$^{-1}$&rate of EM fluence loss (all nuclear interactions)\\
$p_4$&$-0.231$&&$z_\mathrm{n}$ coefficient of $p_3$\\
$p_5$&5.948&mm&incident beam $\sigma_x=\sqrt{A_2}$\\
$p_6$&31.55&mrad\quad&incident beam $\sigma_\theta=\sqrt{A_0}$\\
$p_7$&$-0.1462$&&emittance or $A_1$ driver (see text)\\
\noalign{\vspace{8pt}}
$p_8$&0.01269&&$D_\mathrm{el}$ multiplier\\
$p_9$&0.1043&&$r_\mathrm{el}=$ 1/e parameter, elastic scattering ($r_\mathrm{n}$ units)\\
$p_{10}$&$-0.3751$&&$dp_9/dr_\mathrm{n}$\\
$p_{11}$&$-0.1970$&&$dp_9/dr_\mathrm{n}^2$\\
$p_{12}$&0.06406&&$d/dr_\mathrm{n}^2$ of Bragg peak in elastic scattering\\
$p_{13}$&0.2562&&$d/dr_\mathrm{n}$ of Bragg peak width in elastic scattering\\ 
\noalign{\vspace{8pt}}
$p_{14}$&0.004172&&$D_\mathrm{qe}$ multiplier\\
$p_{15}$&3.127&&$d/dr_\mathrm{n}$ of $D_\mathrm{qe}$ multiplier\\
$p_{16}$&0.1812&&mean $z_\mathrm{n}$ of QE bump\\
$p_{17}$&1.7042&&$d/dr_\mathrm{n}$ of mean $z_\mathrm{n}$ of QE bump\\
$p_{18}$&0.5357&&halfwidth ($z_\mathrm{n}$ units) of QE bump\\
$p_{19}$&$-0.5460$&&$d/dr_\mathrm{n}$ of $p_{18}$\\
$p_{20}$&0.1972&&$r_\mathrm{qe}=$ 1/e parameter, QE bump ($r_\mathrm{n}$ units)\\
$p_{21}$&$-0.4402$&&$d/dr_\mathrm{n}$ of $p_{20}$\\
$p_{22}$&0.2082&&$d/dr_\mathrm{n}^2$ of $p_{20}$\\
\noalign{\vspace{8pt}}
$p_{23}$&0.0008981&&$D_\mathrm{au}$ multiplier\\
$p_{24}$&0.8092&&$r_\mathrm{au}=$ 1/e parameter, $D_\mathrm{au}$ ($r_\mathrm{n}$ units)\\
$p_{25}$&$-0.1431$&&$d/dr_\mathrm{n}^2$ of $p_{24}$
\end{tabular}
\end{center}
\caption{Parameter values corresponding to MD fit shown in Figs.\,\ref{fig:dflgMD} and \ref{fig:dlinMD}.\label{tbl:parms}}
\end{table}
\clearpage

\begin{table}[p]
\setlength{\tabcolsep}{5pt}
\begin{center}
\begin{tabular}{lclllllll}
\multicolumn{1}{c}{1$^\mathrm{st}$ author}&           
\multicolumn{1}{c}{year}&           
\multicolumn{1}{c}{ref.}&           
\multicolumn{1}{c}{institution}&           
\multicolumn{1}{c}{measured}&           
\multicolumn{1}{c}{core}&           
\multicolumn{1}{c}{halo}&           
\multicolumn{1}{c}{TPS}&
\multicolumn{1}{c}{MC}\\           
\noalign{\vspace{2pt}}
\noalign{\begin{center}Measurements\end{center}}
\noalign{\vspace{6pt}}
Pedroni&2005&\cite{pedroniPencil}&PSI&hollow frames&G&G&in-house\\
Sawakuchi&2010&\cite{Sawakuchi2010}&MDACC&transverse, FSF&2G\\
Clasie&2012&\cite{Clasie2012}&MGH&circles, corr BPC&G&G&in-house&G4\\
Anand&2012&\cite{Anand2012}&MDACC&transverse, BPC\\
\noalign{\vspace{2pt}}
\noalign{\begin{center}Monte Carlo Studies\end{center}}
\noalign{\vspace{6pt}}
Sawakuchi&2010&\cite{SawakuchiMC2010}&MDACC&&&&&MCNPX\\
Sawakuchi&2010&\cite{SawakuchiMCNPX}&MDACC&&&&&MCNPX\\
Grevillot&2011&\cite{GrevillotPhD2011}&Lyon (PhD)&$\sigma_x$, FSF&&&XiO&GATE/G4\\
Peeler&2012&\cite{Peeler2012}&MDACC&&&&&MCNPX\\
\noalign{\vspace{2pt}}
\noalign{\begin{center}Parameterizations\end{center}}
\noalign{\vspace{8pt}}
Soukup&2005&\cite{soukup}&T\"ubingen&&G&G&XiO&G4, VMC\\
Zhang&2011&\cite{Zhang2011}&MDACC&corr BPC&&&Eclipse\\
Zhu&2013&\cite{Zhu2013}&MDACC&corr BPC&2G&G&Eclipse&MCNPX\\
\noalign{\vspace{2pt}}
\noalign{\begin{center}Alternative Transverse Profiles\end{center}}
\noalign{\vspace{6pt}}
Fuchs&2012&\cite{Fuchs2012}&Vienna&&G&Voigt&in-house&GATE/G4\\
Knutson&2012&\cite{Knutson}&LSU (M.A.)&&G&G+CL&&MCNPX\\
Li&2012&\cite{Li2012}&MDACC&&G&G+CL&Eclipse\\
\end{tabular}
\end{center}
\caption{Summary of literature reviewed here. Most papers span several categories. G = Gaussian, CL = Cauchy-Lorentz, FSF = field size factor, BPC = Bragg peak chamber, G4 = Geant4, TPS = treatment planning system.\label{tbl:lit}}
\end{table}

\begin{table}[p]
\setlength{\tabcolsep}{5pt}
\begin{center}
\begin{tabular}{lrrrrrrrrrr}
\multicolumn{1}{c}{material}&           
\multicolumn{1}{c}{$T_1$}&           
\multicolumn{1}{c}{$R_1$}&           
\multicolumn{1}{c}{$z$}&           
\multicolumn{1}{c}{$\sigma_\mathrm{G4}$}&           
\multicolumn{1}{c}{$\sigma_\mathrm{G4C}$}&           
\multicolumn{1}{c}{$\sigma(R_1)$}&           
\multicolumn{1}{c}{$z/R_1$}&           
\multicolumn{1}{c}{$f_\mathrm{PK}$}&           
\multicolumn{1}{c}{$\sigma_\mathrm{FE}$}&           
\multicolumn{1}{c}{$\sigma_\mathrm{G4C}/\sigma_\mathrm{FE}$}\\          
\multicolumn{1}{c}{}&           
\multicolumn{1}{c}{MeV}&           
\multicolumn{1}{c}{cm}&           
\multicolumn{1}{c}{cm}&           
\multicolumn{1}{c}{cm}&           
\multicolumn{1}{c}{cm}&           
\multicolumn{1}{c}{cm}&           
\multicolumn{1}{c}{}&           
\multicolumn{1}{c}{}&           
\multicolumn{1}{c}{cm}&           
\multicolumn{1}{c}{}\\          
\noalign{\vspace{7pt}}
 PMMA &   210.6&    24.3&     0.0&   0.314&        &   0.510&   0.000&         &        &       \\
      &        &        &    18.6&   0.468&   0.264&        &   0.766&    0.622&   0.317&   0.83\\
      &        &        &    22.6&   0.551&   0.393&        &   0.931&    0.874&   0.445&   0.88\\
\noalign{\vspace{3pt}}
 water&   230.0&    33.2&     0.0&   0.300&        &   0.747&   0.000&         &        &       \\
      &        &        &    10.0&   0.310&   0.078&        &   0.301&    0.141&   0.105&   0.74\\
      &        &        &    30.0&   0.620&   0.542&        &   0.904&    0.829&   0.619&   0.88\\
      &        &        &    32.0&   0.690&   0.621&        &   0.964&    0.932&   0.696&   0.89\\
\end{tabular}
\end{center}
\caption{Worksheet for Geant4\,/\,Fermi-Eyges comparison. $\sigma_\mathrm{G4}$ was obtained from \cite{Grevillot2010}. For PMMA it was estimated from the 61\% level in Fig.\,10 assuming the $x$-axis `mm' means `cm', and for water, from the MS column of Table\,5.
\label{tbl:G4oFE}}
\end{table}
\clearpage
\listoffigures
\clearpage



\clearpage

\clearpage

\clearpage

\clearpage

\clearpage

\clearpage

\clearpage

\clearpage

\clearpage

\clearpage

\clearpage


\clearpage


\end{document}